\documentclass[a4paper,11pt]{article}
\pdfoutput=1 

\usepackage{jcappub} 

\usepackage[T1]{fontenc} 

\usepackage[dvipsnames]{xcolor}

\usepackage{placeins}

\usepackage[nameinlink]{cleveref}
\crefname{figure}{figure}{figures}
\crefname{section}{section}{sections}
\crefname{equation}{equation}{}

\usepackage{soul}
\usepackage{lineno}

\def\setsymbol#1#2{\expandafter\def\csname #1\endcsname{#2}}
\def\getsymbol#1{\csname #1\endcsname}

\newbox\tablebox    \newdimen\tablewidth
\def\leaderfil{\leaders\hbox to 5pt{\hss.\hss}\hfil}
\def\tablenote#1 #2\par{\begingroup \parindent=0.8em
    \abovedisplayshortskip=0pt\belowdisplayshortskip=0pt
    \noindent
    $$\hss\vbox{\hsize\tablewidth \hangindent=\parindent \hangafter=1 \noindent
    \hbox to \parindent{$^#1$\hss}\strut#2\strut\par}\hss$$
    \endgroup}

\def\L2{\ifmmode L_2\else $L_2$\fi}

\def\DeltaT{\ifmmode \Delta T\else $\Delta T$\fi}
\def\deltat{\ifmmode \Delta t\else $\Delta t$\fi}
\def\fknee{\ifmmode f_{\rm knee}\else $f_{\rm knee}$\fi}
\def\Fmax{\ifmmode F_{\rm max}\else $F_{\rm max}$\fi}
\def\solar{\ifmmode{\rm M}_{\mathord\odot}\else${\rm M}_{\mathord\odot}$\fi}
\def\Msolar{\ifmmode{\rm M}_{\mathord\odot}\else${\rm M}_{\mathord\odot}$\fi}
\def\Lsolar{\ifmmode{\rm L}_{\mathord\odot}\else${\rm L}_{\mathord\odot}$\fi}
\def\inv{\ifmmode^{-1}\else$^{-1}$\fi}
\def\mo{\ifmmode^{-1}\else$^{-1}$\fi}
\def\sup#1{\ifmmode ^{\rm #1}\else $^{\rm #1}$\fi}
\def\expo#1{\ifmmode \times 10^{#1}\else $\times 10^{#1}$\fi}
\def\,{\thinspace}
\def\lsim{\mathrel{\raise .4ex\hbox{\rlap{$<$}\lower 1.2ex\hbox{$\sim$}}}}
\def\gsim{\mathrel{\raise .4ex\hbox{\rlap{$>$}\lower 1.2ex\hbox{$\sim$}}}}

\def\simprop{\mathrel{\raise .4ex\hbox{\rlap{$\propto$}\lower 1.2ex\hbox{$\sim$}}}}
\def\deg{\ifmmode^\circ\else$^\circ$\fi}
\def\pdeg{\ifmmode $\setbox0=\hbox{$^{\circ}$}\rlap{\hskip.11\wd0 .}$^{\circ}
          \else \setbox0=\hbox{$^{\circ}$}\rlap{\hskip.11\wd0 .}$^{\circ}$\fi}
\def\arcs{\ifmmode {^{\scriptstyle\prime\prime}}
          \else $^{\scriptstyle\prime\prime}$\fi}
\def\arcm{\ifmmode {^{\scriptstyle\prime}}
          \else $^{\scriptstyle\prime}$\fi}
\newdimen\sa  \newdimen\sb
\def\parcs{\sa=.07em \sb=.03em
     \ifmmode \hbox{\rlap{.}}^{\scriptstyle\prime\kern -\sb\prime}\hbox{\kern -\sa}
     \else \rlap{.}$^{\scriptstyle\prime\kern -\sb\prime}$\kern -\sa\fi}
\def\parcm{\sa=.08em \sb=.03em
     \ifmmode \hbox{\rlap{.}\kern\sa}^{\scriptstyle\prime}\hbox{\kern-\sb}
     \else \rlap{.}\kern\sa$^{\scriptstyle\prime}$\kern-\sb\fi}
\def\ra[#1 #2 #3.#4]{#1\sup{h}#2\sup{m}#3\sup{s}\llap.#4}
\def\dec[#1 #2 #3.#4]{#1\deg#2\arcm#3\arcs\llap.#4}
\def\deco[#1 #2 #3]{#1\deg#2\arcm#3\arcs}
\def\rra[#1 #2]{#1\sup{h}#2\sup{m}}

\def\dots{\relax\ifmmode \ldots\else $\ldots$\fi}
\def\WHzsr{\ifmmode $W\,Hz\mo\,sr\mo$\else W\,Hz\mo\,sr\mo\fi}
\def\mHz{\ifmmode $\,mHz$\else \,mHz\fi}
\def\GHz{\ifmmode $\,GHz$\else \,GHz\fi}
\def\mKs{\ifmmode $\,mK\,s$^{1/2}\else \,mK\,s$^{1/2}$\fi}
\def\muKs{\ifmmode \,\mu$K\,s$^{1/2}\else \,$\mu$K\,s$^{1/2}$\fi}
\def\muKRJs{\ifmmode \,\mu$K$_{\rm RJ}$\,s$^{1/2}\else \,$\mu$K$_{\rm RJ}$\,s$^{1/2}$\fi}
\def\muKHz{\ifmmode \,\mu$K\,Hz$^{-1/2}\else \,$\mu$K\,Hz$^{-1/2}$\fi}
\def\MJysr{\ifmmode \,$MJy\,sr\mo$\else \,MJy\,sr\mo\fi}
\def\MJysrmK{\ifmmode \,$MJy\,sr\mo$\,mK$_{\rm CMB}\mo\else \,MJy\,sr\mo\,mK$_{\rm CMB}\mo$\fi}
\def\microns{\ifmmode \,\mu$m$\else \,$\mu$m\fi}

\def\muK{\ifmmode \,\mu$K$\else \,$\mu$\hbox{K}\fi}
\def\microK{\ifmmode \,\mu$K$\else \,$\mu$\hbox{K}\fi}
\def\muW{\ifmmode \,\mu$W$\else \,$\mu$\hbox{W}\fi}
\def\kms{\ifmmode $\,km\,s$^{-1}\else \,km\,s$^{-1}$\fi}
\def\kmsMpc{\ifmmode $\,\kms\,Mpc\mo$\else \,\kms\,Mpc\mo\fi}
\providecommand{\sorthelp}[1]{}

\def\lb{\textit{LiteBIRD}}
\def\pl{\textit{Planck}}

\def\creff@jnl#1{{\rm#1\/}}

\def\aj{\creff@jnl{AJ}}                  
\def\araa{\creff@jnl{ARA\&A}}            
\def\apj{\creff@jnl{ApJ}}                
\def\apjl{\creff@jnl{ApJ}}               
\def\apjs{\creff@jnl{ApJS}}              
\def\ao{\creff@jnl{Appl.Optics}}         
\def\apss{\creff@jnl{Ap\&SS}}            
\def\aap{\creff@jnl{A\&A}}               
\def\aapr{\creff@jnl{A\&A~Rev.}}         
\def\aaps{\creff@jnl{A\&AS}}             
\def\azh{\creff@jnl{AZh}}                        
\def\baas{\creff@jnl{BAAS}}              
\def\jcap{\creff@jnl{JCAP}}              
\def\jrasc{\creff@jnl{JRASC}}            
\def\memras{\creff@jnl{MmRAS}}           
\def\mnras{\creff@jnl{MNRAS}}            
\def\pra{\creff@jnl{Phys.Rev.A}}         
\def\prb{\creff@jnl{Phys.Rev.B}}         
\def\prc{\creff@jnl{Phys.Rev.C}}         
\def\prd{\creff@jnl{Phys.Rev.D}}         
\def\prl{\creff@jnl{Phys.Rev.Lett}}      
\def\physrep{\creff@jnl{Phys.Rep.}}      
\def\pasp{\creff@jnl{PASP}}              
\def\pasj{\creff@jnl{PASJ}}              
\def\qjras{\creff@jnl{QJRAS}}            
\def\skytel{\creff@jnl{S\&T}}            
\def\solphys{\creff@jnl{Solar~Phys.}}    
\def\sovast{\creff@jnl{Soviet~Ast.}}     
 \def\ssr{\creff@jnl{Space~Sci.Rev.}}    
\def\zap{\creff@jnl{ZAp}}                
\def\nat{\creff@jnl{Nature}}             

\title{\boldmath LiteBIRD Science Goals and Forecasts. Mapping the Hot Gas in the Universe}

\author[1]{M.\,Remazeilles,}
\author[2]{M.\,Douspis,}
\author[3,4]{J.\,A.\,Rubiño-Martín,}
\author[5]{A.\,J.\,Banday,}
\author[6]{J.\,Chluba,}
\author[7,8]{P.\,de\,Bernardis,}
\author[7,8]{M.\,De\,Petris,}
\author[3]{C.\,Hernández-Monteagudo,}
\author[9]{G.\,Luzzi,}
\author[10]{J.\,Macias-Perez,}
% ...
\affiliation[1]{Instituto de Fisica de Cantabria (IFCA, CSIC-UC), Avenida los Castros SN, 39005, Santander, Spain}
\affiliation[2]{Université Paris-Saclay, CNRS, Institut d’Astrophysique Spatiale, 91405, Orsay, France}
\affiliation[3]{Instituto de Astrofísica de Canarias, E-38200 La Laguna, Tenerife, Canary Islands, Spain}
\affiliation[4]{Departamento de Astrofísica, Universidad de La Laguna (ULL), E-38206, La Laguna, Tenerife, Spain}
\affiliation[5]{IRAP, Université de Toulouse, CNRS, CNES, UPS, Toulouse, France}
\affiliation[6]{Jodrell Bank Centre for Astrophysics, Alan Turing Building, Department of Physics and Astronomy, School of Natural Sciences, The University of Manchester, Oxford Road, Manchester M13 9PL, UK}
\affiliation[7]{Dipartimento di Fisica, Università La Sapienza, P. le A. Moro 2, Roma, Italy}
\affiliation[8]{INFN Sezione di Roma, P.le A. Moro 2, 00185 Roma, Italy}
\affiliation[9]{Space Science Data Center, Italian Space Agency, via del Politecnico, 00133, Roma, Italy}
\affiliation[10]{Université Grenoble Alpes, CNRS, LPSC-IN2P3, 53, avenue des Martyrs, 38000 Grenoble, France}
% ...

\emailAdd{remazeilles@ifca.unican.es}

\abstract{We assess the capabilities of the \lb\ mission to map the hot gas distribution in the Universe through the thermal Sunyaev-Zeldovich (SZ) effect. Our analysis relies on comprehensive simulations incorporating various sources of Galactic and extragalactic foreground emission, while accounting for the specific instrumental characteristics of the \lb\ mission, such as detector sensitivities, frequency-dependent beam convolution, inhomogeneous sky scanning, and $1/f$ noise. We implement a tailored component-separation pipeline to map the thermal SZ Compton $y$-parameter over 98\,\% of the sky. Despite lower angular resolution for galaxy cluster science, \lb\ provides full-sky coverage and, compared to the \pl\ satellite, enhanced sensitivity, as well as more frequency bands to enable the construction of an all-sky thermal SZ $y$-map, with reduced foreground contamination at large and intermediate angular scales. By combining \lb\ and \pl\ channels in the component-separation pipeline, we also obtain an optimal $y$-map that leverages the advantages of both experiments, with the higher angular resolution of the \pl\ channels enabling the recovery of compact clusters beyond the \lb\ beam limitations, and the numerous sensitive \lb\ channels further mitigating foregrounds. The added value of \lb\ is highlighted through the examination of maps, power spectra, and one-point statistics of the various sky components. After component separation, the $1/f$ noise from \lb's intensity channels is effectively mitigated below the level of the thermal SZ signal at all multipoles. Cosmological constraints on $S_8 = \sigma_8 \left(\Omega_{\rm m}/0.3\right)^{0.5}$ obtained from the \lb-\pl\ combined $y$-map power spectrum exhibits a 15\,\% reduction in uncertainty compared to constraints derived from \pl\ alone. This improvement can be attributed to the increased portion of uncontaminated sky available in the \lb-\pl\ combined $y$-map.}

\begin{document}
\maketitle
\flushbottom

\section{Introduction}
\label{sec:intro}

The warm-hot intergalactic medium (WHIM), which covers vast regions of space between galaxies and galaxy clusters, contains highly ionized gas with temperatures ranging from around $10^5$ to $10^7$\,K, while within the gravitational potential wells of galaxy clusters, the collapsed gas can reach even higher temperatures of $10^7$--$10^8$\,K. As cosmic microwave background (CMB) photons travel through this hot, ionized gas, they are upscattered to higher energies by inverse Compton scattering with energetic free electrons.
This causes a distinctive spectral distortion to the CMB blackbody spectrum, with varying amplitude depending on the line-of-sight direction. This phenomenon, known as the thermal Sunyaev-Zeldovich (SZ) effect and originally theorized over half a century ago \citep{Zeldovich1969,Sunyaev1972}, is now routinely observed by CMB experiments due to its characteristic spectral signature.

Thousands of galaxy clusters, with masses ranging from $10^{14}$ to $10^{15}\,M_\odot$ and redshifts spanning from $z\simeq 0$ to $z\simeq 1.5$, have been detected by means of the thermal SZ effect in submillimetre sky observations \citep{planck2014-a36,ACT-SZ-Cat2021,Melin2021,SPT-SZ-Cat2015,Bleem2020}. Upcoming ground-based CMB experiments \citep{SO2019,CMBS4-2019} are expected to extend current cluster catalogues by more than one order of magnitude, revealing clusters with masses as low as $M_{500}\simeq 10^{14}\,M_\odot$ out to redshifts $z\gtrsim 2$. 

Aside from cluster catalogues, the distinct frequency dependence of the thermal SZ effect has enabled the extraction and mapping of the entire hot gas distribution over the sky, including diffuse unbound gas between clusters, from multi-frequency observations through dedicated component separation methods \citep{Remazeilles2011a,Remazeilles2013,Hurier2013,Bourdin2020}. This has led to Compton parameter $y$-maps that probe the full thermal SZ emission \citep{planck2014-a28,Aghanim2019,Madhavacheril2020,Tanimura2022,Bleem2022,Chandran2023,McCarthy2024,Coulton2024}, in contrast to cluster catalogues that exclusively capture the emission from well-resolved, massive clusters. Having $y$-maps that encompass the entire thermal SZ emission, including the diffuse filamentary gas structures between clusters, is especially valuable for endeavours like the search for missing baryons \citep{deGraaff2019} and the tomographic cross-correlations with other tracers of large-scale structure \citep{Vikram2017,Makiya2018,Pandey2019,Koukoufilippas2020,Chiang2020,Yan2021}.

As a redshift-independent probe of large-scale structure, the thermal SZ effect has long been recognized as an important cosmological observable \citep{Birkinshaw1999,Komatsu1999,Carlstrom2002,Komatsu2002,Rubino-Martin2003,Mroczkowski2019}, complementing primary CMB measurements by providing independent constraints on some cosmological parameters. Thermal SZ data from ESA's \pl\ satellite mission \citep{planck2016-l01} utilizing the statistics of cluster number counts from the \pl\ cluster catalogue \citep{planck2013-p15,planck2014-a30} and the one-point, two-point and three-point statistics of the Compton $y$-map \citep{planck2013-p05b,planck2014-a28,Bolliet2018} have yielded the first low-redshift constraints on parameters such as the matter density ($\Omega_{\rm m}$) and the amplitude of dark matter fluctuations ($\sigma_8$). This has brought to light the first indications of tension regarding $\sigma_8$ when compared to measurements from high-redshift CMB observations \citep{planck2014-a15}. 

\lb, the Lite (Light) satellite for the study of $B$-mode polarization and Inflation from cosmic background Radiation Detection \citep{Litebird_ptep2023}, is a fourth-generation space mission dedicated to CMB observations, which was selected by the Japan Aerospace Exploration Agency (JAXA) in May 2019 for a planned launch in the Japanese Fiscal Year 2032. 
It will observe the full sky with three telescopes in 15 frequency bands between $40$ and $402$\,GHz, offering an extensive data set in both temperature and polarization. While \lb's primary scientific objective is to detect the primordial gravitational waves predicted by cosmic inflation through large-scale CMB $B$-mode polarization observations, its unprecedented sensitivity across the full sky and large number of frequency channels can benefit thermal SZ science, as we aim to demonstrate in this study.

This work is part of a series of papers that present the science
achievable by the \lb\ space mission, expanding on the overview
published in ref.~\cite{Litebird_ptep2023}. In particular, this
work focuses on the thermal SZ effect.
Despite lower angular resolution than \pl, \lb\ will provide full-sky temperature maps across a substantially larger number of frequency channels than \pl, with also greater sensitivity to improve foreground removal and component separation, thus positioning it to deliver the next all-sky map of the thermal SZ effect. 

In this study, we demonstrate the capability of \lb\ in mapping the thermal SZ Compton $y$-parameter over the celestial sphere with higher fidelity than \pl. In addition, we propose to combine both \pl\ and \lb\ data sets for thermal SZ $y$-map reconstruction, leveraging the advantages of both experiments. We show how the inclusion of the numerous, sensitive \lb\ channels along with the \pl\ channels over a wide frequency range results in enhanced mitigation of the foreground contamination in the reconstructed $y$-map, while \pl\ channels provide the required angular resolution to reconstruct the thermal SZ signal beyond \lb's beam limitations. Moreover, we illustrate how the \lb-\pl\ combined $y$-map substantially reduces the uncertainty associated with the inferred $\sigma_8$ parameter when compared to using solely \pl\ data. Finally, we explore \lb's potential to detect the subtle thermal SZ signal from patchy reionisation by cross-correlating the \lb\ $y$-map with an inhomogeneous optical depth map derived from a future high-resolution ground-based CMB survey such as CMB-S4 \citep{CMBS4-2019}.

This paper is organized as follows. In \cref{sec:sims}, we present the sky simulations of both \lb\ and \pl\ data sets that we use in our analysis. In \cref{sec:nilc}, we describe the component separation method that we implement for thermal SZ $y$-map reconstruction. Our results from the \lb\ and joint \lb-\pl\ analyses are discussed in \cref{sec:results}. \Cref{subsec:ymaps} assesses the quality of the $y$-maps, while \cref{subsec:spectra,subsec:pdfs} evaluate the power spectrum and one-point statistics of the $y$-maps and their residuals, \cref{subsec:noise1overf} discusses the impact of correlated $1/f$ noise, \cref{subsec:cosmo} focuses on inferring cosmological parameter constraints, and finally, \cref{subsec:reionisation} forecasts the detection of the thermal SZ signal from patchy reionisation. We conclude in \cref{sec:conclusion}.

\section{Sky temperature simulations}\label{sec:sims} 

Various codes are available for simulating sky observations at submillimetre wavelengths, such as WebSky \citep{Stein2020} for extragalactic components, the Python Sky Model \citep{Thorne2017} for Galactic emissions, and the Planck Sky Model \citep{Delabrouille2013,planck2014-a14}, which comprehensively covers both Galactic and extragalactic emission. Each of these tools has its own advantages and limitations. To compare the relative performance of \lb\ and \pl, considering the angular resolution of both experiments, we find it suitable to use the Planck Sky Model (PSM) to generate comprehensive sky simulations that incorporate various Galactic and extragalactic components of emission. Realistic template maps and spectral models from the PSM are used to scale the different components of emission across the frequency channels of both \lb\ ($40$--$402$\,GHz) and \pl\ ($30$--$857$\,GHz). The simulated sky maps at each frequency, resulting from the superposition of all the components, are further convolved with Gaussian beams, with full-width-at-half-maximum (FWHM) values as provided for each channel by the instrument specifications of \lb\ \citep{Litebird_ptep2023} and \pl\  \citep{planck2016-l01}. 

Gaussian white noise maps are generated for each frequency channel using the quoted \pl\ 2018 sensitivities in temperature \citep{planck2016-l01} and the \lb\ Instrument Model (IMO)  sensitivities in polarization \citep{Litebird_ptep2023}, that we rescaled by a factor of $\sqrt{2}$ for temperature. These noise maps are added to the sky maps at each frequency to complete the simulation. 

In addition, independent realistic noise maps are simulated to accommodate the inhomogeneous scan strategy of \lb\ and account for the expected $1/f$ noise in temperature, as discussed in \cref{subsec:instruments}. These extra simulations are utilized in \cref{subsec:noise1overf} to study the impact of \lb's $1/f$ noise on the reconstructed thermal SZ $y$-map.

The simulated maps have a \texttt{HEALPix}\footnote{\url{https://healpix.jpl.nasa.gov/}} format \citep{gorski2005} with pixel resolution of $N_{\rm side}=512$ for the \lb\ channels and $N_{\rm side}=2048$ for the \pl\ channels. \Cref{subsec:extragal,subsec:gal} provide detailed descriptions of the models employed to simulate the different sky components.

\subsection{Extragalactic components}\label{subsec:extragal}

\begin{figure}[tbp]
\centering 
\includegraphics[width=0.5\textwidth,clip]{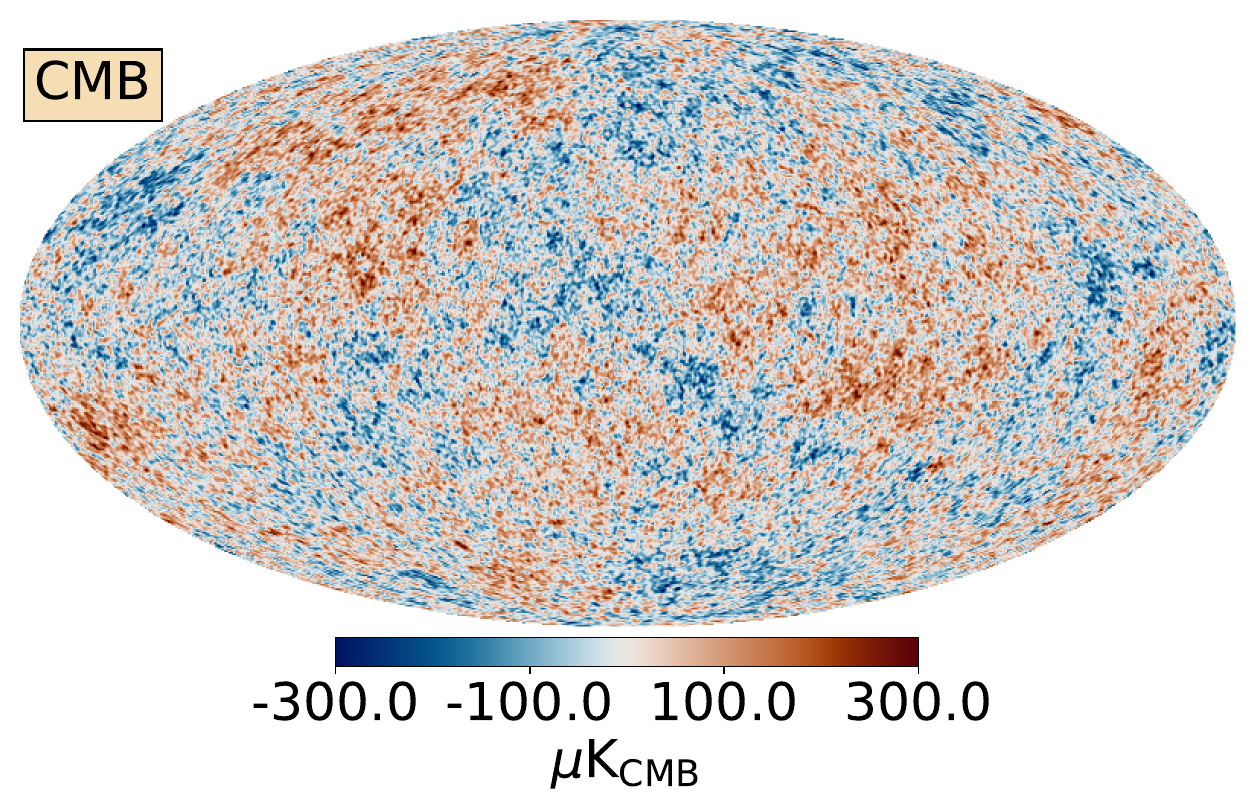}~
\includegraphics[width=0.5\textwidth,clip]{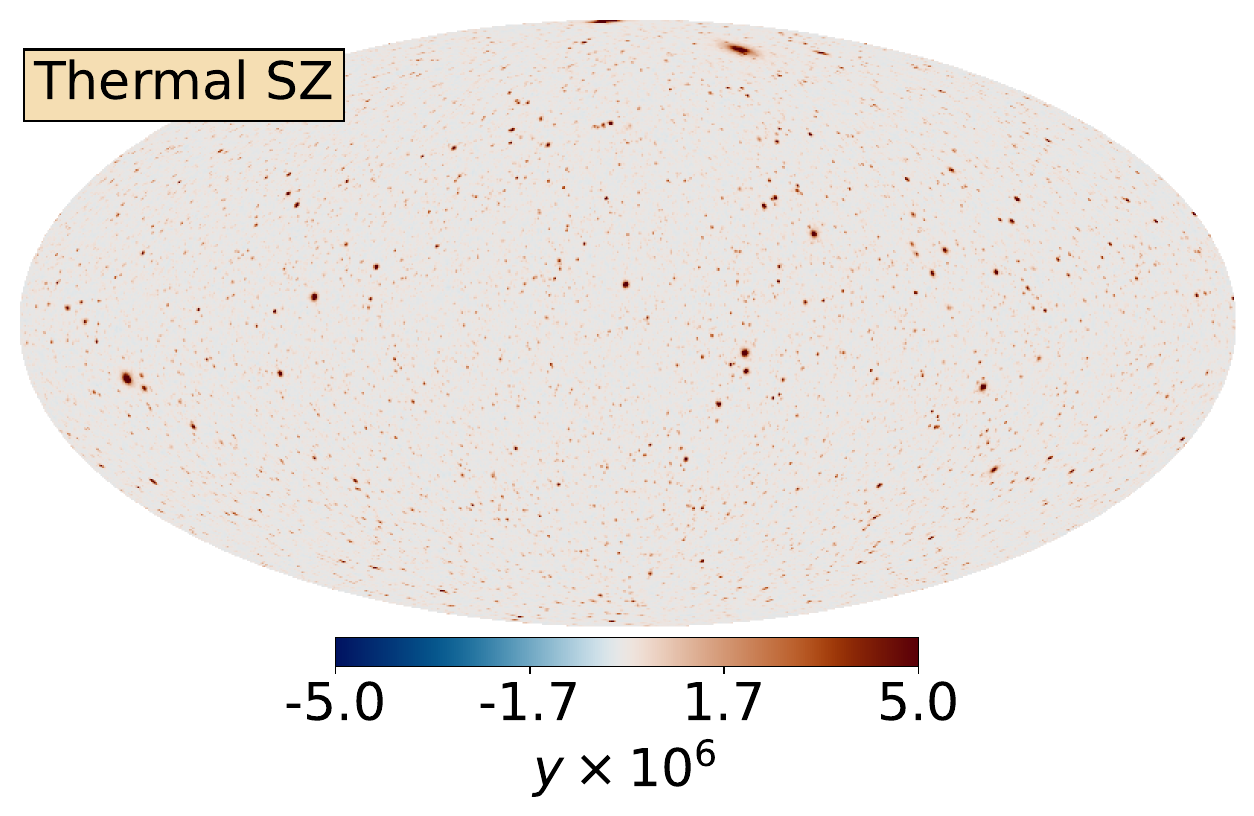}
\hfill
\includegraphics[width=0.5\textwidth,clip]{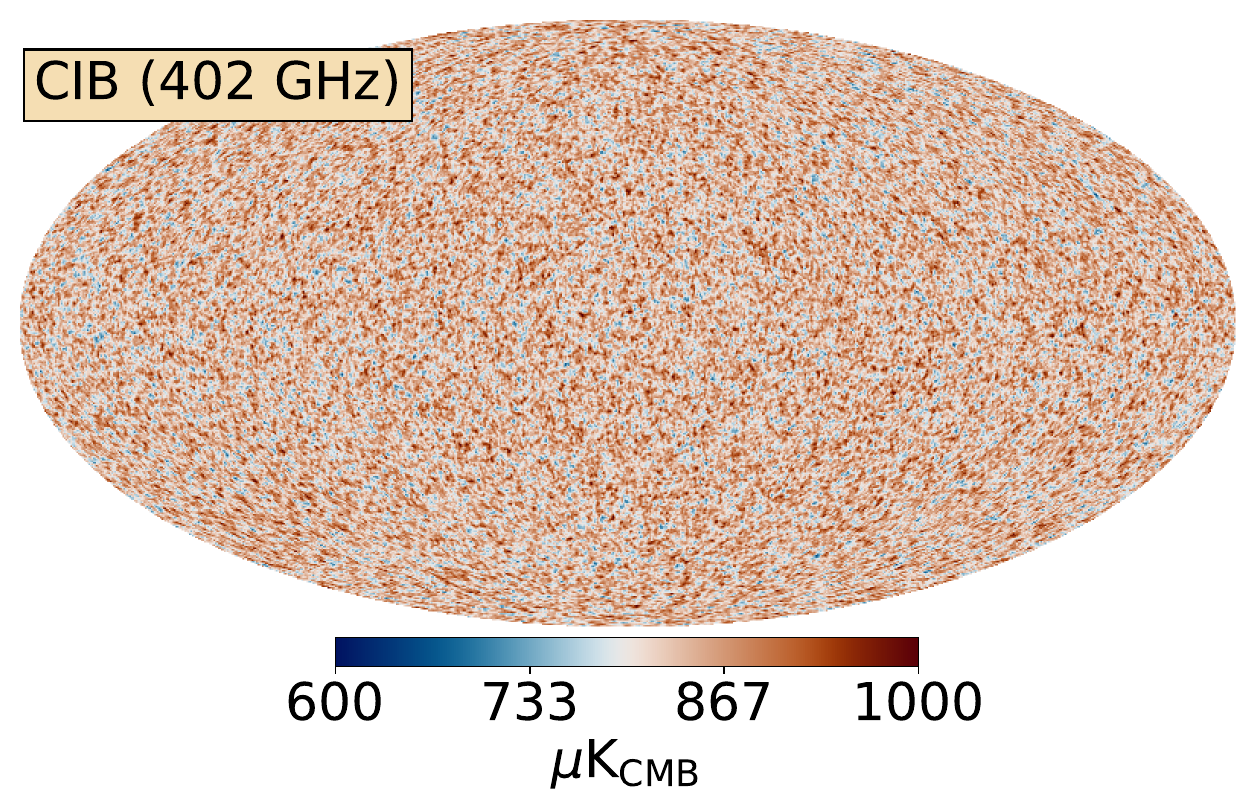}~
\includegraphics[width=0.5\textwidth,clip]{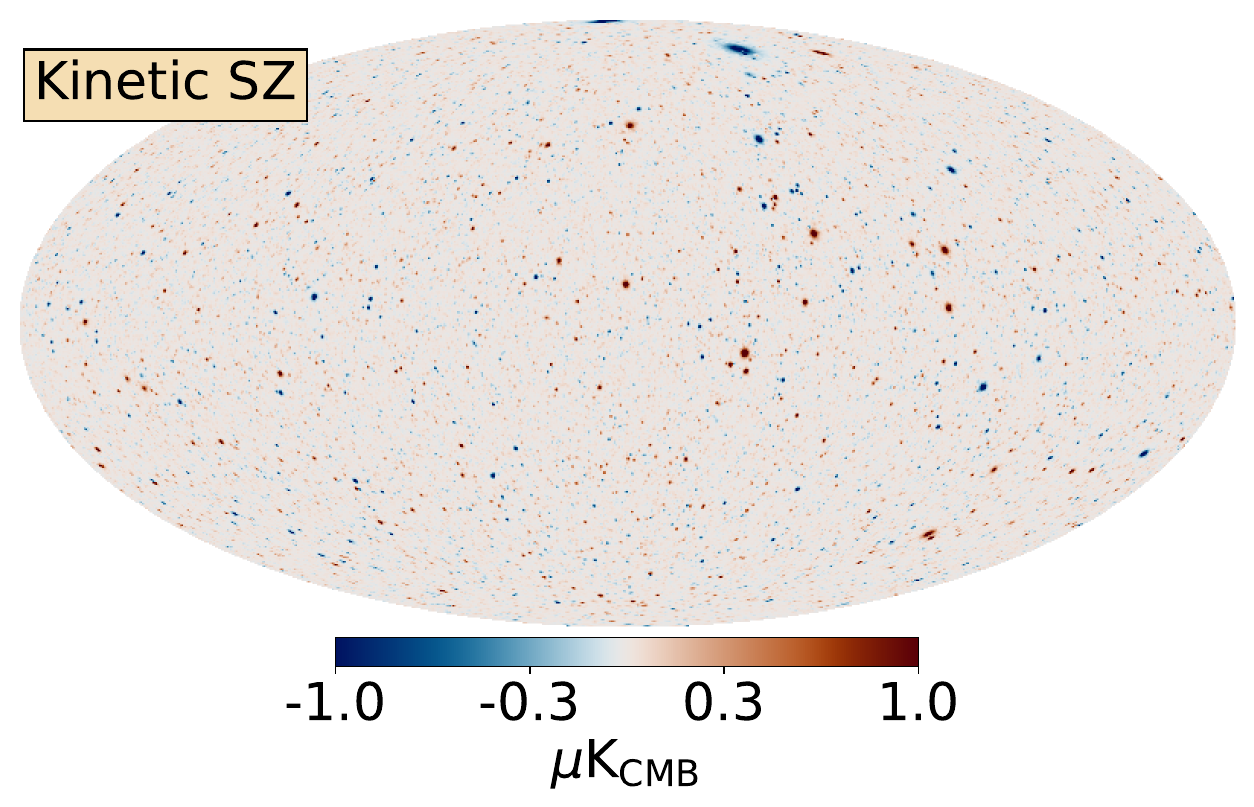}
\hfill
\includegraphics[width=0.5\textwidth,clip]{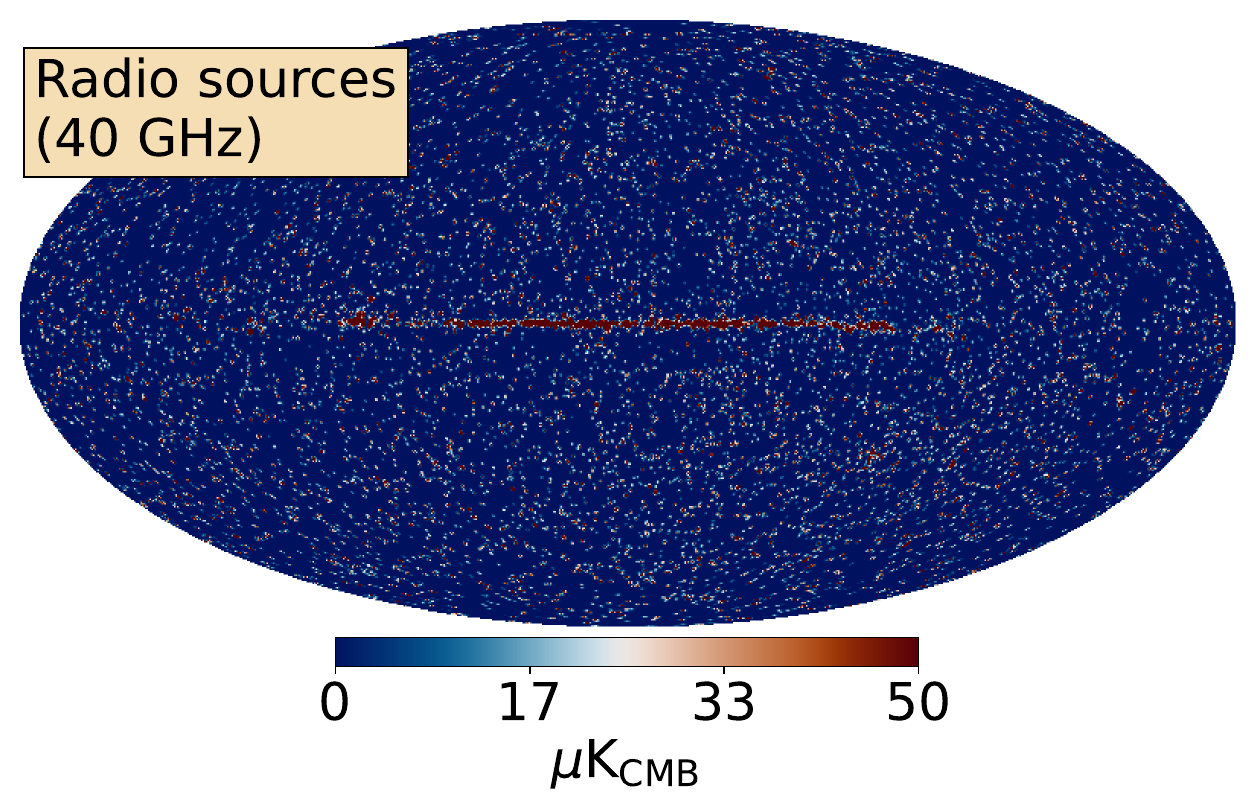}~
\includegraphics[width=0.5\textwidth,clip]{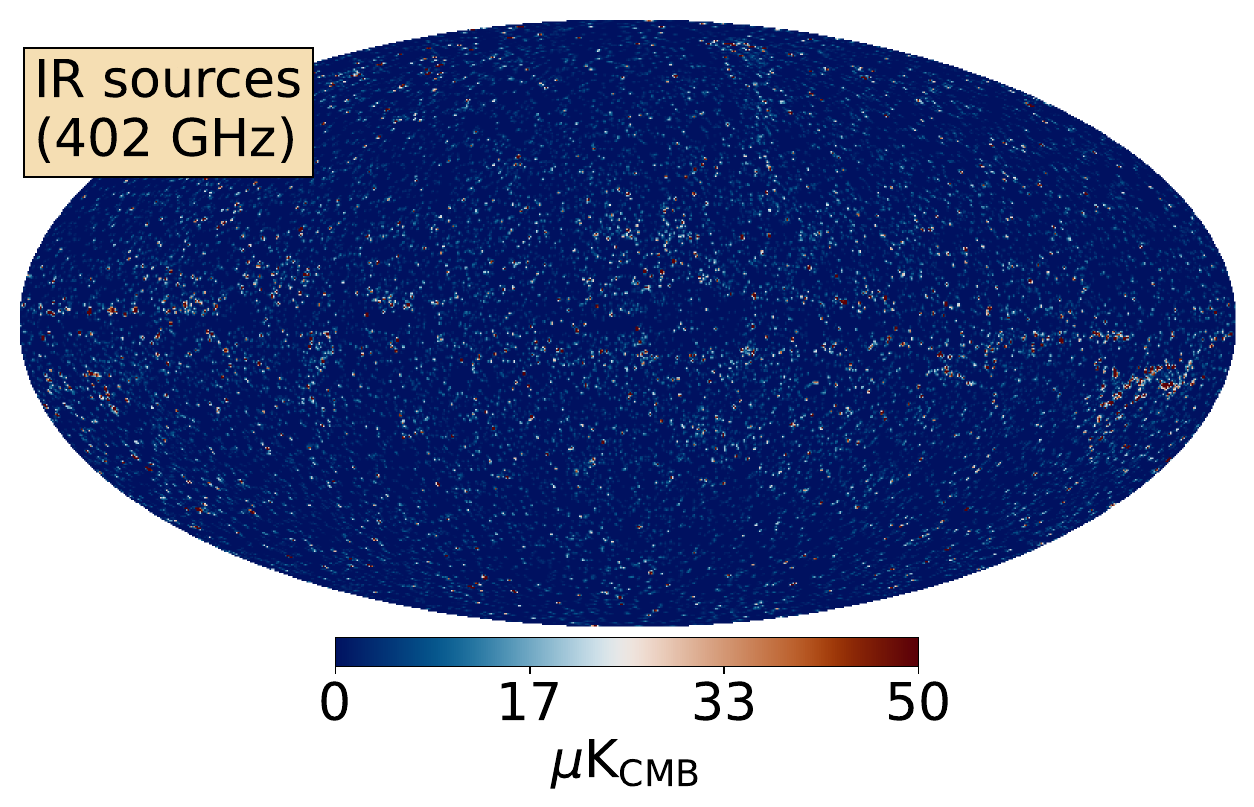}
\caption{\label{fig:sim1} \lb\ sky temperature simulation of extragalactic components. \emph{Left column}: CMB (\emph{top}), CIB at $402$\,GHz (\emph{middle}), compact radio sources at $40$\,GHz (\emph{bottom}). \emph{Right column}: thermal SZ (\emph{top}), kinetic SZ (\emph{middle}), compact infrared sources at $402$\,GHz (\emph{bottom}). 
}
\end{figure}

The extragalactic components of the simulation include thermal SZ emission, the primary component of interest that we aim at extracting, kinetic SZ emission, cosmic microwave background (CMB) temperature anisotropies, cosmic infrared background (CIB) anisotropies, compact radio sources and compact infrared sources. The simulated maps of the extragalactic components are shown in \cref{fig:sim1}. 

\subsubsection{Thermal SZ}\label{subsubsec:tsz}

The thermal SZ emission from galaxy clusters is simulated using both real and random cluster catalogues. A first set of fake clusters randomly distributed over the sky is generated, using a Poisson distribution of the Tinker halo mass function \citep{Tinker2008} to sample masses and redshifts and a uniform distribution to assign Galactic coordinates. The sampled cluster masses range from $10^{14}$ to $10^{16}\,M_\odot$. The Compton $y$-parameter of the random clusters is modelled using the universal pressure profile of ref.~\citep{Arnaud2010}. Furthermore, real clusters of mass larger than $10^{14}\,M_\odot$ are included in the thermal SZ Compton $y$-map using information from the \pl, ACT, SPT and ROSAT cluster catalogues \citep{planck2014-a36,Hilton2021,Bleem2015,Piffaretti2011}.  Simulated clusters within the same redshift and mass range as the injected real clusters are removed from the random catalogue to avoid double-counting. In addition, diffuse thermal SZ emission is simulated from theoretical thermal SZ angular power spectrum \citep{Komatsu2002} assuming the same halo mass function and pressure profile as above and cosmological parameters from the \pl\ 2018 $\Lambda\text{CDM}$ best-fit model \citep{planck2016-l06}. The simulated thermal SZ Compton $y$-parameter map is shown in the top right panel of \cref{fig:sim1}. Its angular power spectrum is consistent with the \pl\ 2015 SZ best-fit model \citep{planck2014-a28}, within the 20\,\% uncertainty range allowed by \pl\ observations, across the multipole range $\ell \simeq 20$--$1000$ relevant to \pl\ measurements.

The Compton $y$-map is scaled across the frequency channels using the non-relativistic limit of the thermal SZ spectral energy distribution (SED) given in thermodynamic temperature units by \citep{Zeldovich1969}
\begin{align}
\label{eq:sed}
a(\nu)  = T_{\rm CMB}\left[ x \coth\left( \frac{x}{2} \right) - 4  \right]\,,
\end{align}
where $x=h\nu/k T_\text{CMB}$, with $h$ being the Planck constant, $k$ the Boltzmann constant, $T_\text{CMB}$ the CMB temperature and $\nu$ the frequency. We currently neglect relativistic corrections to the thermal SZ effect arising from electron gas temperature \citep{Rephaeli1995,Challinor1998,Itoh1998,Chluba2012,Chluba2013}, which can result in $\gtrsim 10$\,\% spectral distortion of the SZ intensity at $353$\,GHz for cluster temperatures $\gtrsim 5$\,keV \citep{Erler2018,Remazeilles2019}. This aspect is left for future work, anticipating that \lb\ should have the sensitivity needed to measure the relativistic SZ temperature of the most massive clusters \citep{Remazeilles2020,Litebird_ptep2023}.

\subsubsection{Kinetic SZ}\label{subsubsec:ksz}

The kinetic SZ emission arises from the Doppler boost of the CMB photons caused by the bulk velocities of the clusters along the line of sight. The velocities for both simulated and real clusters are randomly sampled from a normal distribution whose standard deviation is derived from the power spectrum of density fluctuations using the continuity equation \citep{Delabrouille2013}. The simulated kinetic SZ map is shown in the middle right panel of \cref{fig:sim1}, showing overall consistency with the thermal SZ map, but either positive or negative temperatures depending on the sign of the radial velocity of the clusters.

Like CMB temperature anisotropies, the kinetic SZ effect is independent of the frequency when expressed in thermodynamic temperature units. Therefore, the simulated kinetic SZ map remains unchanged across all frequency channels.

\subsubsection{Cosmic microwave background}\label{subsubsec:cmb}

The map of CMB temperature anisotropies is a random Gaussian realisation on the sphere generated from the theoretical CMB temperature power spectrum computed with CAMB \citep{Lewis2000} using cosmological parameters from the \pl\ 2018 $\Lambda\text{CDM}$ best-fit model \citep{planck2016-l06}. The simulated CMB map is shown in the top left panel of \cref{fig:sim1}. 
The CMB anisotropies are independent of frequency when expressed in thermodynamic temperature units, such that the simulated CMB map remains unchanged across all frequency channels. 

\subsubsection{Cosmic infrared background}\label{subsubsec:cib}

Following ref.~\cite{planck2014-a14}, the CIB anisotropies from dusty star-forming galaxies are simulated by the Planck Sky Model assuming three different populations of spiral, starburst and proto-spheroidal galaxies \citep{Cai2013}, which are distributed across redshift shells according to the dark matter distribution generated by CLASS \citep{Blas2011}. Galaxies from a given type of population have the same spectral energy distribution, which is redshifted accordingly. The simulated maps from each population and each redshift shell are added together to generate CIB maps at each frequency. The resulting CIB map for the \lb\ $402$-GHz frequency channel is shown in the middle left panel of \cref{fig:sim1}. Correlations between the CIB and thermal SZ components \citep{Addison2012,planck2014-a29,Maniyar2021} are neglected in this simulation.\footnote{The portion of the thermal SZ signal correlated with the CIB may be suppressed alongside the CIB through variance minimization by the needlet internal linear combination (NILC) method used for thermal SZ reconstruction, resulting in a minor, percent-level loss of thermal SZ RMS. However, such loss can be avoided in principle by \emph{deprojecting} the CIB using a \emph{constrained} ILC for component separation \citep{Remazeilles2011a,Madhavacheril2020,Remazeilles2021,Carones2024,McCarthy2024,Coulton2024}.} Further details on the modelling of the CIB in the Planck Sky Model can be found in refs.~\cite{planck2014-a14,Cai2013}.

\subsubsection{Radio and infrared sources}\label{subsubsec:sources}

Our simulation includes extragalactic radio (active galactic nuclei) and infrared sources (dusty galaxies), as well as unresolved Galactic sources. Radio sources are obtained from existing catalogues of radio surveys at  $4.85$\,GHz \citep{Wright1994,Gregory1996}, $1.4$\,GHz \citep{Condon1998} and $0.843$\,GHz  \citep{Mauch2003}, while infrared sources are provided by the IRAS point-source catalogue \citep{Beichman1988,Moshir1992}. 

Infrared sources are extrapolated to \lb\ and \pl\ frequencies assuming modified-blackbody spectra, while radio sources are extrapolated assuming four power-law spectra depending on the frequency range, with each radio source being assigned either a steep or flat spectral index that is randomly drawn from a Gaussian distribution \citep{Delabrouille2013}. 
The map of compact radio sources at $40$\,GHz is shown in the bottom left panel of \cref{fig:sim1}, while the map of compact infrared sources at $402$\,GHz is displayed in the bottom right panel.

\subsection{Galactic components}\label{subsec:gal}

The Galactic foregrounds include thermal dust emission, synchrotron emission, free-free emission and anomalous microwave emission (AME). The simulated maps of each Galactic component are shown in \cref{fig:sim2}.

\subsubsection{Thermal dust}\label{subsubsec:dust}

The \pl\ GNILC dust template maps \citep{planck2016-XLVIII} are used to simulate Galactic thermal dust emission, due to their reduced contamination from the CIB thanks to filtering by the GNILC pipeline \citep{Remazeilles2011b}. The use of the GNILC dust templates is particularly important for temperature analysis to prevent an overestimation of CIB contamination in the simulated data and the reconstructed SZ map. 

The thermal dust emission is scaled across the frequencies $\nu$ assuming a modified blackbody spectrum with variable spectral index and temperature across the directions $\hat{n}$ on the sky:
\begin{align}
\label{eq:mbb}
I_{\rm dust}(\nu,\hat{n})  = \tau_{353}^{\rm GNILC}(\hat{n})\left(\frac{\nu}{353\,\text{GHz}}\right)^{\beta^{\rm GNILC}(\hat{n})}B_\nu\left(T^{\rm GNILC}(\hat{n})\right)\,,
\end{align}
where $\tau_{353}^{\rm GNILC}(\hat{n})$ is the \pl\ GNILC dust optical depth map at $353$\,GHz given in intensity units, $\beta^{\rm GNILC}(\hat{n})$ is the \pl\ GNILC dust spectral index map and $T^{\rm GNILC}(\hat{n})$ is the \pl\ GNILC dust temperature map, while $B_\nu(T)$ is the Planck's blackbody function. The simulated thermal dust amplitude map at $353$\,GHz is shown in the bottom right panel of \cref{fig:sim2}.

\subsubsection{Synchrotron}\label{subsubsec:sync}

The Galactic synchrotron emission, due to cosmic-ray electrons accelerated by the Galactic magnetic field, is simulated from the Reprocessed Haslam $408$-MHz map \citep{Remazeilles2015}, in which extragalactic radio sources have been subtracted from the original Haslam map \citep{haslam1982}. The use of the source-subtracted Haslam map as a Galactic synchrotron template is particularly important for temperature analysis to prevent double-counting the contamination from radio sources in the simulated data and the reconstructed SZ map.

\begin{figure}[tbp]
\centering 
\includegraphics[width=0.5\textwidth,clip]{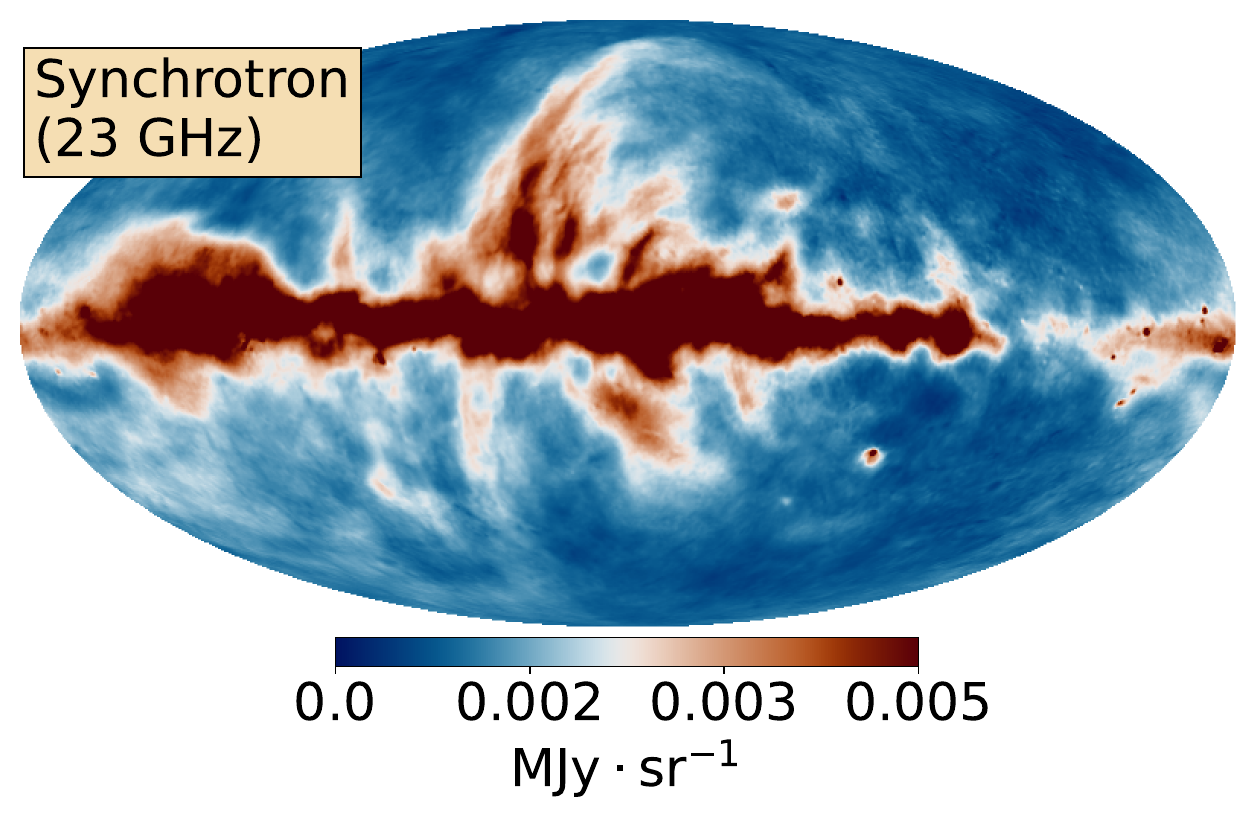}~
\includegraphics[width=0.5\textwidth,clip]{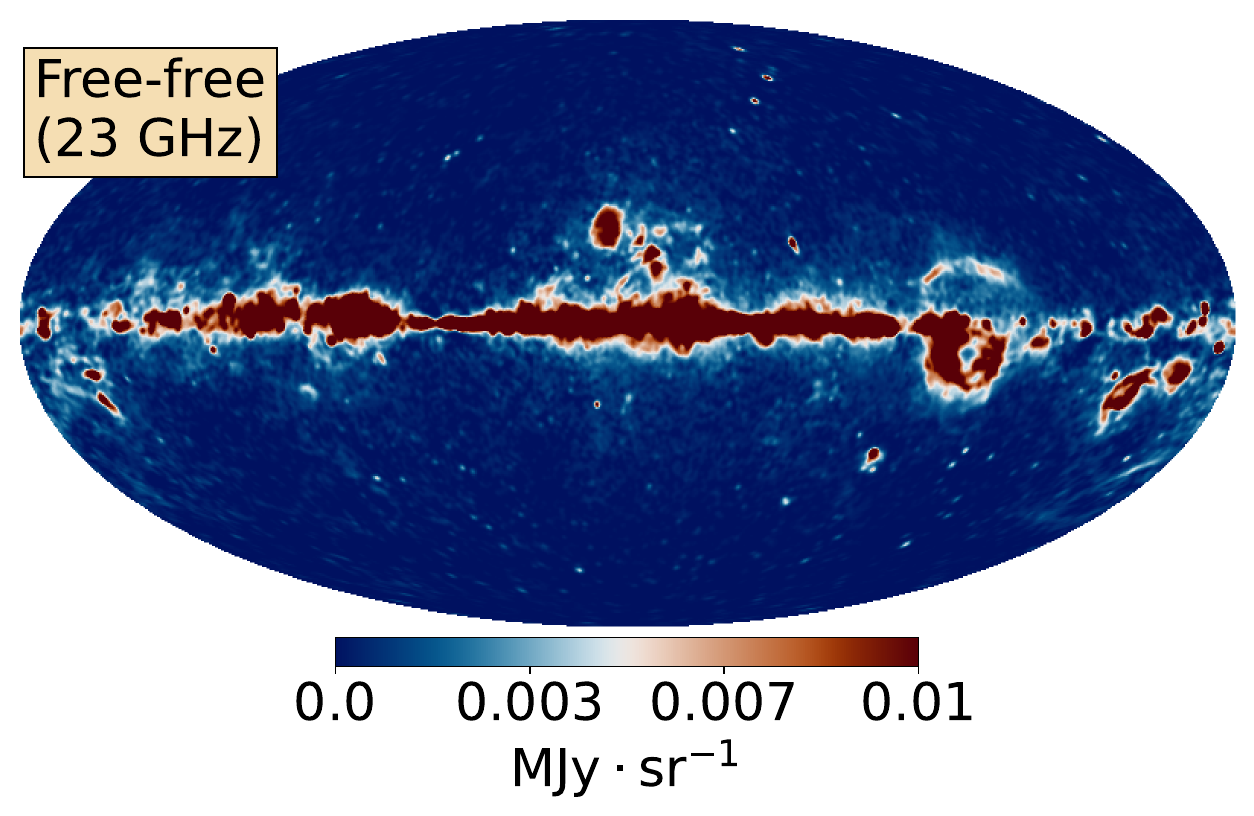}
\hfill
\includegraphics[width=0.5\textwidth,clip]{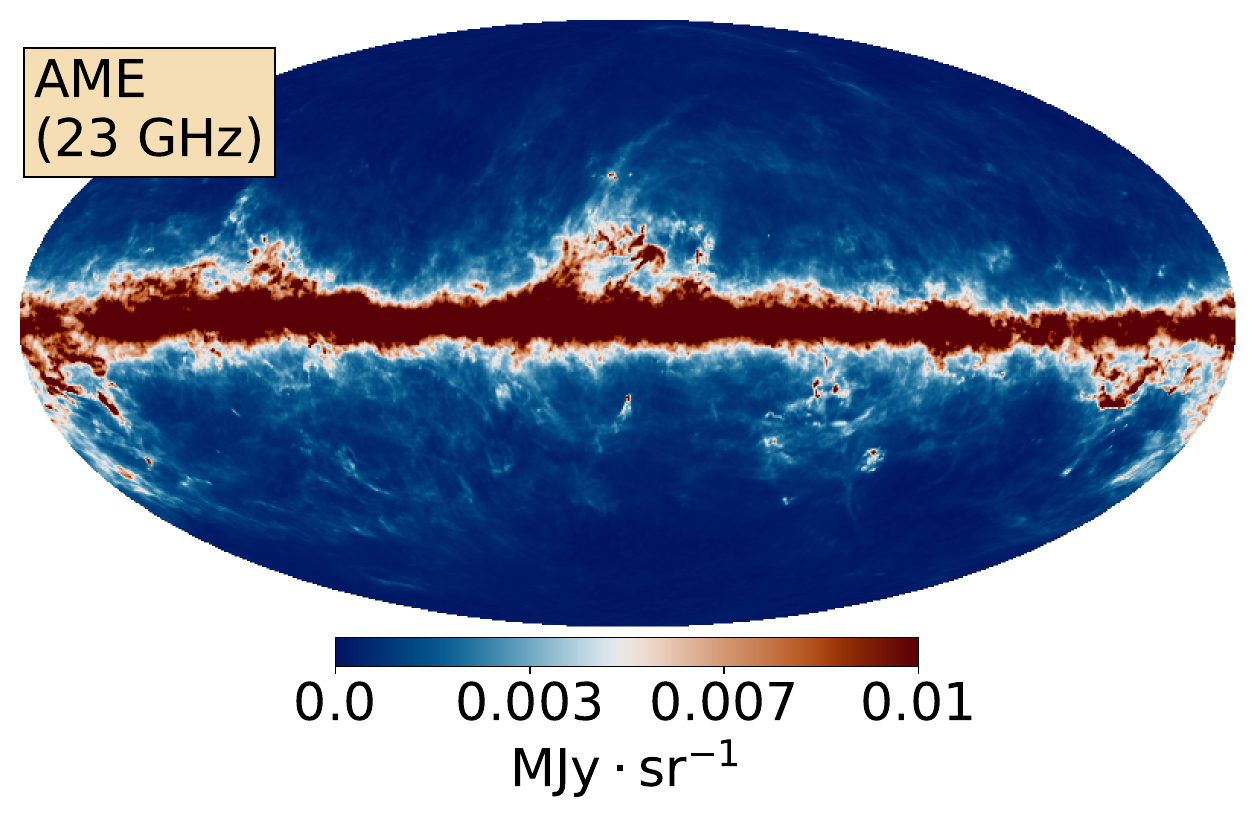}~
\includegraphics[width=0.5\textwidth,clip]{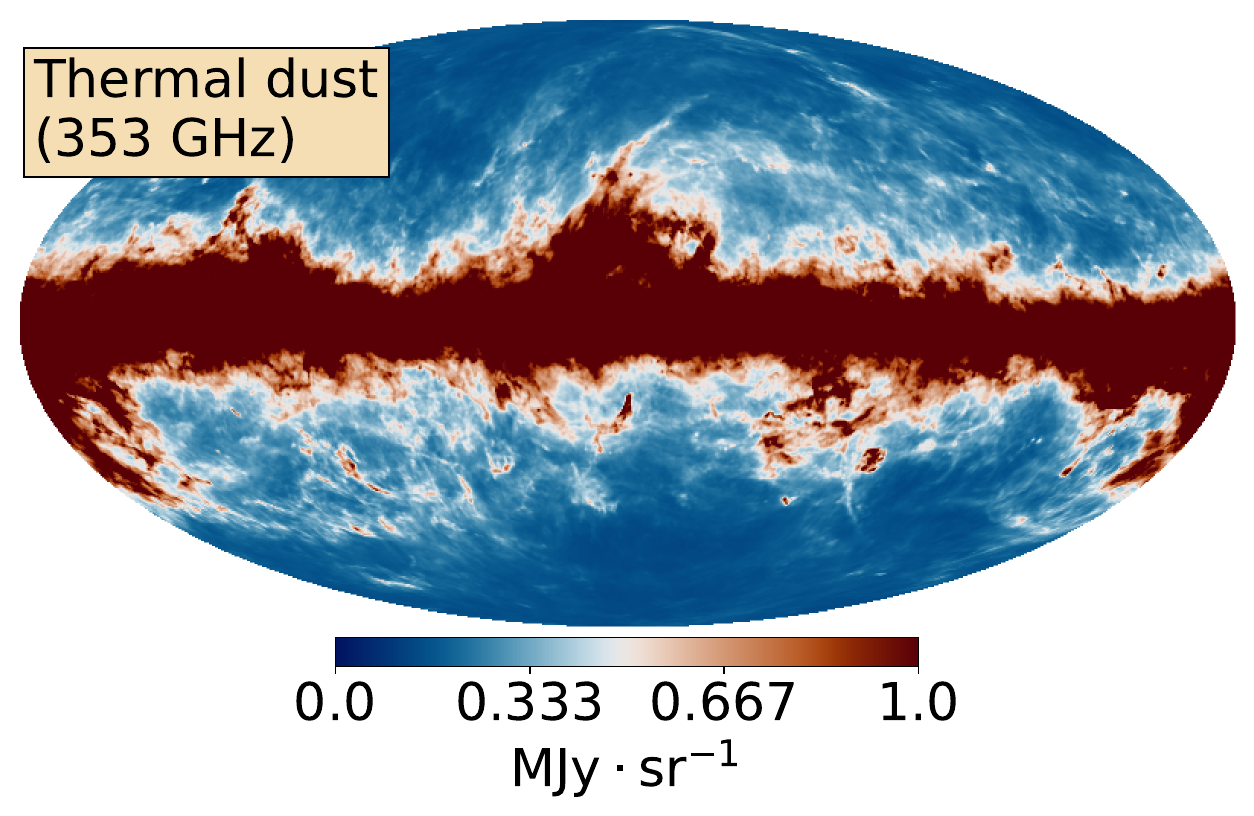}
\caption{\label{fig:sim2} \lb\ sky temperature simulation of Galactic components in intensity units.  \emph{Left column}: synchrotron at $23$\,GHz (\emph{top}), AME at $23$\,GHz (\emph{bottom}). \emph{Right column}: free-free at $23$\,GHz (\emph{top}), thermal dust at $353$\,GHz (\emph{bottom}).}
\end{figure}

The Galactic synchrotron template at $408$\,MHz is extrapolated to \lb\ and \pl\ frequencies using a power-law spectrum with variable spectral index over the sky:
\begin{align}
\label{eq:pl}
I_{\rm sync}(\nu,\hat{n})  = I_{408\,{\rm MHz}}(\hat{n})\left(\frac{\nu}{408\,\text{MHz}}\right)^{\beta_s(\hat{n})}\,,
\end{align}
where $I_{408\,{\rm MHz}}(\hat{n})$ is the Reprocessed Haslam $408$-MHz map and $\beta_s(\hat{n})$ is the synchrotron spectral index template map from ref.~\citep{Miville2008}. The simulated synchrotron amplitude map at $23$\,GHz is shown in the top left panel of \cref{fig:sim2}.

\subsubsection{Free-free}\label{subsubsec:ff}

The \pl\ Commander free-free template maps \citep{planck2014-a12} are used to model the Galactic free-free emission (thermal bremsstrahlung) that is due to Coulomb interactions between free electrons and positively charged nuclei in ionised star-forming regions. Free-free emission is scaled across the frequencies as \citep{Draine2011}
\begin{align}
\label{eq:ff}
I_{\rm ff}(\nu,\hat{n})  & = 10^6\,T_{\rm e}(\hat{n})\left(1-e^{-\tau_{\rm ff}(\nu,\hat{n})}\right)\,,\\
\tau_{\rm ff}(\nu,\hat{n}) & = 0.05468\,T_{\rm e}(\hat{n})^{-3/2}\,\left({\nu\over 10^9}\right)^{-2}\,{\rm EM}(\hat{n})\,g_{\rm ff}(\hat{n},\nu)\,,\\
g_{\rm ff}(\hat{n},\nu) &=1+\log\left(1+ {\rm e}^{4.960+{\sqrt{3}\over \pi}\log\left[\left({\nu\over 10^9}\right)^{-1}\left({T_{\rm e}(\hat{n})\over 10^4\,{\rm K}}\right)^{3/2}\right]} \right)\,,
\end{align}
where ${\rm EM}(\hat{n})$ is the \pl\ Commander free-free emission measure (EM) map and $T_{\rm e}(\hat{n})$ is the \pl\ Commander electron temperature map. The simulated Galactic free-free map at $23$\,GHz is shown in the top right panel of \cref{fig:sim2}.

\subsubsection{AME}\label{subsubsec:ame}

Previous studies have demonstrated a strong correlation between anomalous microwave emission (AME) at low frequencies and thermal dust emission at higher frequencies \cite{Davies2006,planck2014-a31,Fernandez-Torreiro2023}. This phenomenon is believed to be the result of electric dipole radiation emitted by spinning dust grains within our Galaxy \cite{Draine1998}. For these reasons, the AME map at $23$\,GHz is simulated by rescaling the thermal dust template at $353$\,GHz using a cross-correlation factor of $0.91\,{\rm K/K}$ in Rayleigh-Jeans temperature units, as measured by ref.~\citep{planck2014-a31}. 

The AME template is extrapolated to \lb\ and \pl\ frequency channels using the spinning dust spectral model of ref.~\citep{Draine1998}. The AME map at $23$\,GHz is displayed in intensity units in the bottom left panel of \cref{fig:sim2}.

\subsection{Instruments}\label{subsec:instruments}

In our simulations, we model both the \lb\ \citep{Litebird_ptep2023} and \pl\ \citep{planck2016-l01} space-borne experiments, whose instrumental specifications are described hereafter. While \pl\ provides higher angular resolution and wider frequency coverage ($30$--$857$\,GHz) than \lb\ ($40$--$402$\,GHz), \lb\ benefits from a larger number of frequency bands and higher sensitivity. By considering both \lb\ and \pl\ in our simulations, we can leverage their unique instrumental capabilities and combine their data to enhance SZ analyses.

\subsubsection{\lb\ Instrument Model}\label{subsubsec:lb} 

The \lb\ Instrument Model (IMo) provides a quantitative description of the entire \lb\ experiment, encompassing the three telescopes -- LFT (Low-Frequency Telescope), MHT (Mid-Frequency Telescope), and HFT (High-Frequency Telescope) -- as well as the payload and observational strategy, as detailed in ref.~\cite{Litebird_ptep2023}. In our simulations, we adhere to the instrumental specifications outlined in the \lb\ IMo to generate full-sky temperature maps across 15 frequency bands ranging from $40$ to $402$\,GHz. To simulate instrumental white noise in each frequency channel, we use the IMo sensitivities per channel quoted in table~13 of  ref.~\cite{Litebird_ptep2023}, which we rescale by a factor of $\sqrt{2}$ to obtain noise RMS values in temperature. We also rely on the quoted IMo beam FWHM values per channel to perform beam convolution of the simulated sky maps. We omit the IMo top-hat frequency bandpasses from our simulations because their impact on non-parametric component-separation methods, such as the one employed in this analysis, which rely solely on knowledge of the thermal SZ spectrum rather than uncertain foreground models, is well characterized.

Being equipped with a rotating half-wave plate (HWP) modulator, the \lb\ mission must allow for significant reduction of the low-frequency  $1/f$ noise in the polarization data, but not in the temperature data. Therefore, additional simulations accounting for \lb's inhomogeneous scan strategy and $1/f$ noise in temperature have been implemented in order to study the impact of these systematics on the $y$-map in \cref{subsec:noise1overf}. To simulate \lb\ $1/f$ noise, a standard maximum-likelihood map-making algorithm \citep{Borrill1999} was employed on time-ordered data (TOD) simulations, with both realistic and pessimistic knee frequencies of respective values $f_{\rm knee}=30$\,mHz and $f_{\rm knee}=100$\,mHz (LiteBIRD Collaboration, in prep.).

\subsubsection{\pl\ LFI and HFI instruments}\label{subsubsec:planck} 

The \pl\ satellite mission, launched by the European Space Agency (ESA) in May 2009, was equipped with two instruments: the Low-Frequency Instrument (LFI) and the High-Frequency Instrument (HFI). These instruments allowed for the observation of the full sky in nine frequency bands between $30$ and $857$\,GHz with an average sensitivity of $50\,\mu{\rm K}\cdot\text{arcmin}$ and relatively high angular resolution.

For the simulation of noise maps and beam convolution in the \pl\  frequency channels, we have adopted the instrumental specifications (sensitivities and beam FWHM values) as provided in the table~4 of ref.~\citep{planck2016-l01}. We generate Gaussian white noise maps across the nine \pl\ frequency channels using the noise RMS values per channel from ref.~\citep{planck2016-l01}. Despite these simplified noise assumptions, as shown later in \cref{subsec:spectra}, the power spectrum of noise residuals in the recovered $y$-map from the analysis of white-noise simulations is consistent across multipoles with that from the \textit{Planck} release 2 (PR2) SZ analysis \citep{planck2014-a28}, which accounts for real noise.

\section{Component separation for thermal SZ effect}\label{sec:nilc} 

To extract the thermal SZ Compton $y$-parameter signal out of the multi-frequency sky maps of the simulation, we implement the Needlet Internal Linear Combination (NILC) method \citep{delabrouille2009,Remazeilles2011a,Remazeilles2013}, using a pipeline that is similar to the one utilised for \pl\ SZ data analysis \citep{planck2014-a28,Chandran2023}. Needlets \citep{Narcowich2006,Marinucci2008} are a family of spherical wavelets which allow for localized data processing in both pixel domain and harmonic domain. This is particularly useful for component separation as the relative contribution of the various components of emission to the data varies both across different sky regions and across different regimes of angular scales. Furthermore, the multi-resolution feature of the wavelet-based component separation method NILC enables seamless integration of diverse data sets from different experiments with varying resolution and sky coverage \citep{Remazeilles2013}. 

In this study, our NILC pipeline is applied to the set of multi-frequency sky maps from either the \lb\ simulation, the \pl\  simulation or the combination of both simulations, so that three different $y$-maps are produced, namely: a \lb\ $y$-map; a \pl\ $y$-map; and a \lb-\pl\ combined $y$-map. Given that the average angular resolution is about $30'$ for \lb\ channels and $10'$ for \pl\ channels, the \lb\ $y$-map is reconstructed at $30'$ resolution, while the \pl\ $y$-map and the  \lb-\pl\ combined $y$-map are both reconstructed at $10'$ resolution. Our NILC pipeline proceeds as follows. 

The multi-frequency sky maps, $d_\nu(p)$ for frequency $\nu$ with pixel $p$ dependence, from either \lb, \pl, or both \lb\ and \pl\ simulations are spherical-harmonic transformed into harmonic coefficients $a_{\ell m,\nu}$:
\begin{align}
\label{eq:sht}
d_\nu(p) = \sum_{\ell,m} a_{\ell m,\nu}\,Y_{\ell m}(p)\,,
\end{align}
where $Y_{\ell m}(p)$ denote the spherical harmonics. Each $a_{\ell m,\nu}$ is divided by their native beam transfer function at that frequency, $b_{\ell,\nu}$, and multiplied by a Gaussian beam transfer function, $b_\ell^{\rm out}$, associated with the desired output resolution, i.e.~$30'$ for the \lb\ analysis, and $10'$ for the \pl\ and the \lb-\pl\ analyses:
\begin{align}
\label{eq:rebeaming}
 a_{\ell m,\nu}  
 \rightarrow
 a_{\ell m,\nu}\,\frac{b_\ell^{\rm out}}{b_{\ell,\nu}}\,.
\end{align}

\begin{figure}[tbp]
\centering 
\includegraphics[width=0.8\textwidth,clip]{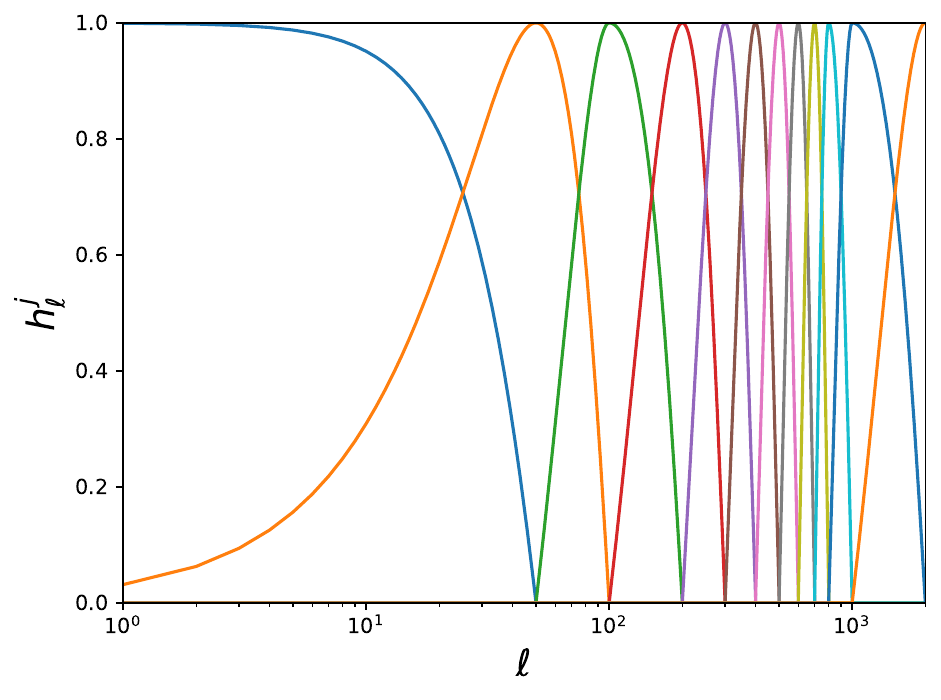}
\caption{\label{fig:needlets} Needlet bandpass functions in harmonic domain.}
\end{figure}
 
 The multi-frequency harmonic coefficients $a_{\ell m,\nu}$ then undergo bandpass filtering in the harmonic domain using cosine-shaped needlet windows, $h_\ell^{j}$, as shown in \cref{fig:needlets}, which partition the signal at each frequency into separate contributions from different ranges of angular scales. The functional form of the needlet bandpasses is chosen as
 \begin{align}
\label{eq:needlets}
h_\ell^{j} = 
\begin{cases}
\cos\left( \frac{\ell_{\rm peak}^j  - \ell}{\ell_{\rm peak}^j  - \ell_{\rm peak}^{j-1}}\frac{\pi}{2}\right)   & \text{if}\ \ell_{\rm peak}^{j-1}\leq \ell < \ell_{\rm peak}^j\,, \\[0.2cm]
\\
\cos\left( \frac{\ell - \ell_{\rm peak}^j}{\ell_{\rm peak}^{j+1}- \ell_{\rm peak}^j}\frac{\pi}{2}\right)   & \text{if}\ \ell_{\rm peak}^j\leq \ell < \ell_{\rm peak}^{j+1}\,,
\end{cases}
\end{align}
where $\ell_{\rm peak} = 0, 50, 100, 200, 300, 400, 500, 600, 700, 800, 1000, 2000$ for $j=1,\cdots,12$,
so that 
\begin{align}
\label{eq:cond}
\sum_j \left(h_\ell^{j}\right)^2 = 1 \quad \forall \ell\,.
\end{align}
\Cref{eq:cond} ensures that the total power in the data is preserved after forward and inverse needlet transformations.
 The first ten needlet window functions ($j=1,\cdots,10$) are used to process both \lb\ and \pl\ channels, while the last two needlet windows ($j=11,12$) are used to process only the \pl\ channels due to the limited resolution of the \lb\ maps. The resulting bandpass-filtered coefficients, ${a_{\ell m,\nu}^j = h_\ell^{j}a_{\ell m,\nu}}$, finally undergo inverse spherical harmonic transformation. As a result, for each frequency channel we obtain ten (or twelve) needlet maps,
\begin{align}
\label{eq:needlet_map}
d_\nu^j(p) = \sum_{\ell,m} h_\ell^{j}\,a_{\ell m,\nu}\,Y_{\ell m}(p)\,,
\end{align}
which capture temperature fluctuations  at specific angular scales, as determined by the corresponding needlet window $h_\ell^{j}$ used. 

For each needlet scale $j$, we estimate the elements $C_{\nu\nu'}^j(p)$ of the data covariance matrix $C^j(p)$ in each pixel $p$  for a pair of frequencies as the convolution in the pixel domain 
\begin{align}
C_{\nu\nu'}^j(p)  = \sum_{p'} K^j(p,p')\,d_\nu^j(p')\,d_{\nu'}^j(p')\,,
\end{align}
where 
\begin{align}
\label{eq:kernel}
K^j(p,p') = \frac{1}{ 2\pi\left(\sigma^j\right)^2 } \exp \left( - \frac{ \| \vec{n}(p) - \vec{n}(p') \|^2 } { 2 \left(\sigma^j\right)^2 } \right)\,,
\end{align}
is the bidimensional symmetric Gaussian function used for the convolution and $\vec{n}(p)$ is the three-dimensional vector on the \texttt{HEALPix} sphere associated with pixel $p$. The width $\sigma^j$ of the Gaussian function determines the effective size of the pixel domain surrounding pixel $p$ within which the local covariance of the data is computed for needlet scale $j$. The degree of localisation in pixel domain given by $\sigma^j$ depends on the needlet scale $j$ considered. As we move from the first to the last needlet band, probing smaller and smaller angular scales, $\sigma^j$ becomes progressively smaller.

For each needlet scale $j$, we then obtain an estimate, $\widehat{y}^{\,j}$, of the thermal SZ Compton parameter signal by assigning ILC weights to the frequency maps:
\begin{align} 
\label{eq:ilc}
\widehat{y}^{\,j} (p) = \sum_\nu \frac{ \sum_{\nu'} a_{\nu'}\, \left[C^j(p)\right]^{-1}_{\nu\nu'} }{ \sum_{\nu'}\sum_{\nu''} a_{\nu'}\, \left[C^j(p)\right]^{-1}_{\nu'\nu''}\, a_{\nu''} }\,d_\nu^j(p)\,,
\end{align}
 where $a_\nu$ is the SED of the thermal SZ effect given by \cref{eq:sed}. By construction, \cref{eq:ilc} ensures the preservation of the thermal SZ signal at needlet scale $j$ without any multiplicative error, while minimizing the variance of the additive error arising from residual foreground and noise contamination.
 
Finally, the estimated maps $\widehat{y}^{\,j} (p)$ are transformed into spherical harmonic coefficients, $\widehat{y}^{\,j}_{\ell m}$, and these coefficients are combined to reconstruct the complete thermal SZ map, $\widehat{y} (p)$, incorporating contributions from all angular scales, as follows:
\begin{align} 
\label{eq:nilc}
\widehat{y} (p) = \sum_{\ell, m}\left(\sum_j h_\ell^j\,\widehat{y}^{\,j}_{\ell m}\right)Y_{\ell m}(p)\,.
\end{align}
 \Cref{eq:nilc} provides the NILC map of the thermal SZ Compton parameter for any combination of data, whether it be from \lb, \pl, or the combined \lb-\pl\ data sets.

\section{Results}\label{sec:results}

In this section, we assess the quality of the reconstructed thermal SZ Compton $y$-parameter maps from the \pl, \lb, and combined \lb-\pl\ data sets. 
The three different $y$-maps are compared through map visualization (\cref{subsec:ymaps}), power spectrum calculation for signal and residuals (\cref{subsec:spectra}), one-point statistics estimation for signal and residuals (\cref{subsec:pdfs}), and cosmological parameter constraints (\cref{subsec:cosmo}). \lb's responses to $1/f$ noise (\cref{subsec:noise1overf}) and the patchy reionisation SZ effect (\cref{subsec:reionisation}) are also discussed.

\subsection{Reconstructed thermal SZ \texorpdfstring{$y$}{y}-maps}\label{subsec:ymaps} 

\begin{figure}[tbp]
\centering 
\includegraphics[width=0.5\textwidth,clip]{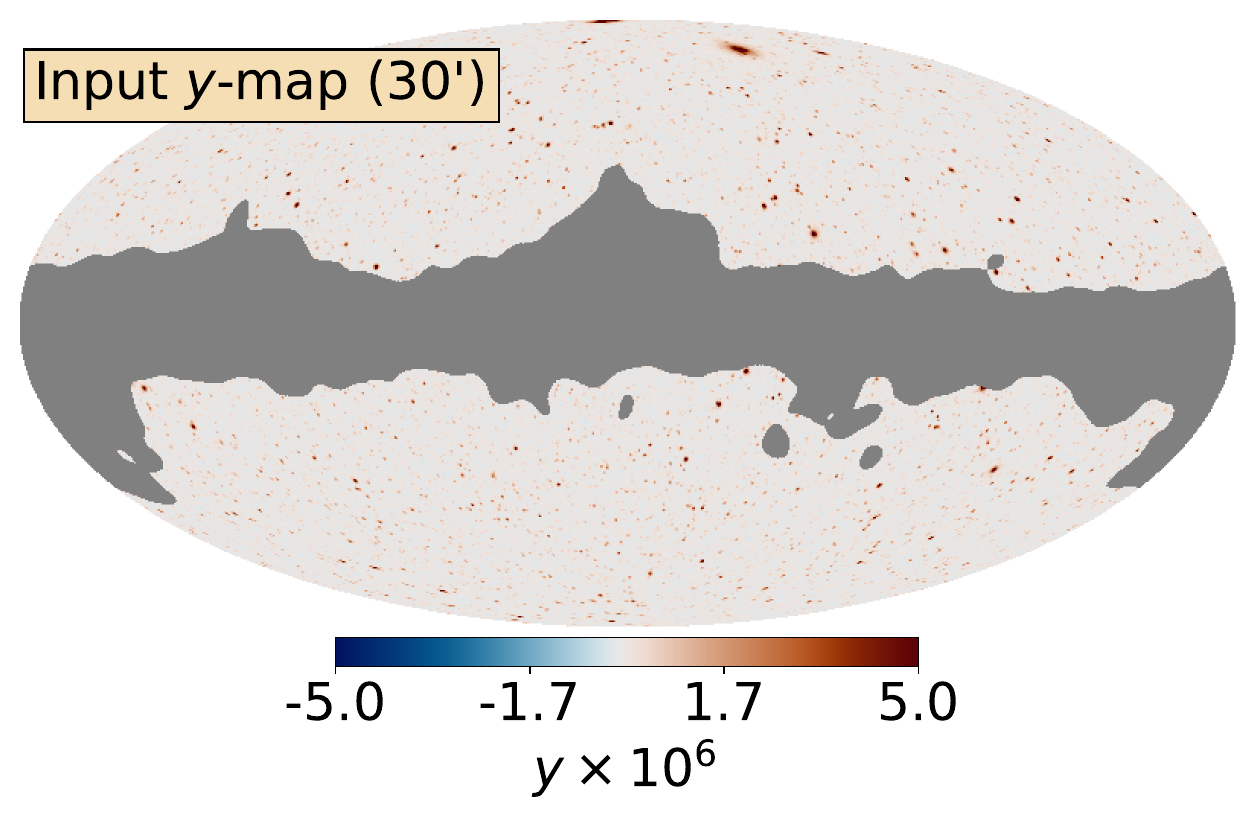}
\hfill
\includegraphics[width=0.5\textwidth,clip]{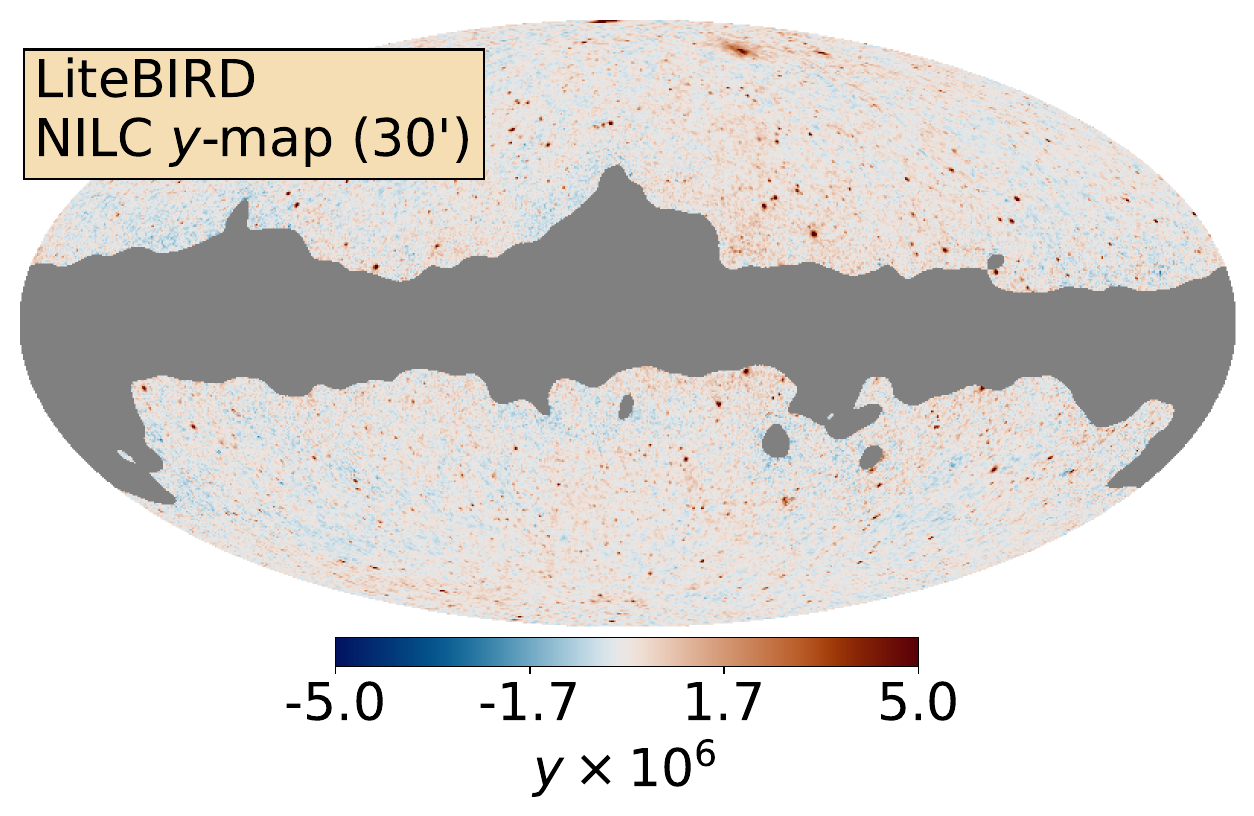}~
\includegraphics[width=0.5\textwidth,clip]{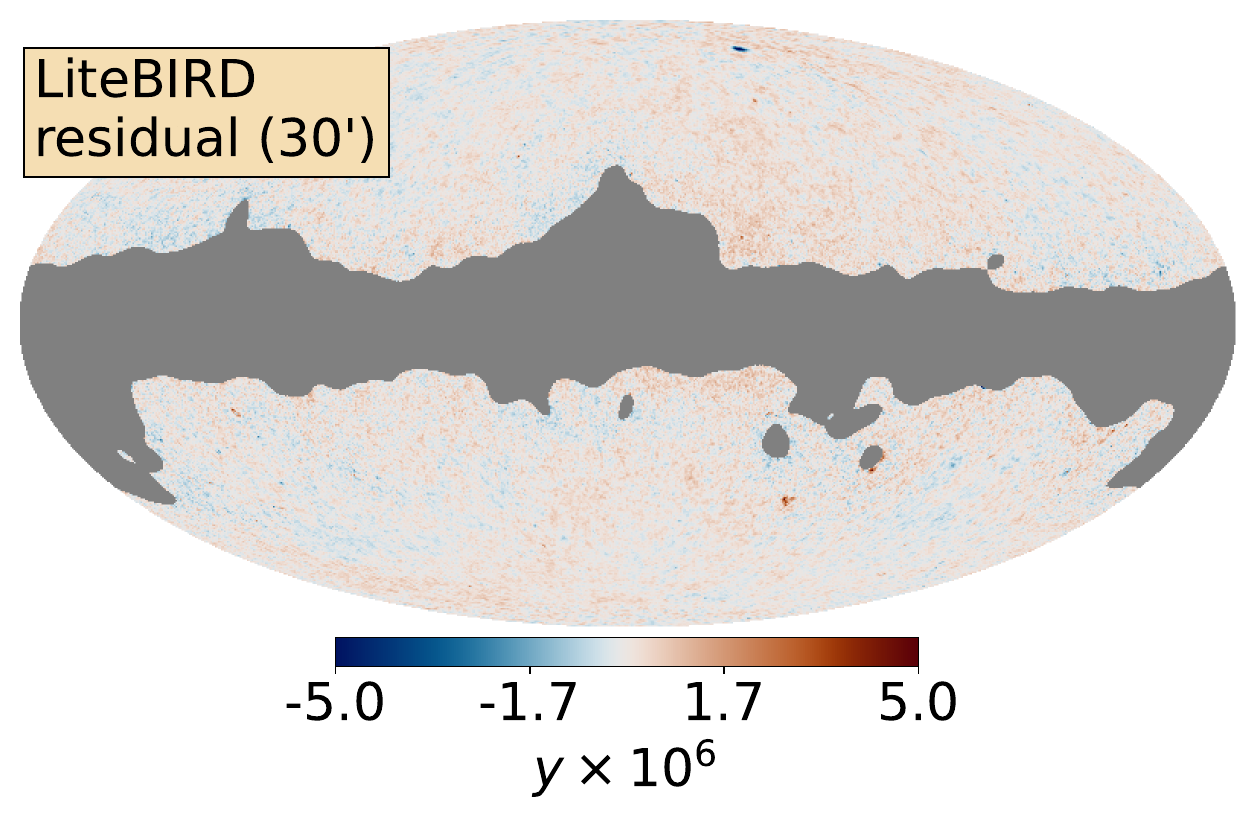}
\hfill
\includegraphics[width=0.5\textwidth,clip]{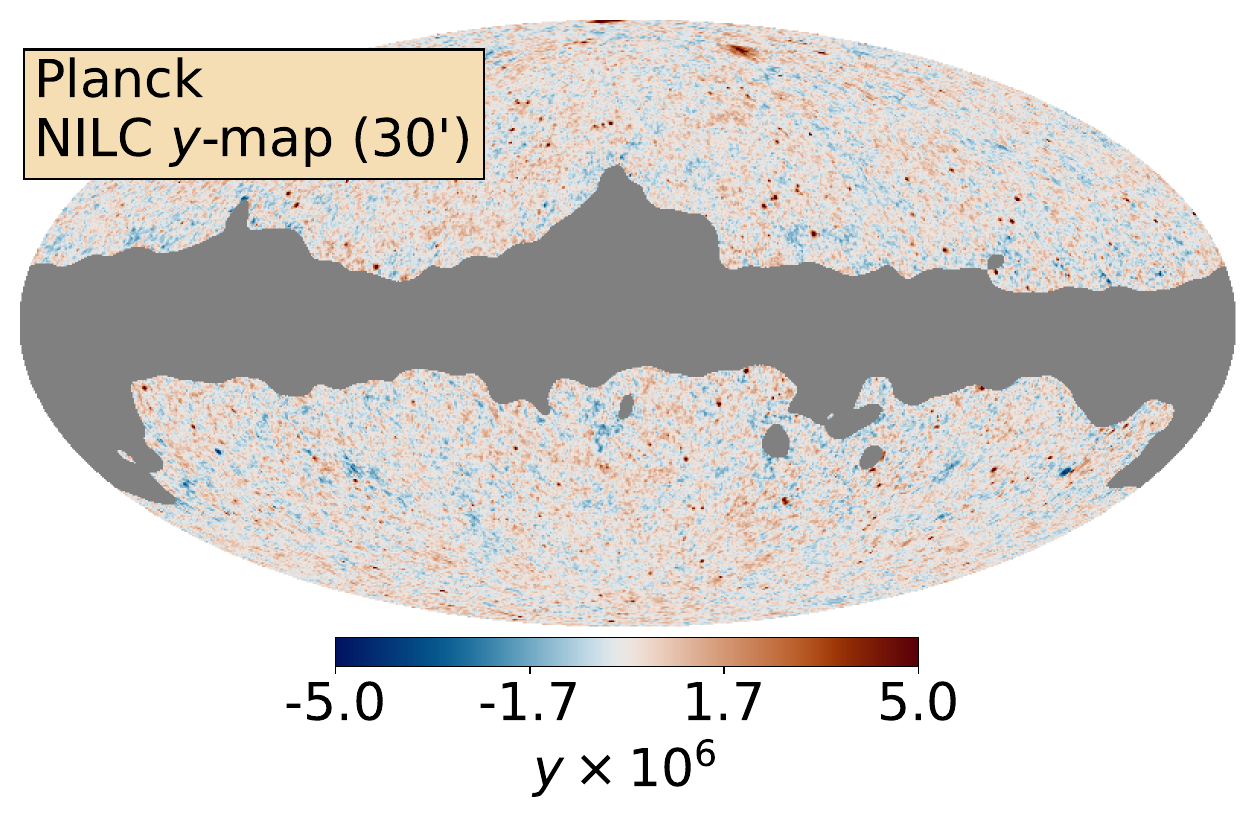}~
\includegraphics[width=0.5\textwidth,clip]{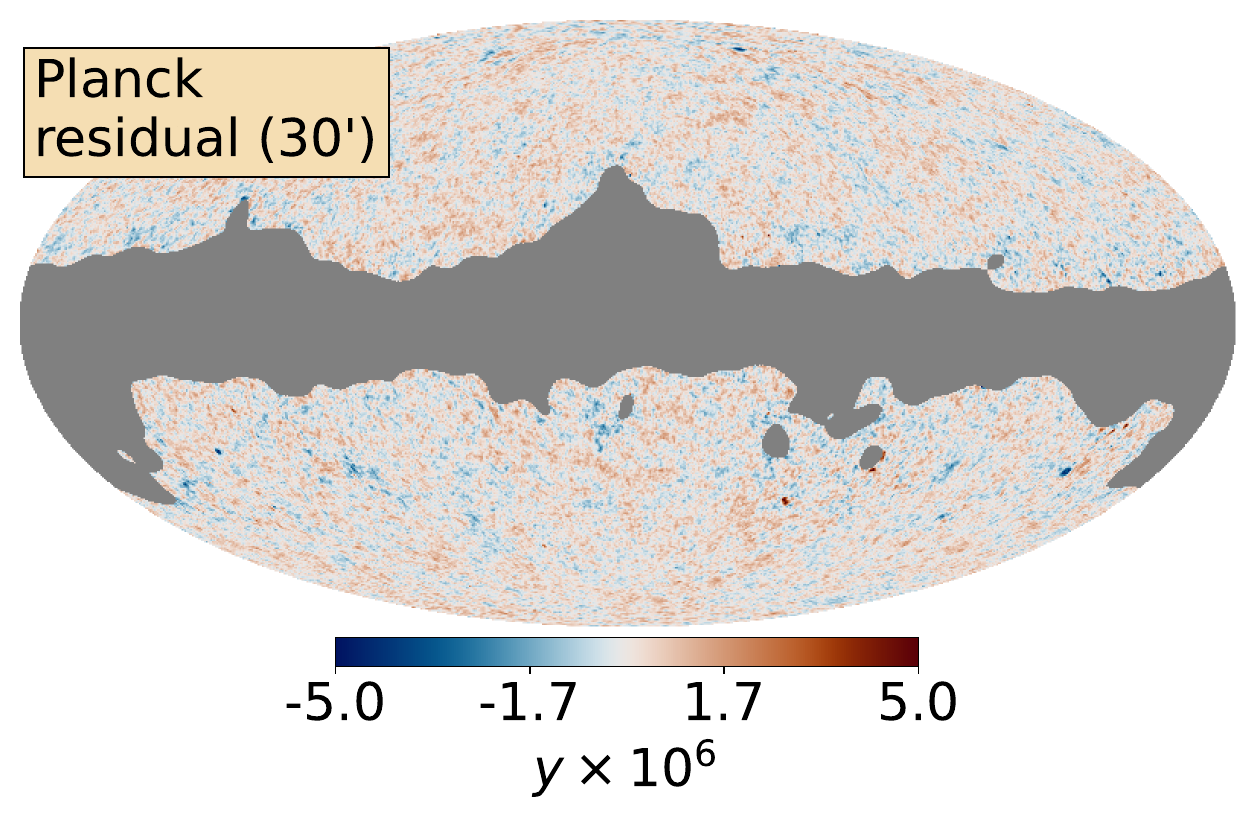}
\caption{\label{fig:ymap} \emph{Top}: Input thermal SZ $y$-map of the sky simulation smoothed to $30'$ angular resolution. \emph{Middle}: Recovered \lb\ thermal SZ $y$-map at $30'$ angular resolution after foreground cleaning with NILC (\emph{left}) and residuals (\emph{right}). \emph{Bottom}: Recovered \pl\ thermal SZ $y$-map after foreground cleaning with NILC (\emph{left}) and residuals (\emph{right}), smoothed to $30'$ angular resolution (originally, $10'$) for comparison. The $y$-map obtained from \lb\ exhibits significantly lower residual contamination compared to the \pl\ $y$-map.} 
\end{figure}

\Cref{fig:ymap} shows the reconstructed \lb\ $y$-map at $30'$ angular resolution over $67\,\%$ of the sky after component separation using NILC (middle left panel). 
It is compared with the input $y$-map of the simulation that has been smoothed to the same $30'$ resolution over the same sky area (top panel). The difference between the \lb\ and input $y$-maps is shown in the middle right panel, highlighting residual foreground and noise contamination.
The depicted mask, leaving $67\,\%$ of the sky, is used to mitigate residual foregrounds for power spectrum calculations in \cref{subsec:spectra}, even though component separation is performed over $98\,\%$ of the sky.
Additionally, the reconstructed \pl\ $y$-map, smoothed to $30'$ resolution from its original $10'$, is presented in the bottom left panel, while the difference between the \pl\ and input $y$-maps is shown in the bottom right panel. Most of the galaxy clusters of the input $y$-map are clearly recovered in both \lb\ and \pl\ $y$-maps after component separation. However, the \lb\ $y$-map demonstrates notably higher fidelity to the true $y$-map in contrast to the \pl\ $y$-map, which exhibits more significant residual contamination. This improvement of \lb\ over \pl\ can be attributed to the combination of higher sensitivity in \lb's channels and the availability of a greater number of frequency bands for efficient foreground mitigation and component separation.

\begin{figure}[tbp]
\centering 
\includegraphics[width=0.5\textwidth,clip]{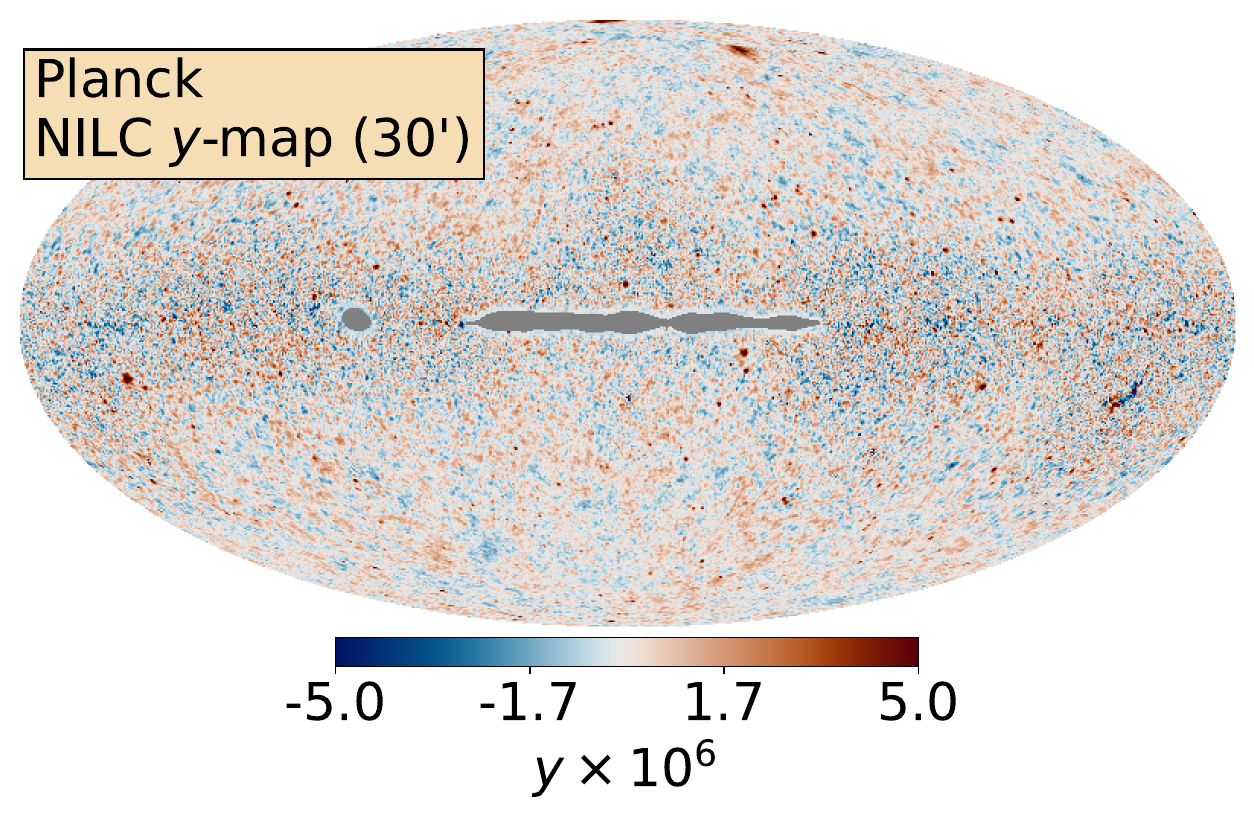}~
\includegraphics[width=0.5\textwidth,clip]{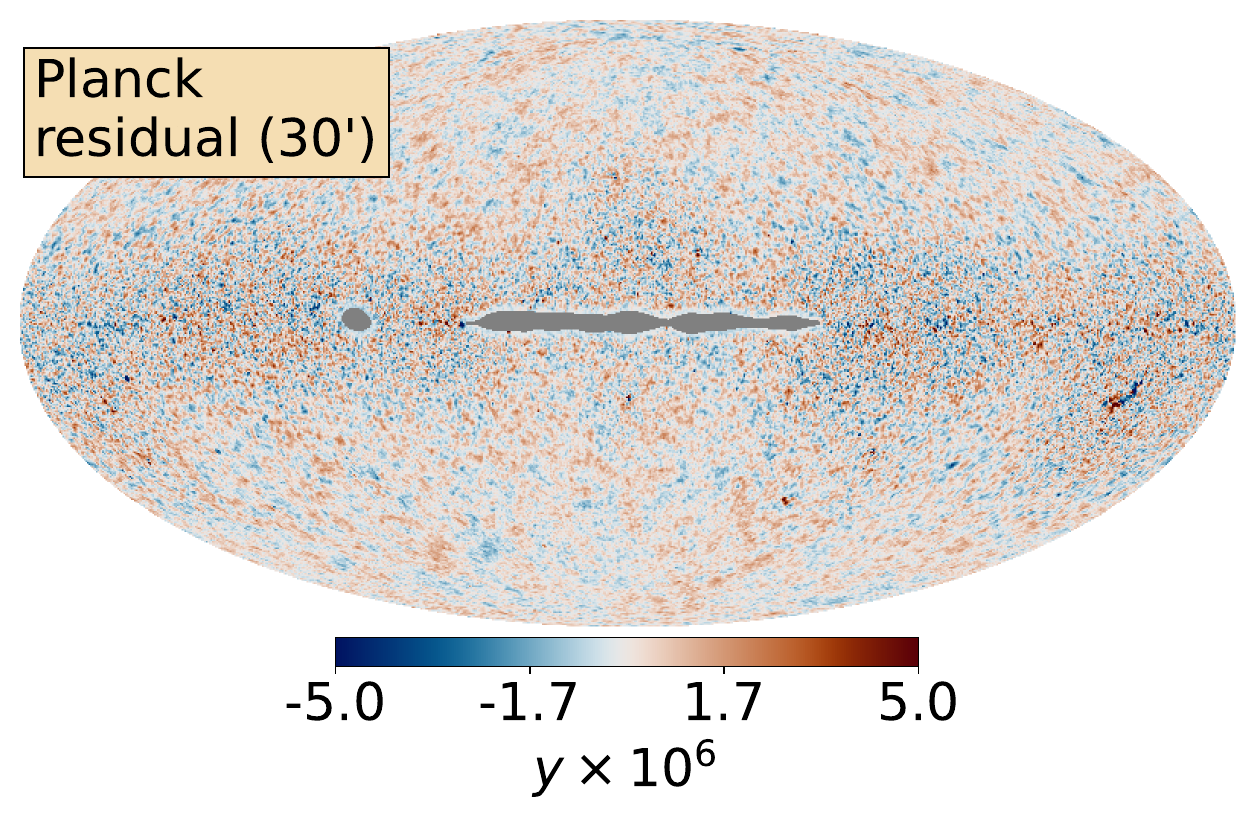}
\hfill
\includegraphics[width=0.5\textwidth,clip]{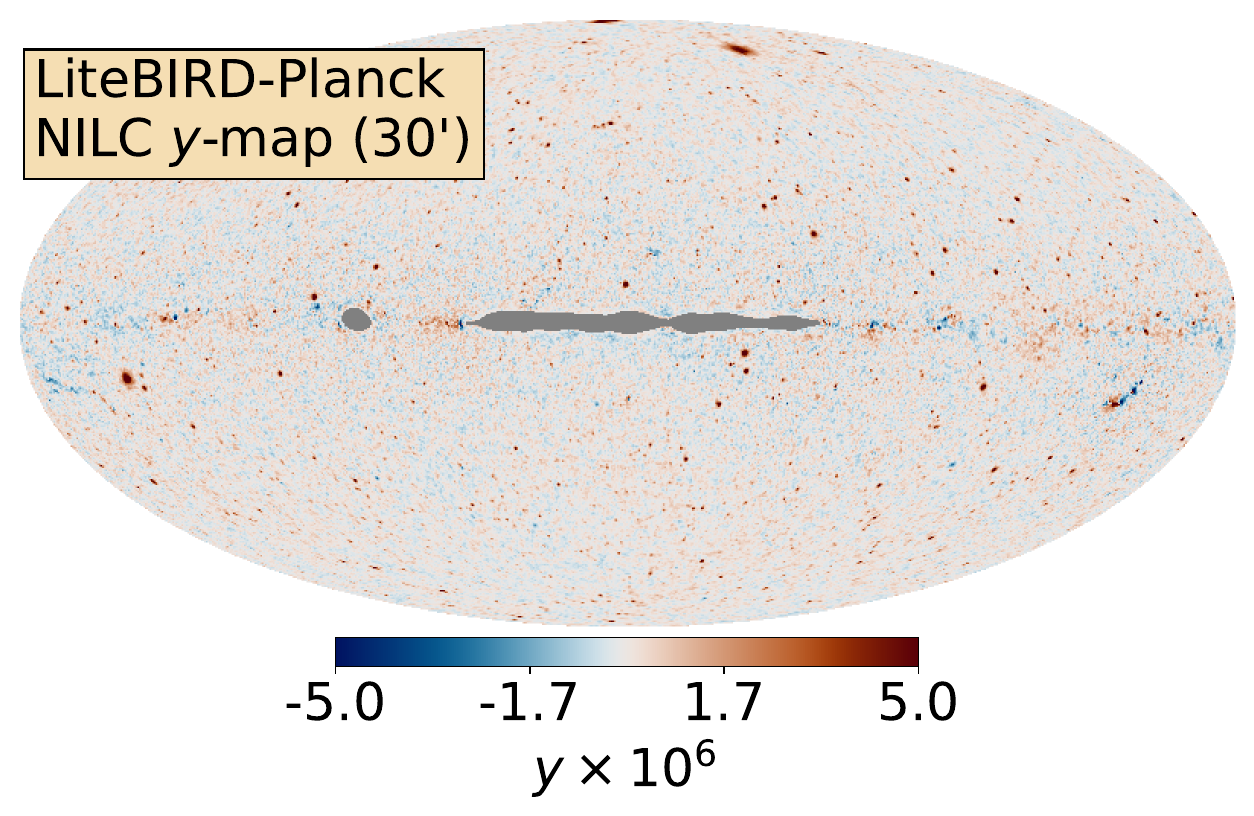}~
\includegraphics[width=0.5\textwidth,clip]{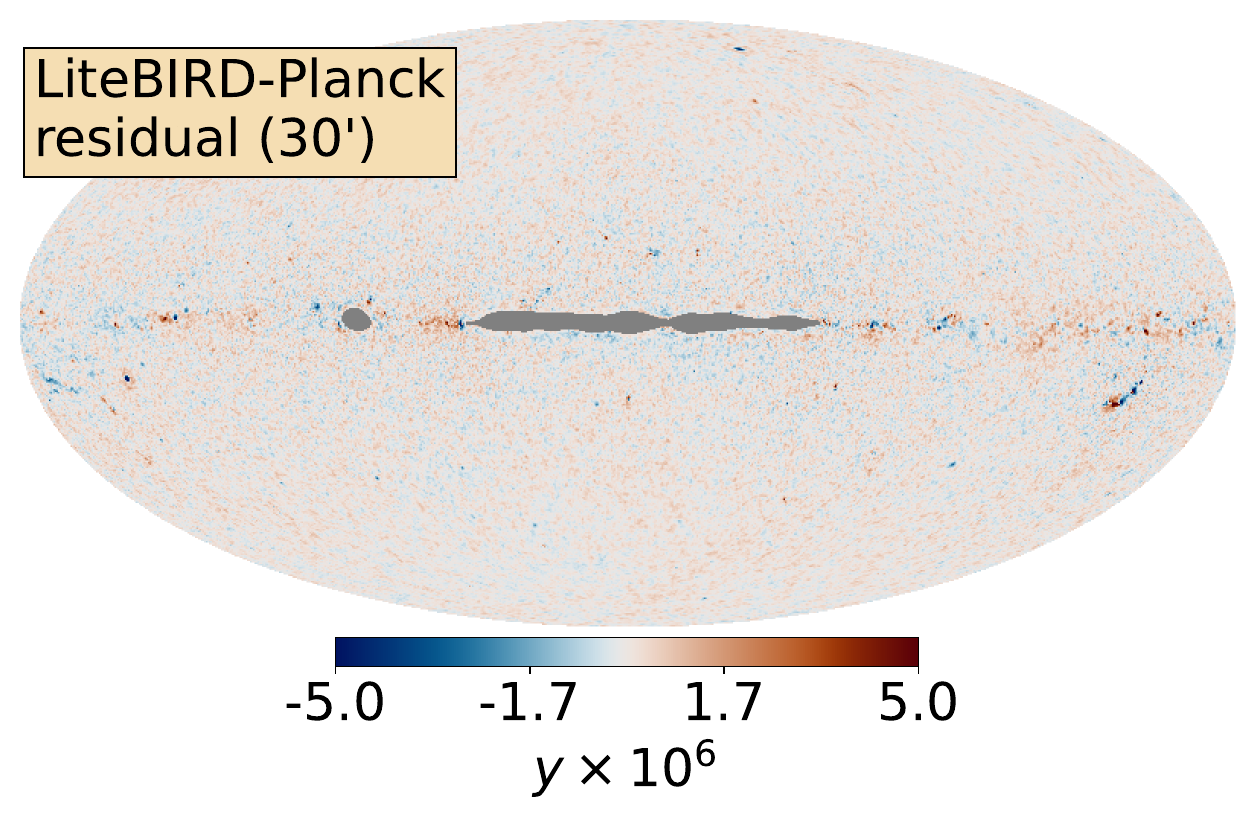}~
\caption{\label{fig:ymap_fsky98pc} Comparison of the \pl\ $y$-map (\emph{top left}) and its associated residuals (\emph{top right}) with the \lb-\pl\ combined $y$-map (\emph{bottom left}) and its associated residuals (\emph{bottom right}) over $f_{\rm sky}=98\,\%$ of the sky. The inclusion of \lb\ channels significantly reduces the residual foreground contamination, especially around the Galactic plane, compared to \pl\ alone. Maps are smoothed to $30'$ resolution (originally, $10'$) for rendering.}
\end{figure}

By combining both \lb\ and \pl\ simulated data sets with NILC, the quality of the reconstructed $y$-map can be further improved, thanks to the additional increase in the number of available frequency channels for component separation. \Cref{fig:ymap_fsky98pc} compares the \pl\ $y$-map (top left panel) with the \lb-\pl\ combined $y$-map (bottom left panel) over $98\,\%$ of the sky, along with their associated residuals (i.e., the difference between the reconstructed and input $y$-maps) in the top right and bottom right panels, respectively. The \pl\ $y$-map exhibits significant residual contamination around the Galactic plane, whereas the \lb-\pl\ combined $y$-map displays uniformly high signal-to-noise throughout the sky, revealing observable galaxy clusters in the Galactic plane region. The presence of \lb\ channels effectively contributes to reducing the residual contamination left by the \pl\ channels in the Compton $y$-map.

\begin{figure}[tbp]
\centering 
\includegraphics[width=0.33\textwidth,clip]{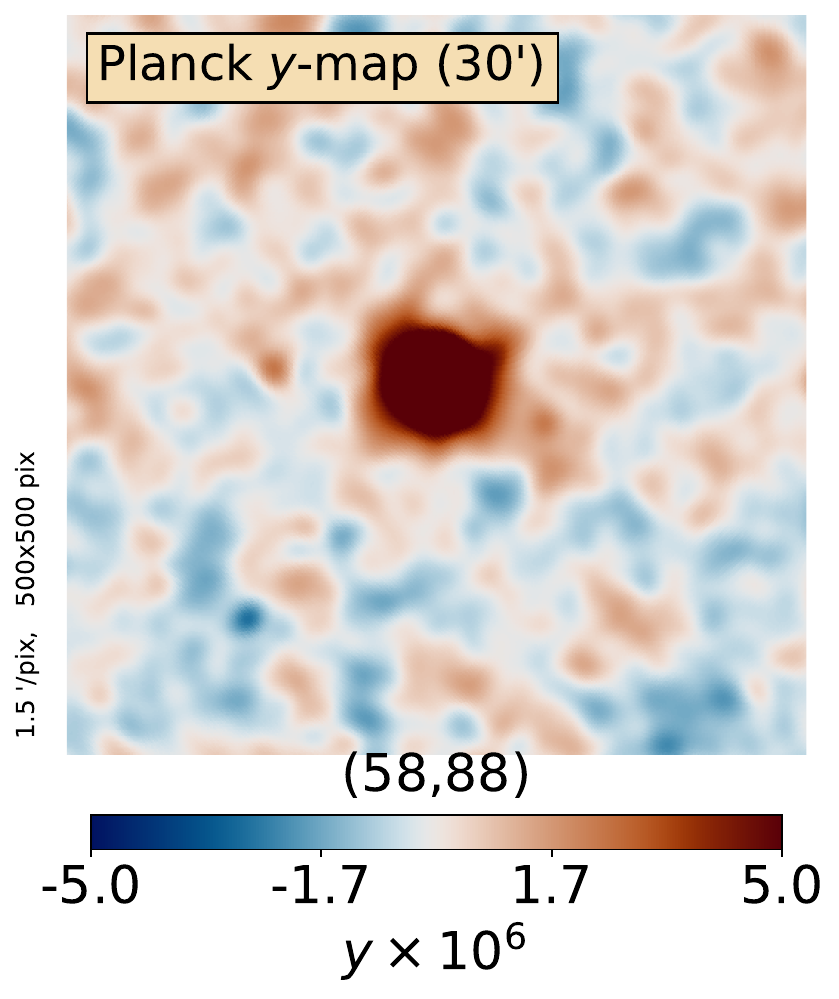}~
\includegraphics[width=0.33\textwidth,clip]{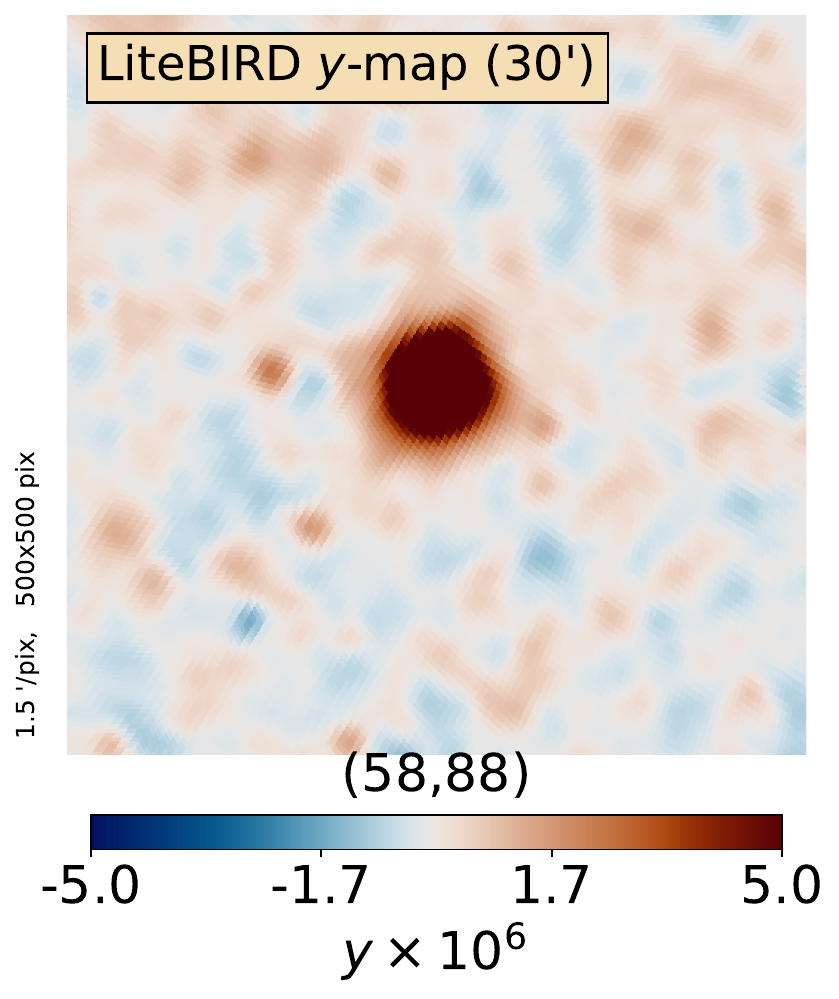}~
\includegraphics[width=0.33\textwidth,clip]{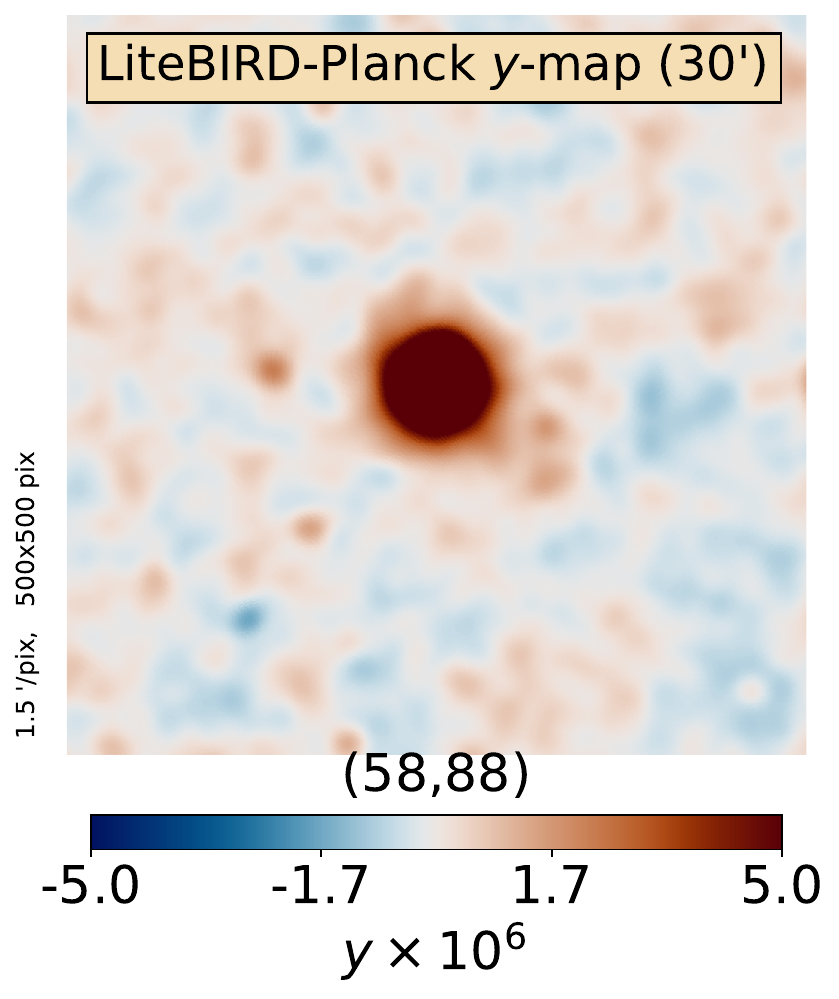}~\\
\includegraphics[width=0.33\textwidth,clip]{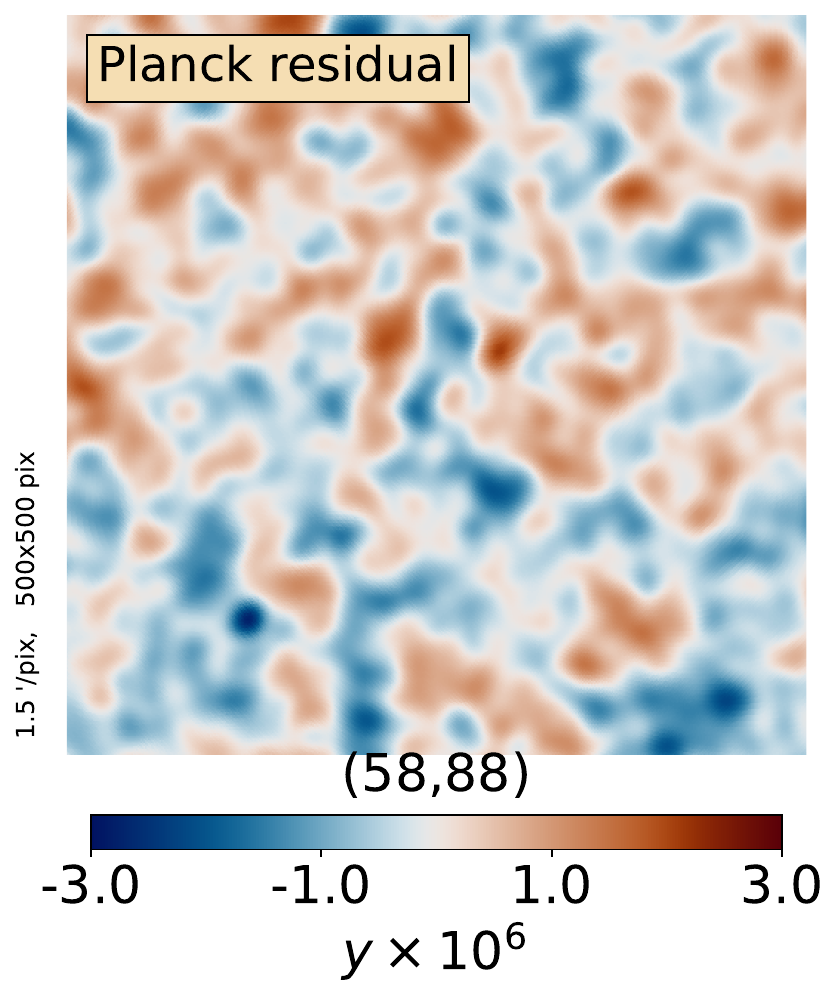}~
\includegraphics[width=0.33\textwidth,clip]{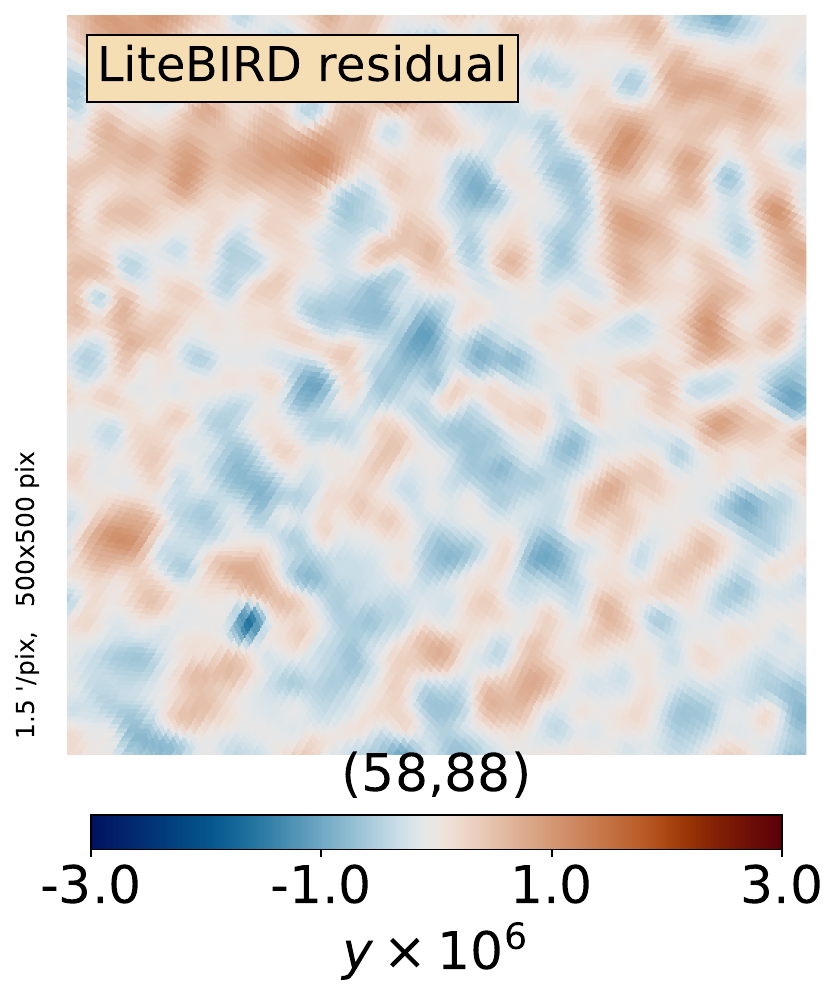}~
\includegraphics[width=0.33\textwidth,clip]{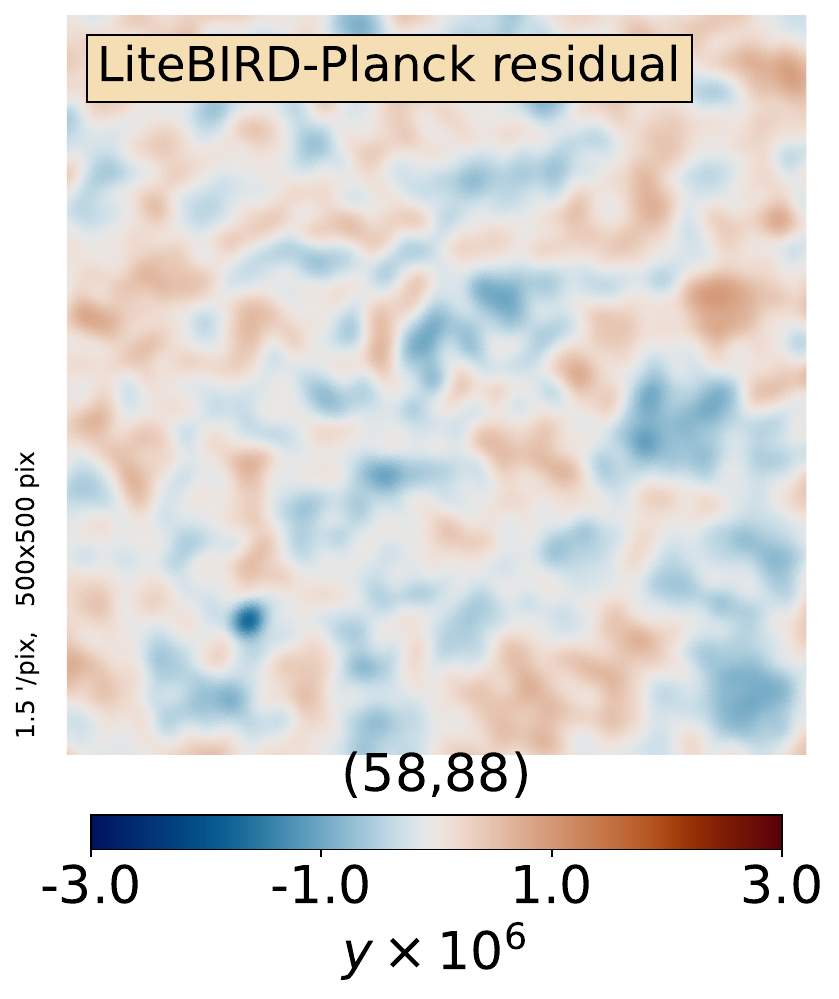}~\\
\caption{\label{fig:coma} \emph{Top}: $12.5^\circ \times 12.5^\circ$ gnomonic projection of the reconstructed $y$-maps (all smoothed to $30'$ resolution) centred on the Coma cluster at Galactic coordinates $(l,b)=(58.1^\circ,88.0^\circ)$. \emph{Left}: \pl\ $y$-map. \emph{Middle}: \lb\ $y$-map. \emph{Right}: \lb-\pl\ combined $y$-map. \emph{Bottom}: Difference between the reconstructed and input $y$-maps (residual) in the same sky area. The residual foreground and noise contamination gets more and more reduced from the left to the right panel, with the RMS value of the residual maps being $0.64\times 10^{-6}$ for \pl, $0.35\times 10^{-6}$ for \lb, and $0.32\times 10^{-6}$ for \lb-\pl\ in this sky area.}
\end{figure}

\Cref{fig:coma} presents a zoomed-in view of the Coma cluster using a $12.5^\circ \times 12.5^\circ$ gnomonic projection for the \pl\ (left), \lb\ (middle), and \lb-\pl\ (right) $y$-maps centred at Galactic coordinates $(l,b)=(58.1^\circ,88.0^\circ)$. To highlight residuals in each $y$-map, the difference between the reconstructed $y$-maps and the input $y$-map is also shown in the bottom row. Evidently, the background contamination diminishes progressively from the left to the right panels, with corresponding RMS values of $0.64\times 10^{-6}$ for \pl, $0.35\times 10^{-6}$ for \lb\ (a $45$\,\% decrease), and $0.32\times 10^{-6}$ for \lb-\pl\ (a $50$\,\% decrease), again emphasising the enhanced performance of \lb\ over \pl\ in mitigating foregrounds and the benefits of combining both data sets for thermal SZ effect reconstruction.

\begin{figure}[tbp]
\centering 
\includegraphics[width=0.27\textwidth,clip]{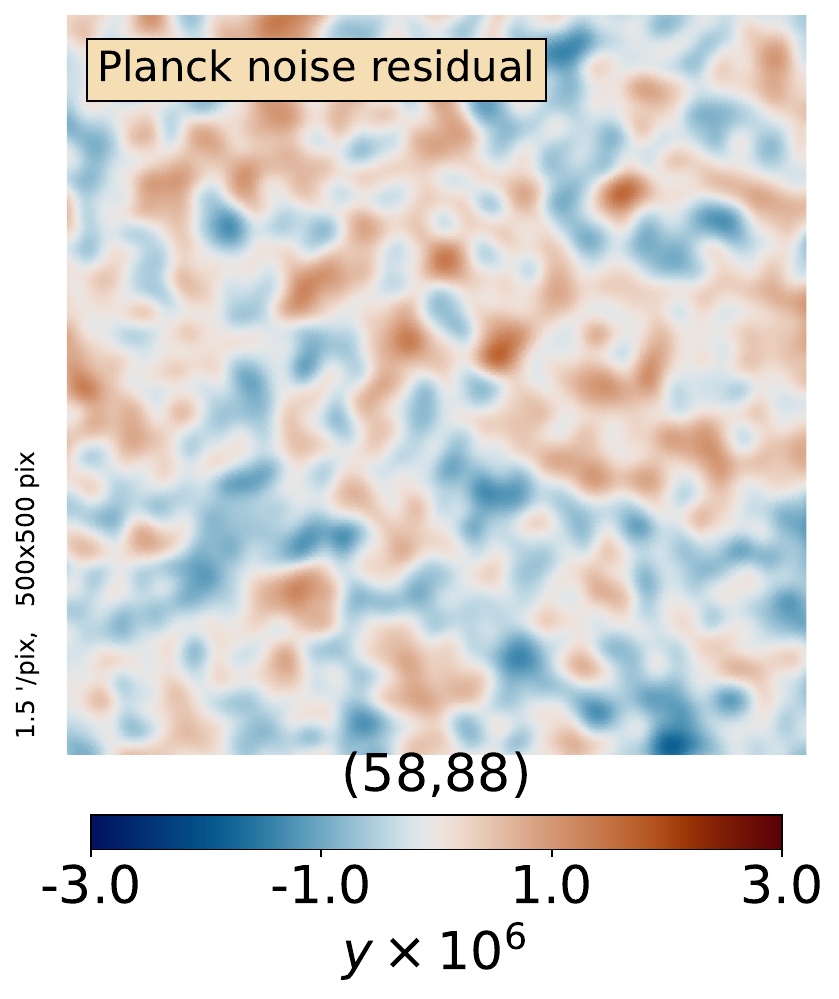}~
\includegraphics[width=0.27\textwidth,clip]{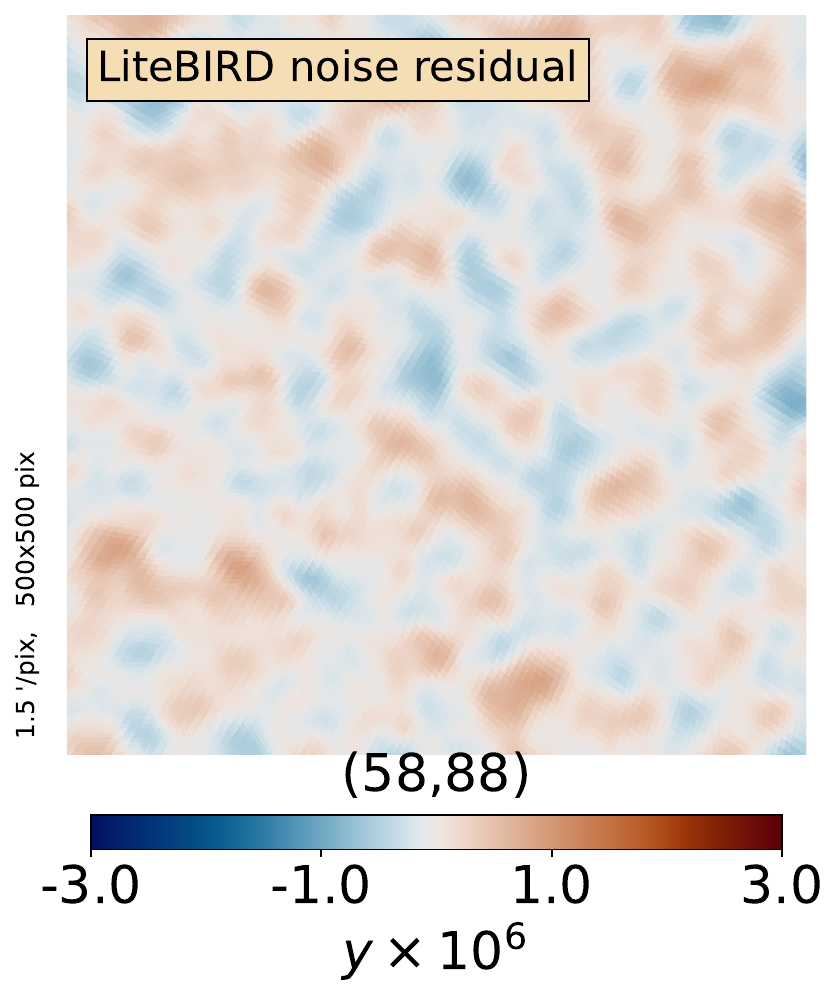}~
\includegraphics[width=0.27\textwidth,clip]{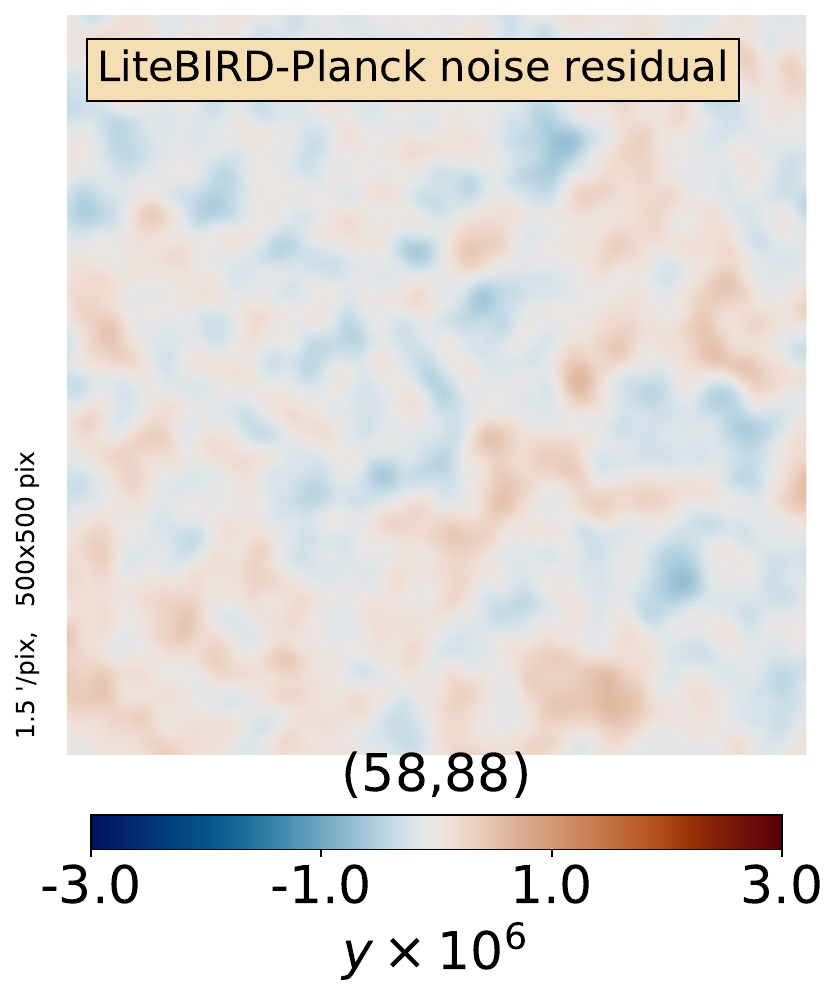}~\\
\includegraphics[width=0.27\textwidth,clip]{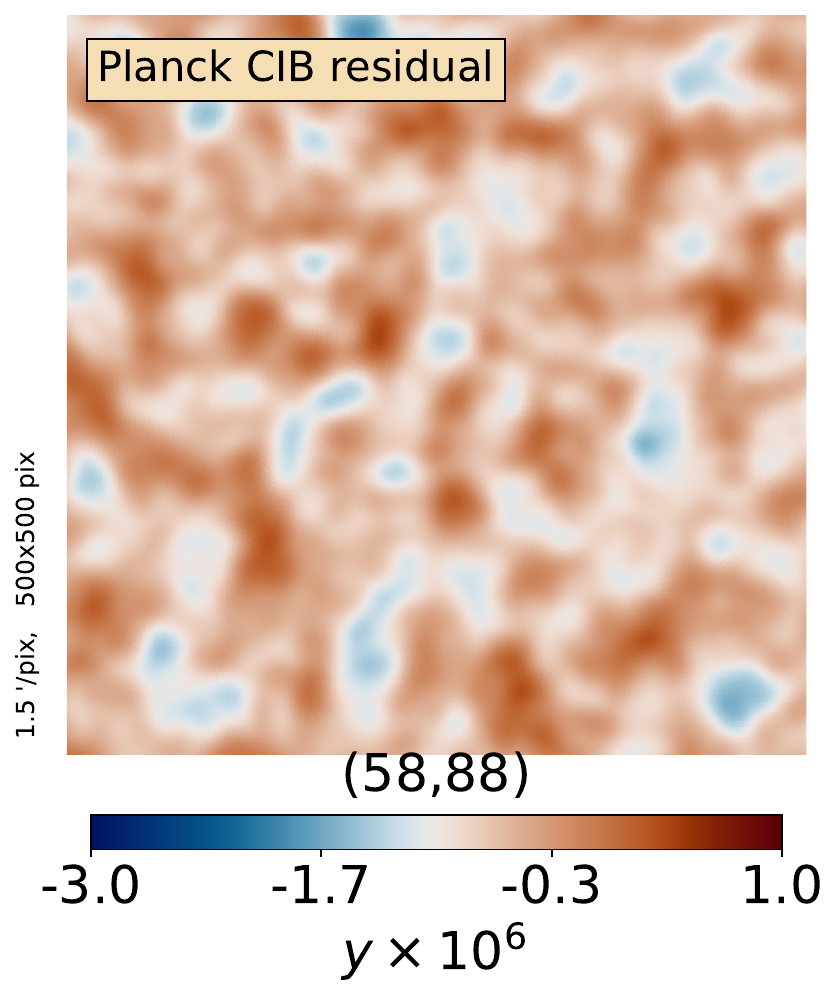}~
\includegraphics[width=0.27\textwidth,clip]{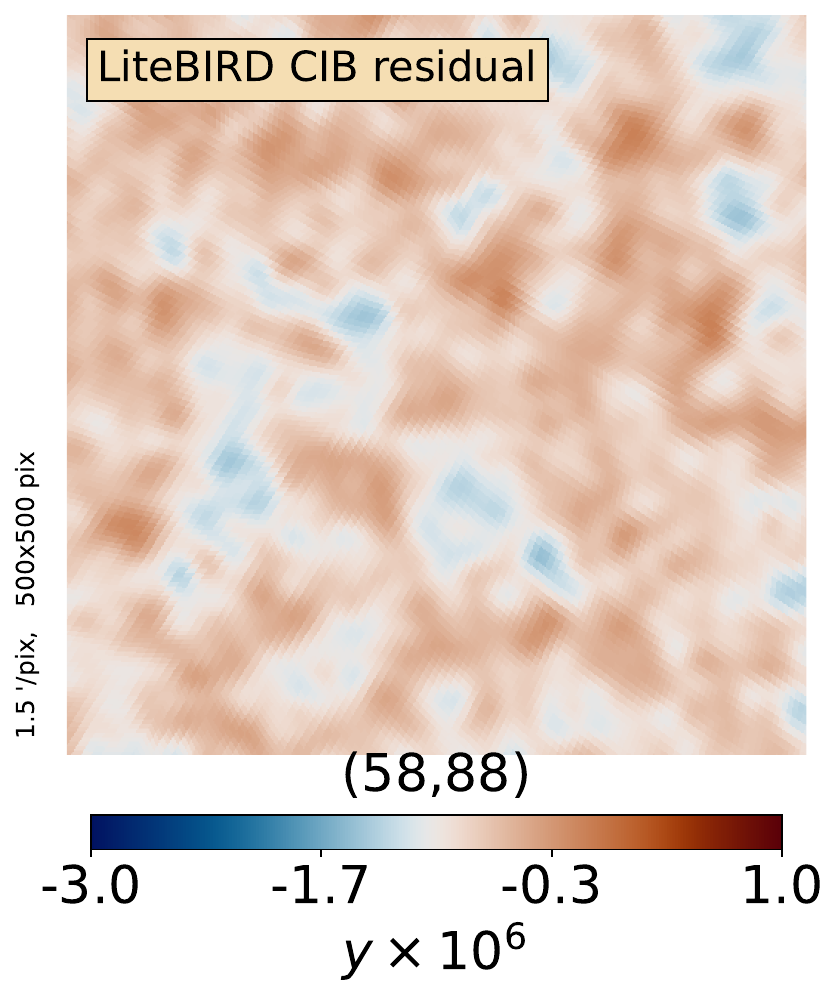}~
\includegraphics[width=0.27\textwidth,clip]{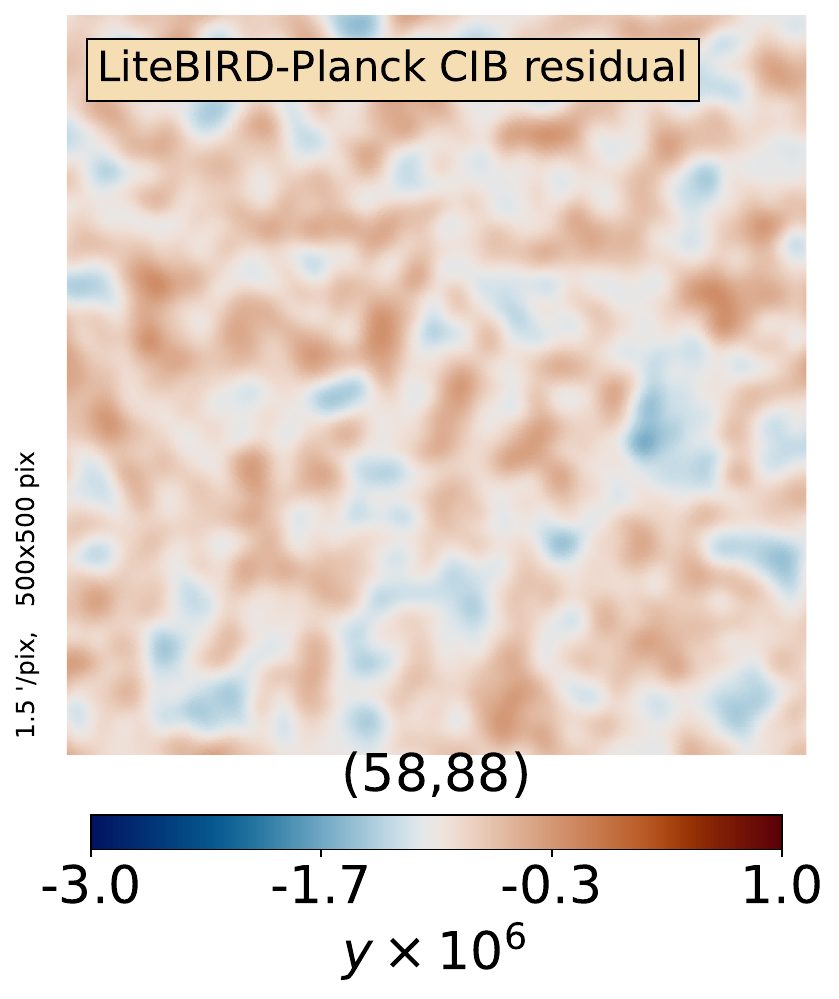}~\\
\includegraphics[width=0.27\textwidth,clip]{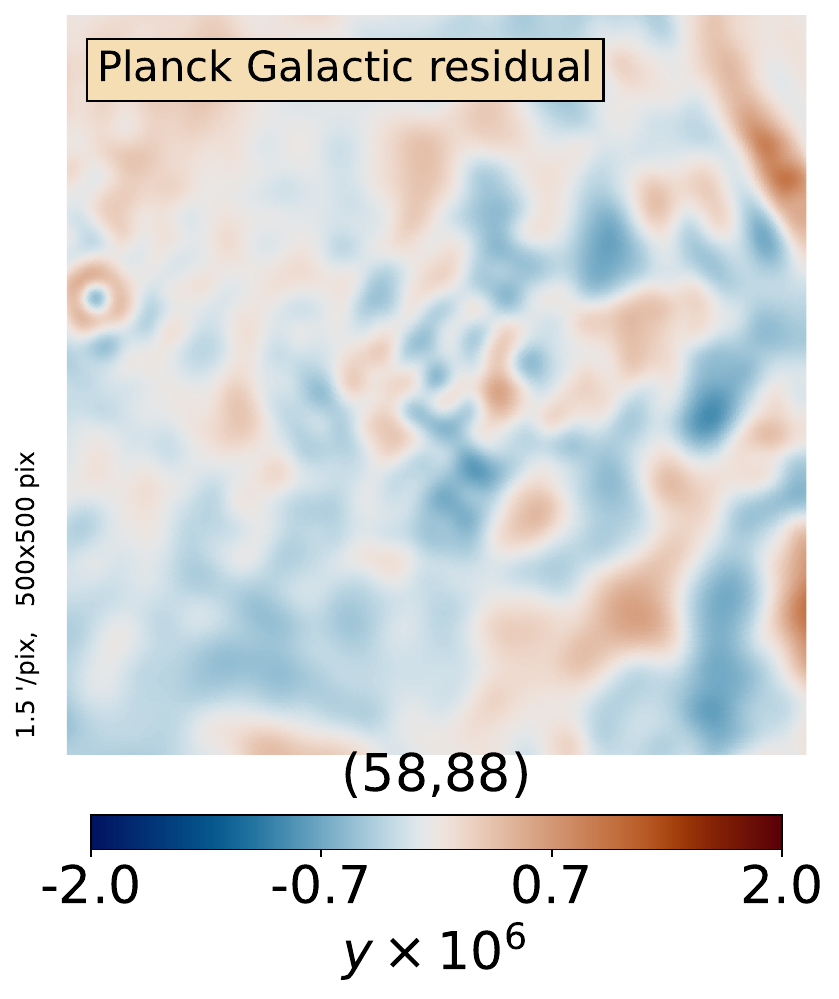}~
\includegraphics[width=0.27\textwidth,clip]{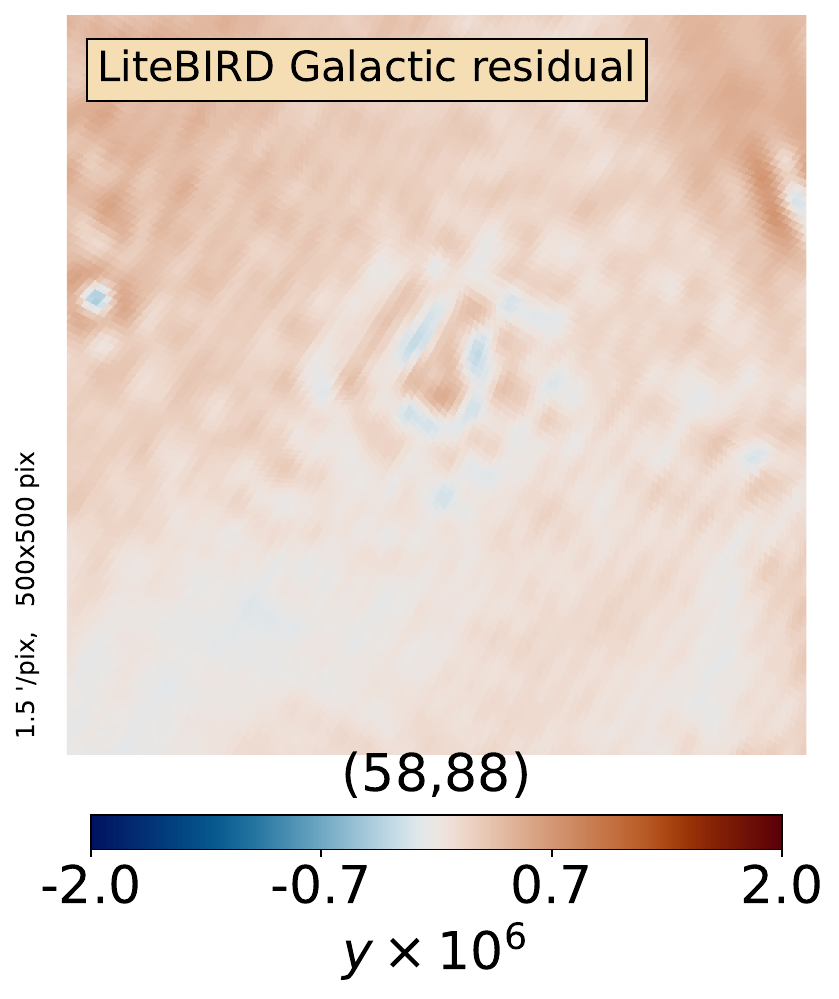}~
\includegraphics[width=0.27\textwidth,clip]{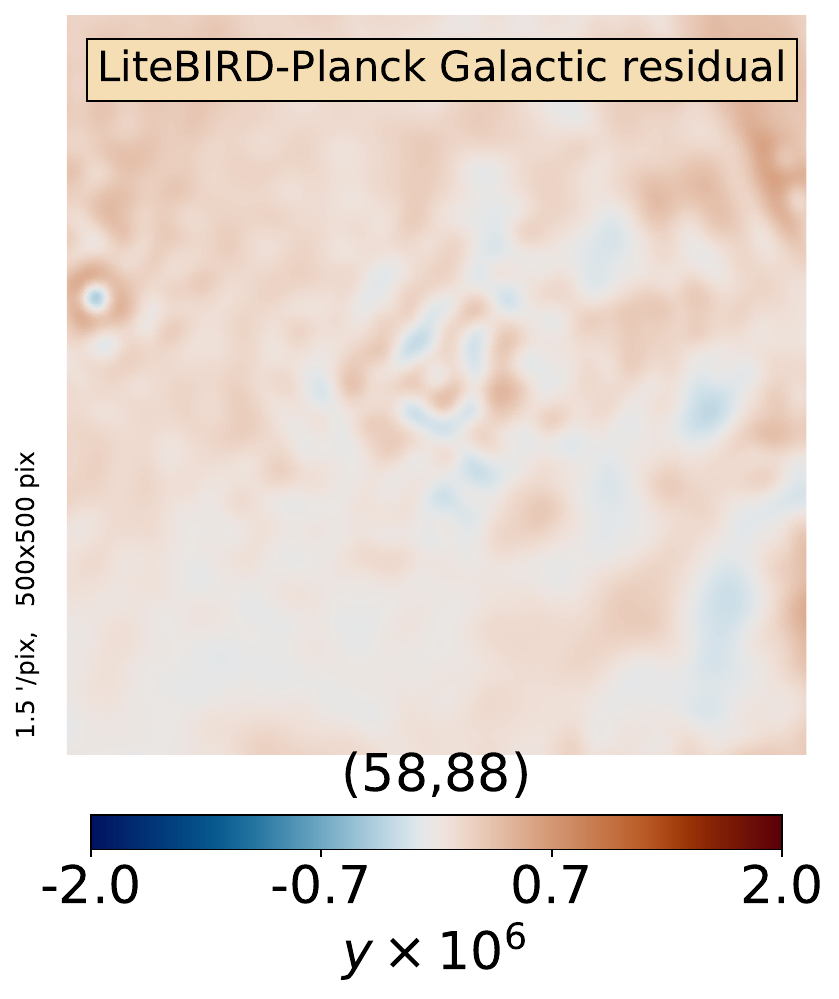}~\\
\includegraphics[width=0.27\textwidth,clip]{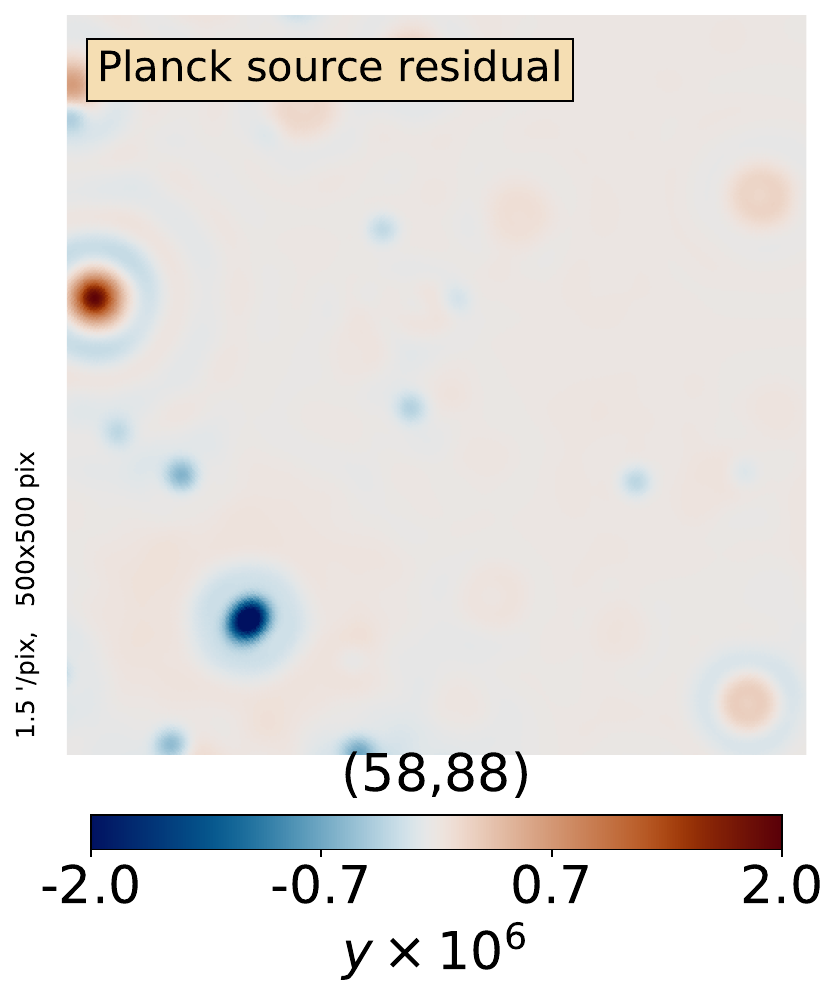}~
\includegraphics[width=0.27\textwidth,clip]{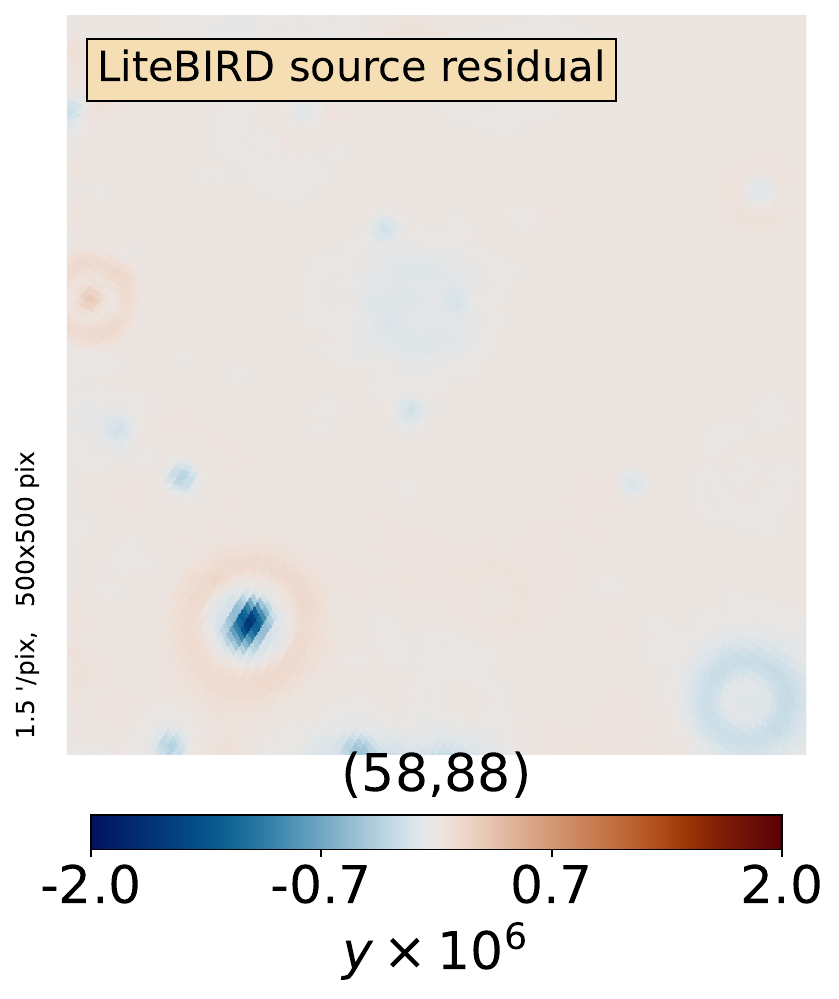}~
\includegraphics[width=0.27\textwidth,clip]{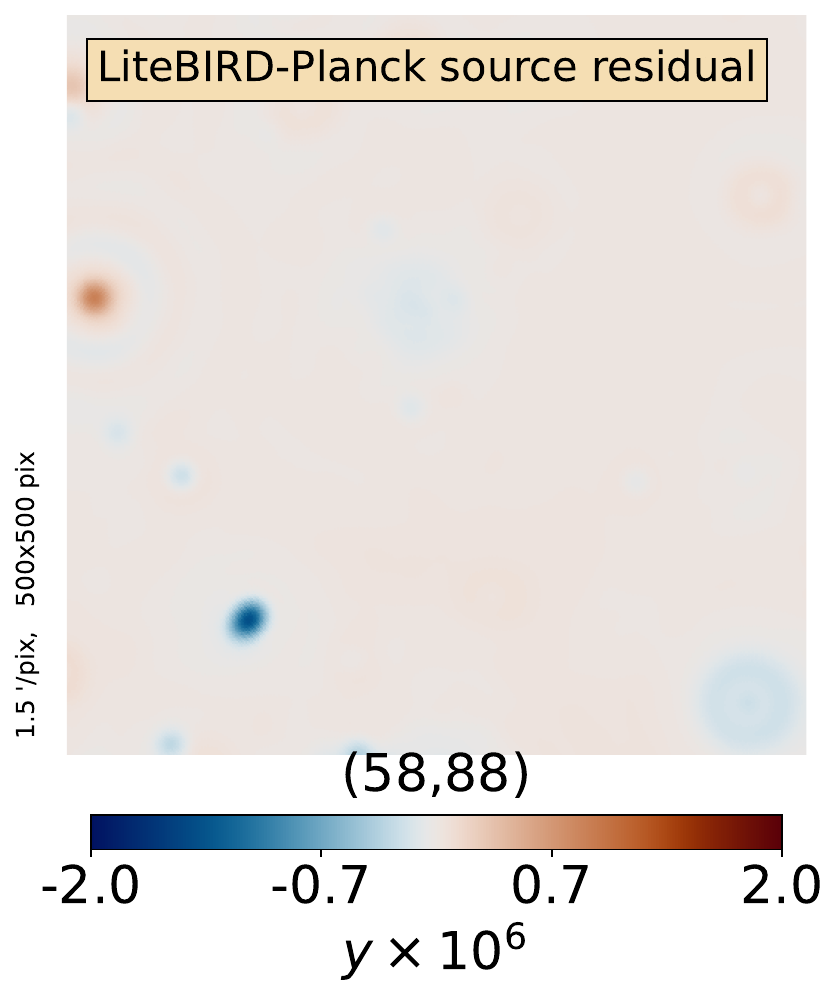}
\caption{\label{fig:coma2} Residual contamination from noise (\emph{first row}), CIB (\emph{second row}), Galactic foregrounds (\emph{third row}), and extragalactic sources (\emph{fourth row}) for the \pl\ (\emph{left}), \lb\ (\emph{middle}), and \lb-\pl\ (\emph{right}) $y$-maps of \cref{fig:coma}.}
\end{figure}

The contribution from different foregrounds to the overall residual contamination of the  \pl\ (\emph{left}), \lb\ (\emph{middle}), and \lb-\pl\ (\emph{right}) $y$-maps is presented in \cref{fig:coma2}, showing noise (top row), CIB (second row), Galactic foregrounds (third row), and extragalactic sources (bottom row). 
These residual foreground maps were generated by applying the same NILC weights to the individual foreground components of the simulation as were used to reconstruct the $y$-map. 
As evident from \cref{fig:coma2}, the residual contamination from either noise, CIB, or Galactic foregrounds is significantly reduced in the \lb\ $y$-map compared to the \pl\ $y$-map, with further suppression observed in the \lb-\pl\ combined $y$-map owing to the increased number of frequency channels for component separation.
 
 As visible from the bottom row of \cref{fig:coma2}, the \lb\ $y$-map also shows much lower contamination by infrared sources,\footnote{In the reconstructed $y$-maps, infrared sources (prominent at high frequencies) appear as positive red spots, while radio sources (prominent at low frequencies) appear as negative blue spots, due to the change in sign of the thermal SZ SED and thus the NILC weights from positive to negative above and below $217$\,GHz, respectively.} with virtually no discernible red spot near the left edge, unlike the \pl\ $y$-map where a distinct red spot is evident at the same angular resolution. The larger contamination by infrared sources in the \pl\ $y$-map can be attributed to the inclusion of high-frequency channels above $545$\,GHz, where infrared sources are more intense. Those channels are absent from \lb, resulting in limited infrared source contamination. However, infrared source contamination is noticeable again in the \lb-\pl\ combined $y$-map due to the reintroduction of  \pl\ channels, albeit with reduced intensity compared to the \pl\ $y$-map. Overall, the residual contamination from both infrared (red spots) and radio (blue spots) sources is lower in the \lb-\pl\ $y$-map compared to the \pl\ $y$-map.  
This improvement is credited to the increased frequency sampling by incorporating \lb\ channels, which enables NILC to better capture source correlations across frequencies and consequently leads to more effective cleaning of source contamination.

\begin{figure}[tbp]
\centering 
\includegraphics[width=0.5\textwidth,clip]{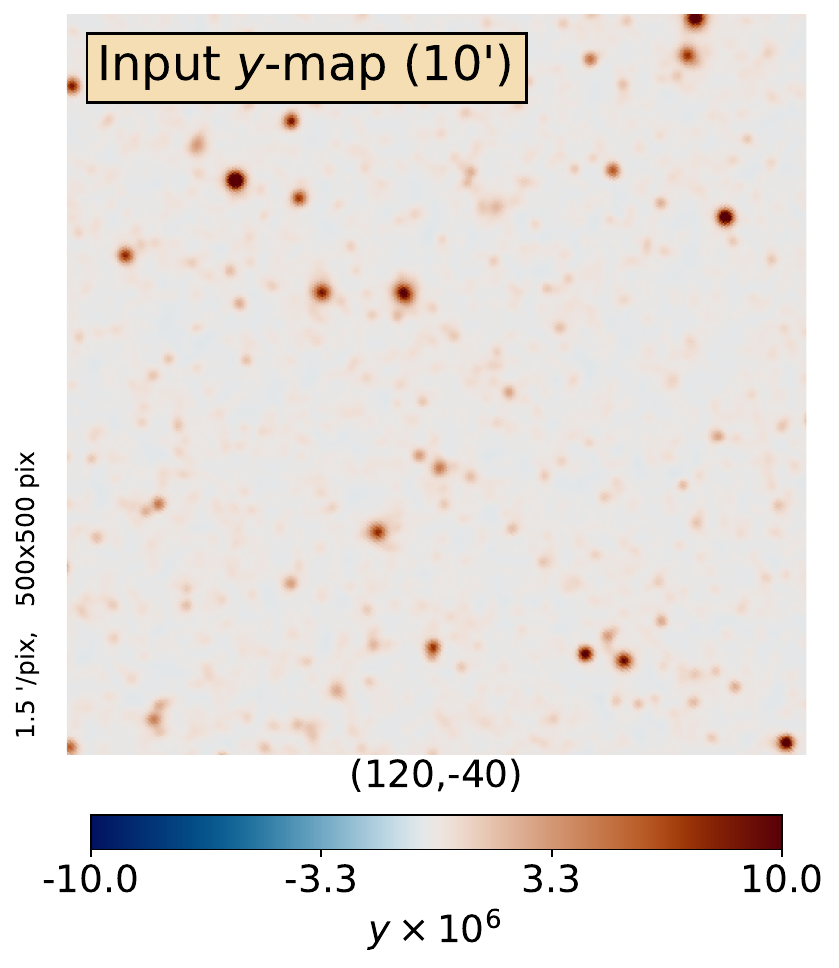}~
\includegraphics[width=0.5\textwidth,clip]{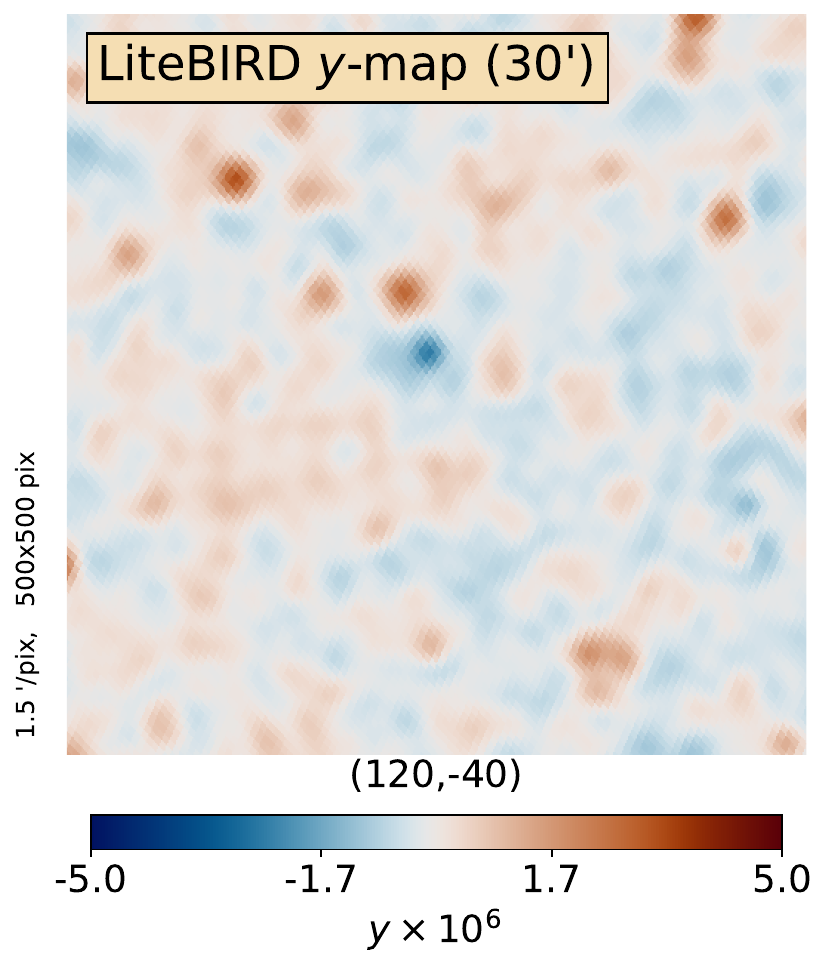}
\hfill
\includegraphics[width=0.5\textwidth,clip]{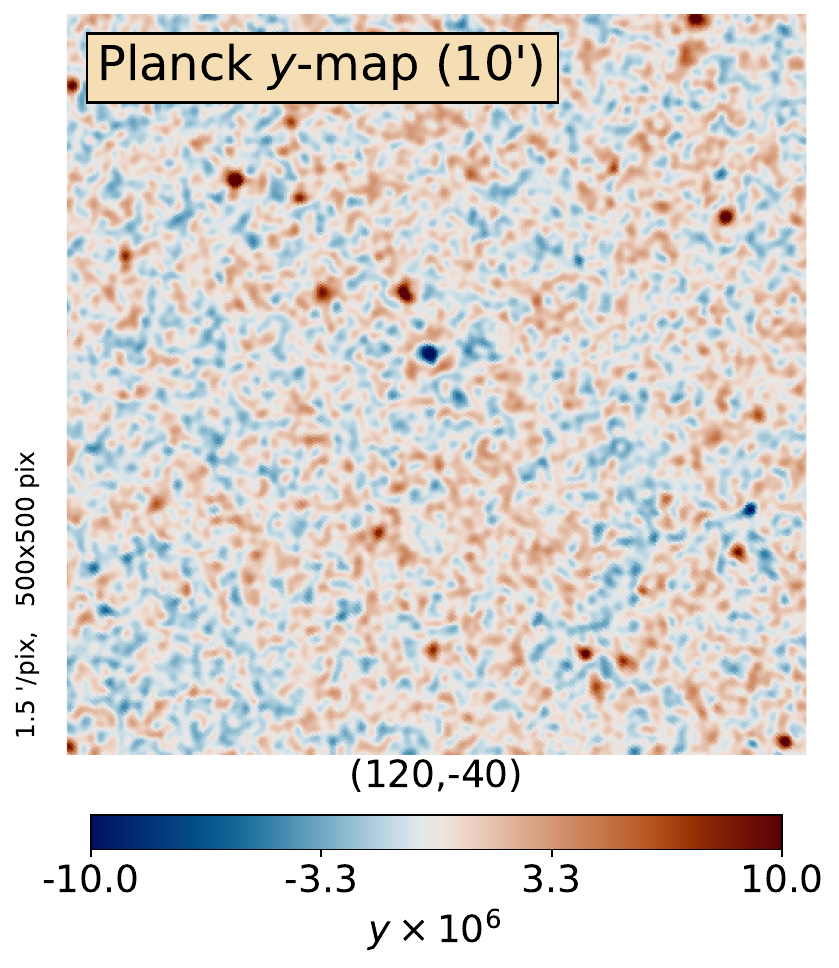}~
\includegraphics[width=0.5\textwidth,clip]{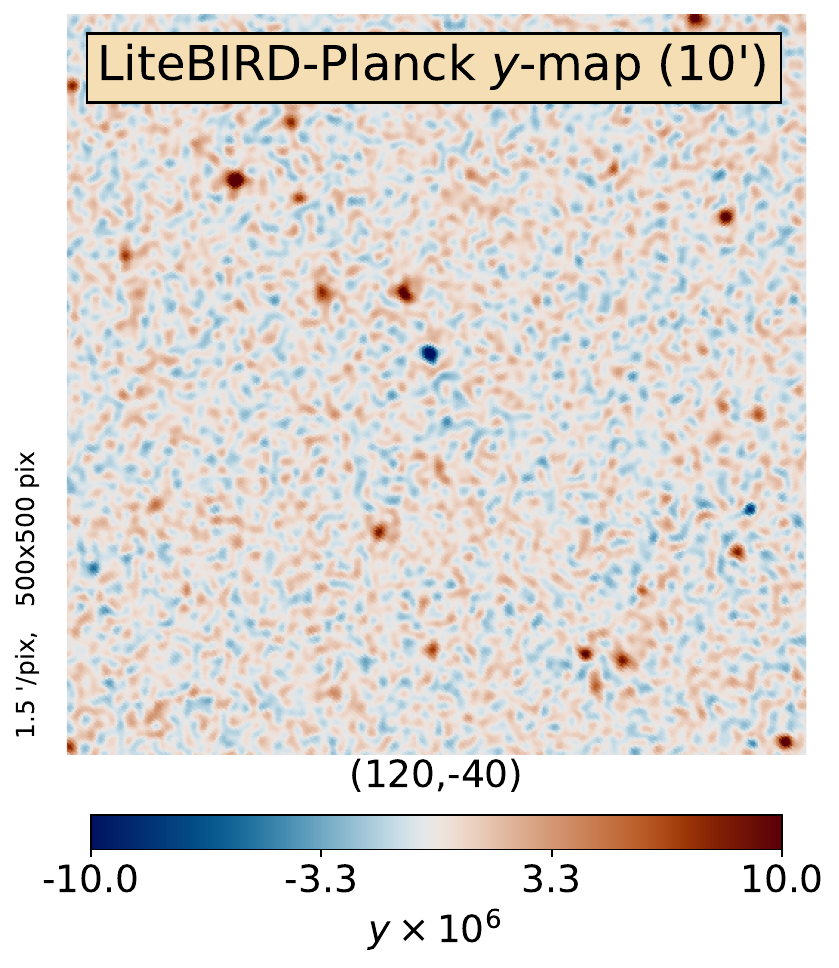}~
\caption{\label{fig:resolution} $12.5^\circ \times 12.5^\circ$ gnomonic projection of the $y$-maps at optimal resolution centred at $(l,b)=(120^\circ,-40^\circ)$. Input $y$-map at $10'$ resolution (\emph{top left}), \lb\ $y$-map at $30'$ resolution (\emph{top right}), \pl\ $y$-map at $10'$ resolution (\emph{bottom left}), and \lb-\pl\ combined $y$-map at $10'$ resolution (\emph{bottom right}). Due to the lower resolution of the \lb\ $y$-map, the colour range is reduced by half compared to other $y$-maps. Most compact clusters cannot be resolved in the \lb\ $y$-map due to beam limitations, while they are detected in the \lb-\pl\ combined $y$-map, with higher signal-to-noise compared to the \pl\ $y$-map.}
\end{figure}

The advantage of the \pl\ channels lies in their capability to provide angular resolution beyond the beam limits of the \lb\ channels, enabling the resolution of more compact galaxy clusters with characteristic radius smaller than the average beam size of \lb. \Cref{fig:resolution} showcases a comparison of the various $y$-maps at their optimal resolution within a $12.5^\circ \times 12.5^\circ$ region centred at Galactic coordinates $(l, b) = (120^\circ, -40^\circ)$. Due to the limited $30'$ angular resolution, the \lb\ $y$-map cannot resolve the most compact galaxy clusters, while they are detectable in both the \pl\ $y$-map and the \lb-\pl\ combined $y$-map, both reconstructed at $10'$ resolution, with the \lb-\pl\ $y$-map exhibiting higher signal-to-noise compared to the \pl\ $y$-map.  

In conclusion, the combination of \lb\ and \pl\ channels in the component-separation pipeline results in an optimal all-sky Compton $y$-map that capitalises on the strengths of both experiments. The high-resolution \pl\ channels allow for the resolution of compact clusters beyond the beam limitations of \lb, while the numerous sensitive \lb\ channels effectively mitigates foreground contamination.

\subsection{SZ power spectrum and residuals}\label{subsec:spectra}  

The $y$-map is obtained by applying the NILC weights to the sky maps across the frequencies. Since these maps are themselves a linear combination of various emission components, by applying the same NILC weights to the maps of a specific foreground component of the simulation across the frequencies, we can assess the residual contamination in the $y$-map attributed to this foreground. The same procedure is applied for assessing projected noise.

\begin{figure}[tbp]
\centering 
\includegraphics[width=0.559\textwidth,clip]{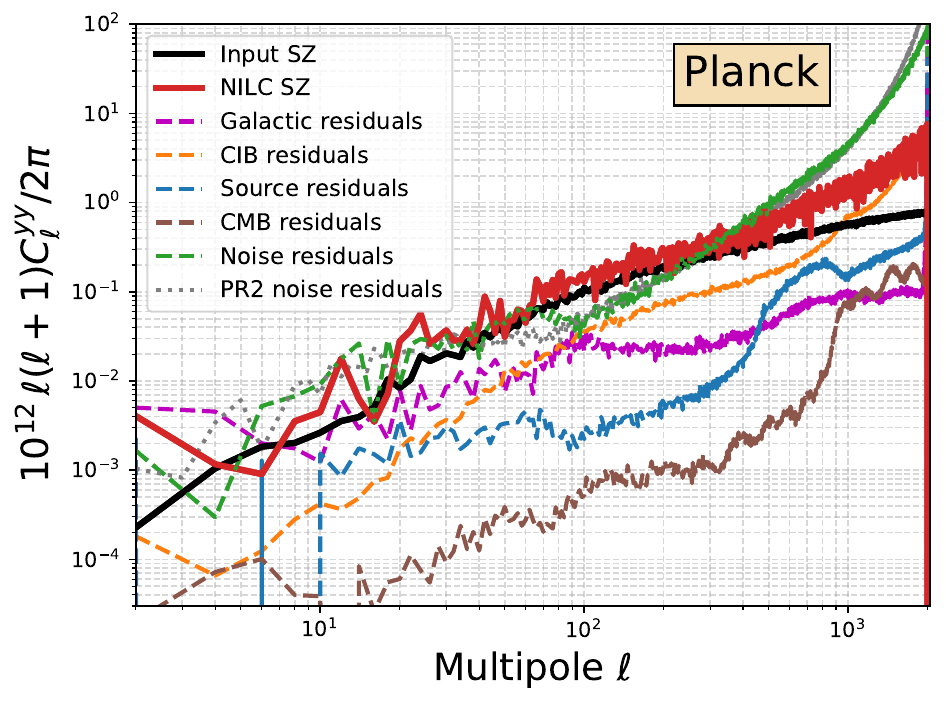}
\hfill
\includegraphics[width=0.559\textwidth,clip]{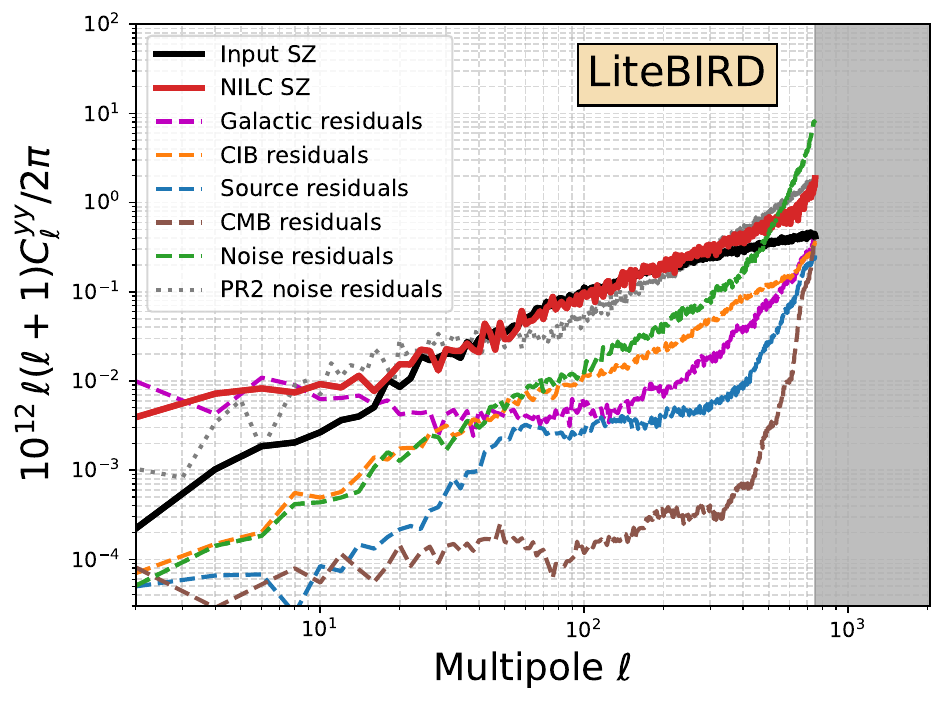}
\hfill
\includegraphics[width=0.559\textwidth,clip]{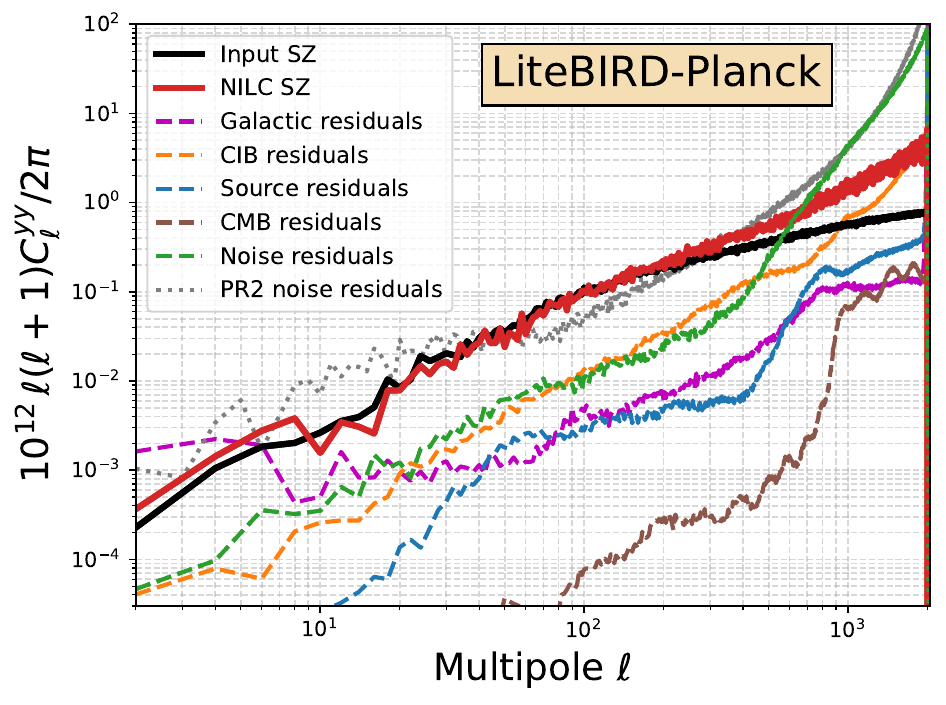}
\caption{\label{fig:ps_comb} \emph{Top}: \pl\ $y$-map power spectrum (solid red) versus input $y$-map power spectrum (solid black), with projected foregrounds (dashed coloured lines) and noise (dashed green) power spectra.  \emph{Middle}: \lb\ $y$-map power spectrum.  \emph{Bottom}:  \lb-\pl\ combined $y$-map power spectrum. The combination of \lb\ and \pl\ channels (\emph{bottom}) leads to lower foreground residuals and noise. }
\end{figure}

Using \texttt{MASTER} \citep{Hivon2002}, we thus compute the angular power spectrum of the $y$-map over $67\,\%$ of the sky, as well as the power spectra of the residual contamination attributed to each foreground component of the simulation over the same portion of the sky. The results are shown in \cref{fig:ps_comb} for the \pl\ (top panel), \lb\ (middle panel) and joint \lb-\pl\ (bottom panel) analyses. The power spectrum of the input $y$-map of the simulation is shown as a solid black line, while the power spectrum of the reconstructed $y$-map corrected for the bias due to the noise power spectrum contribution is shown as a solid red line. The power spectrum contributions from each residual foreground contaminant are shown as dashed coloured lines and the power spectrum of the instrumental noise contamination is shown as a dashed green line. For reference, we include the actual noise power spectrum from the \pl\ NILC full-mission $y$-map of the \pl\ 2015 data analysis \citep{planck2014-a28} as a dotted grey line. This was derived from the half-difference of the so-called first and last half-ring NILC $y$-maps of the Planck Release 2 (PR2). The overlap between the dashed green line and the dotted grey line in the top panel indicates the consistency in the noise levels between the $y$-map reconstruction performed on the \pl\ simulation in this study and the $y$-map reconstruction carried out on real \pl\ data in ref.~\cite{planck2014-a28}.

Starting from the \pl\ simulation results (top panel in \cref{fig:ps_comb}), we see that the residual noise contamination (dashed green) of the \pl\ $y$-map has a level of power as large as the targeted thermal SZ signal (solid black) over a large range of multipoles ($2 \leq \ell \leq 300$), while it significantly exceeds the thermal SZ signal at higher multipoles. After correcting for the noise bias in the $y$-map power spectrum (solid red), the dominant residual foreground contamination arises from Galactic emission at low multipoles $\ell < 100$ (dashed purple) and the CIB at high multipoles $\ell > 100$ (dashed orange). Residual contamination from radio and infrared sources (dashed blue) and CMB (dashed brown) are subdominant.

Comparing with the \lb\ simulation results (middle panel in \cref{fig:ps_comb}), it is evident that the residual noise contamination (dashed green) in the \lb\ $y$-map is approximately one order of magnitude lower than the noise observed in the \pl\  $y$-map for multipoles in the range $2\leq \ell \leq 500$. This reduction can be attributed to the greater number and higher sensitivity of \lb\ channels relative to  \pl\ channels. In contrast, the lower average resolution of \lb\ channels ($30'$) in comparison to  \pl\  channels ($10'$) restricts the capability to investigate the SZ signal at higher multipoles ($\ell > 800$) with an acceptable signal-to-noise ratio. 
The residual contamination due to extragalactic foregrounds also demonstrates significant reduction across all multipoles in the \lb\ $y$-map (middle panel in \cref{fig:ps_comb}) when compared to the \pl\  $y$-map (top panel in \cref{fig:ps_comb}), primarily due to the increased number of frequency channels available for component separation. Additionally, in the case of \lb, the contamination from Galactic foregrounds (dashed purple line) is significantly reduced for multipoles $\ell > 20$ compared to \pl. Nonetheless, at the lowest multipoles ($\ell < 20$), the Galactic foreground contamination is noticeably lower in the \pl\ $y$-map than in the \lb\ $y$-map. This can be explained by recognising that, despite possessing a smaller number of frequency channels than \lb,  \pl\ encompasses a broader frequency coverage. Specifically,  \pl\ included channels at $545$\,GHz and $857$\,GHz that facilitate tracking and mitigating the impact of Galactic thermal dust contamination, which dominates the overall variance in the data at $\ell < 20$ (first needlet band). At higher multipoles ($\ell > 20$), other foregrounds begin to contribute significantly to the overall variance in the data. Consequently, the NILC weights are redistributed to mitigate contamination from several components, prioritising the reduction of contamination from other foregrounds while allowing more dust residuals. In this regime, \lb\ demonstrates better performance than \pl\ due to its larger number of frequency channels.

By incorporating both \lb\ and \pl\ frequency maps into the component-separation process, the reconstructed $y$-map effectively leverages the distinct advantages inherent in each experiment. As illustrated in the bottom panel of \cref{fig:ps_comb}, the noise power spectrum of the \lb-\pl\ combined $y$-map aligns with that of \lb\ for multipoles $\ell < 1000$, beyond which it aligns with \pl\ for those high multipoles where \lb's spatial resolution is limited. The residual foreground contamination, both Galactic and extragalactic, is minimised in the \lb-\pl\ $y$-map across all multipoles. This reduction results from the combination of the maximum number of channels and the broadest achievable frequency coverage through this integrated approach.

\subsection{Other SZ statistics and residuals}\label{subsec:pdfs} 

\begin{figure}[tbp]
\centering 
\includegraphics[width=0.5\textwidth,clip]{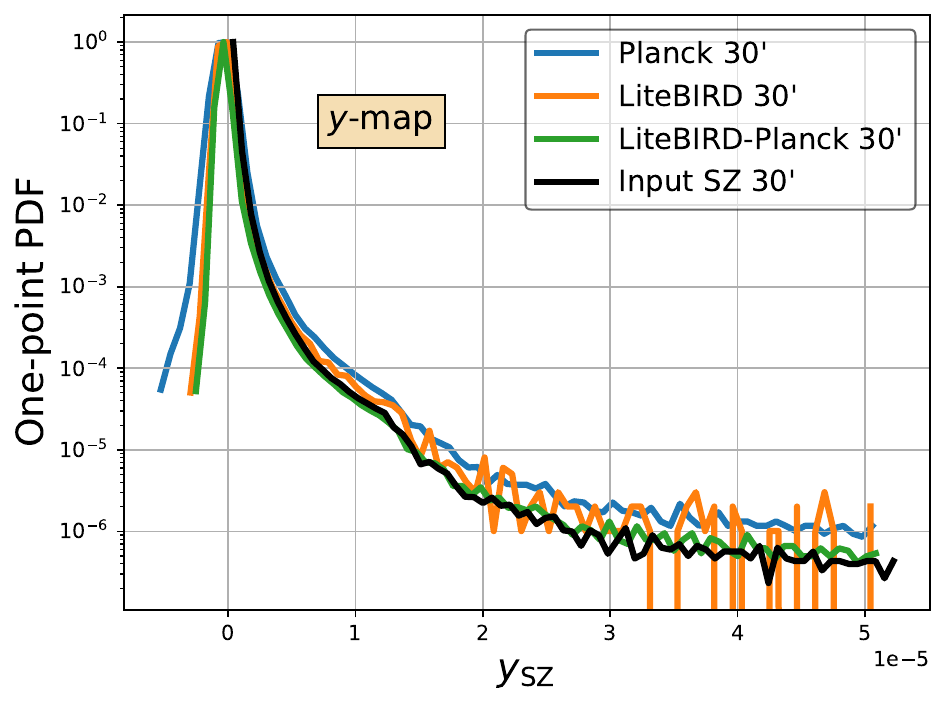}~
\includegraphics[width=0.5\textwidth,clip]{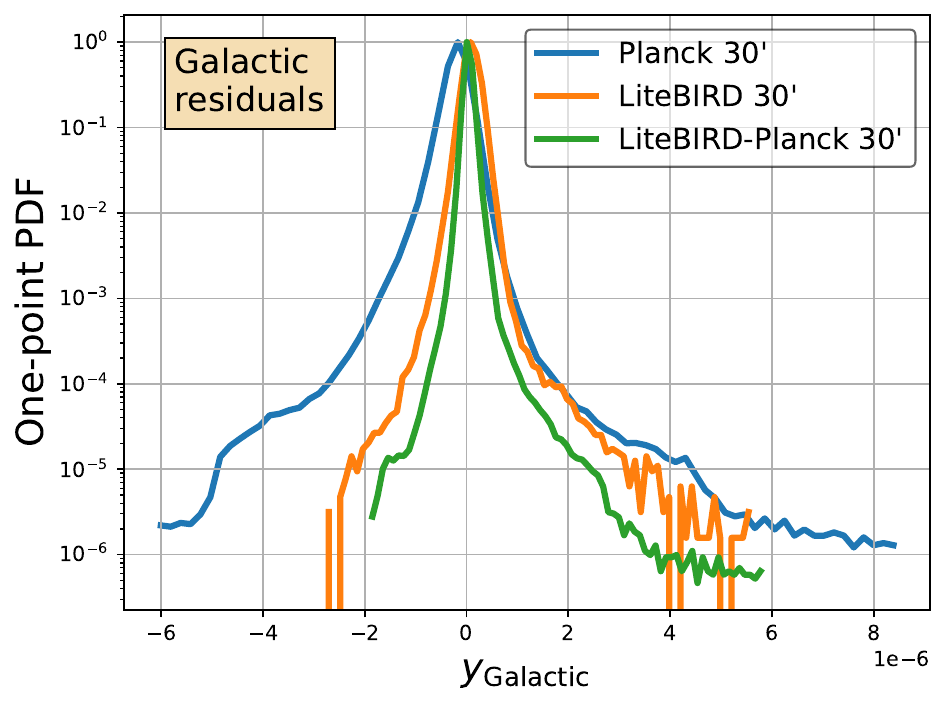}
\hfill
\includegraphics[width=0.5\textwidth,clip]{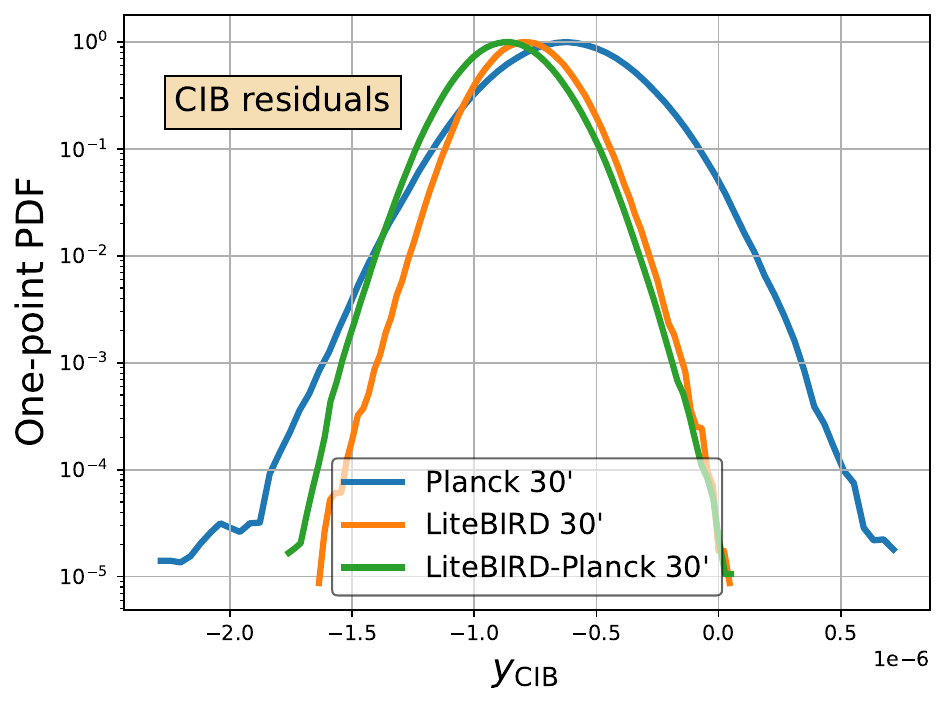}~
\includegraphics[width=0.5\textwidth,clip]{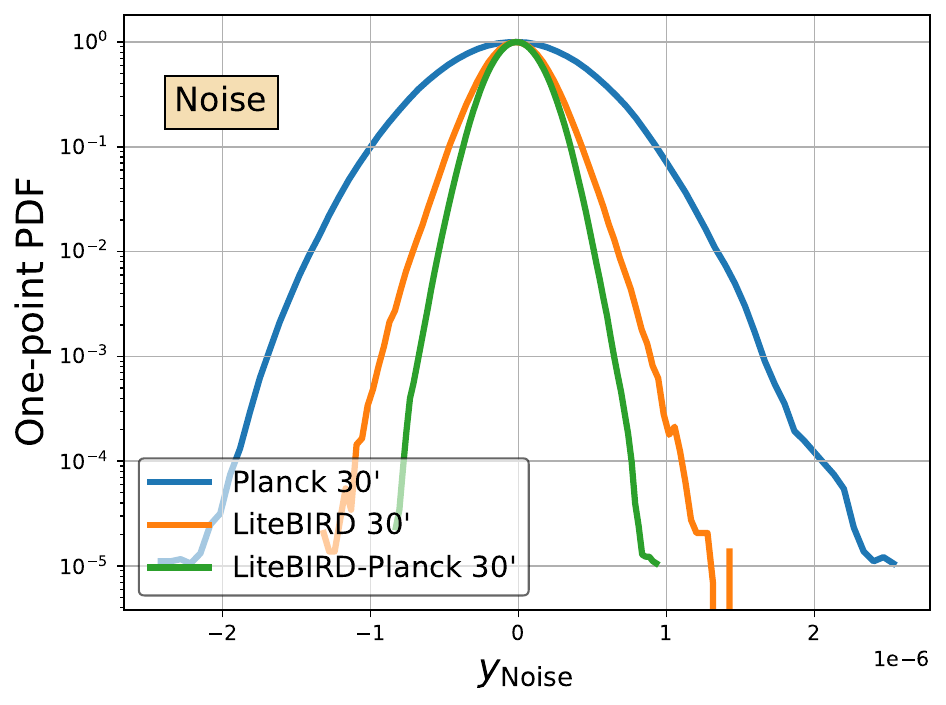}
\caption{\label{fig:onepdf} \emph{Top left}: One-point statistics of the reconstructed NILC $y$-maps over $67\,\%$ of the sky (coloured lines) versus that of the input $y$-map (black line) for \pl\ (blue), \lb\ (orange) and \lb-\pl\ combined (green). \emph{Top right}: One-point statistics of residual Galactic foregrounds for the \pl, \lb\ and \lb-\pl\ combined $y$-maps. \emph{Bottom left}: One-point statistics of residual CIB for the \pl, \lb\ and \lb-\pl\ combined $y$-maps. \emph{Bottom right}: One-point statistics of residual noise for the \pl, \lb\ and \lb-\pl\ combined $y$-maps. 
}
\end{figure}

In \cref{fig:onepdf}, we have computed and displayed the one-point probability density function (PDF) for the reconstructed $y$-maps (top left) as well as their residuals, including Galactic foregrounds (top right), CIB (bottom left), and noise (bottom right) over $67\,\%$ of the sky. All maps have been smoothed down to the same $30'$ angular resolution to allow comparison between histograms. The characteristic positively-skewed distribution of the thermal SZ effect (black line) \citep{Rubino-Martin2003} is observed consistently in all the reconstructed $y$-maps (coloured lines). Notably, the \pl\ $y$-map PDF (blue) displays a larger overall variance when compared to the PDFs of the \lb\ (orange) and \lb-\pl\ (green) $y$-maps. This difference arises from the increased noise and foreground contamination present in the \pl\ $y$-map, as highlighted in the other panels of \cref{fig:onepdf}. As a result of the progressive increase of the number of channels and sensitivity, \lb\ (orange) exhibits narrower distributions in comparison to \pl\ (blue) concerning residual Galactic foreground (top right), CIB (bottom left), and noise (bottom right) contamination. Furthermore, the \lb-\pl\ combination (green) shows even lower foreground and noise contamination levels than \lb, except for the CIB. This is due to \pl's $857$-GHz channel, which is known to increase CIB contamination in the $y$-map and for this reason was discarded from the \pl\ PR2 data analysis at high multipoles for the production of the \pl\ SZ $y$-map (see section 3.1 of ref.~\cite{planck2014-a28} for more details). 

The outcomes of this one-point statistics analysis reinforce the findings from the power spectrum analysis. Once more, they highlight how \lb\ improves over \pl\ in terms of residual Galactic foreground contamination, CIB contamination and noise, and how the combination of \lb\ and \pl\ allows us to even further reduce the residual contamination in the Compton $y$-map.

Moreover, cosmological parameters, in particular $\sigma_8$, can be extracted from a detailed fit to this $y$-map PDF, following the formalism described and used in refs.~\citep{2014arXiv1411.8004H, 2018JCAP...04..019M}. A detailed analysis of the full shape of this PDF is left for a future work. However, one can estimate the relative improvement in the determination of the $\sigma_8$ parameter using the (unnormalised) skewness of the $y$-map. According to ref.~\citep{2013PhRvD..87b3527H} and references therein, the unnormalised skewness scales with $\sigma_8$ as $\langle y^3 \rangle \propto \sigma_8^{9.7\rm{-}11.5}$. By evaluating the skewness in the same analysis mask ($67\,\%$ of the sky), we find that the improvement factor in the relative bias on the determination of $\sigma_8$ is 28\,\%, when comparing the case \pl\ with \lb-\pl, with very little dependence on the choice of the reference scaling exponent within the range ${9.7\rm{-}11.5}$.

\subsection{Impact of \texorpdfstring{$1/f$}{1/f} noise}\label{subsec:noise1overf} 

Although low-frequency correlated $1/f$ noise along the scan direction can be effectively mitigated in polarization maps through the continuously rotating half-wave plates (HWPs) on the \lb\ telescopes \citep{Sakurai2018,KomatsuK2018,Sakurai2020}, it will still impact intensity maps to a larger extent, which requires careful consideration for our purpose.

Here, we repeat our previous analysis on \lb\ simulations, but with the introduction of $1/f$ noise on top of the current simulations, and we assess the impact of $1/f$ noise on the reconstruction of the thermal SZ effect. We consider two scenarios when simulating \lb\ $1/f$ noise, a realistic one with a  knee frequency of ${f_{\rm knee}=30}$\,mHz and a pessimistic one with ${f_{\rm knee}=100}$\,mHz. To be even more conservative, we do not apply any destriping algorithm on the intensity maps.

\begin{figure}[tbp]
\centering 
\includegraphics[width=\textwidth,clip]{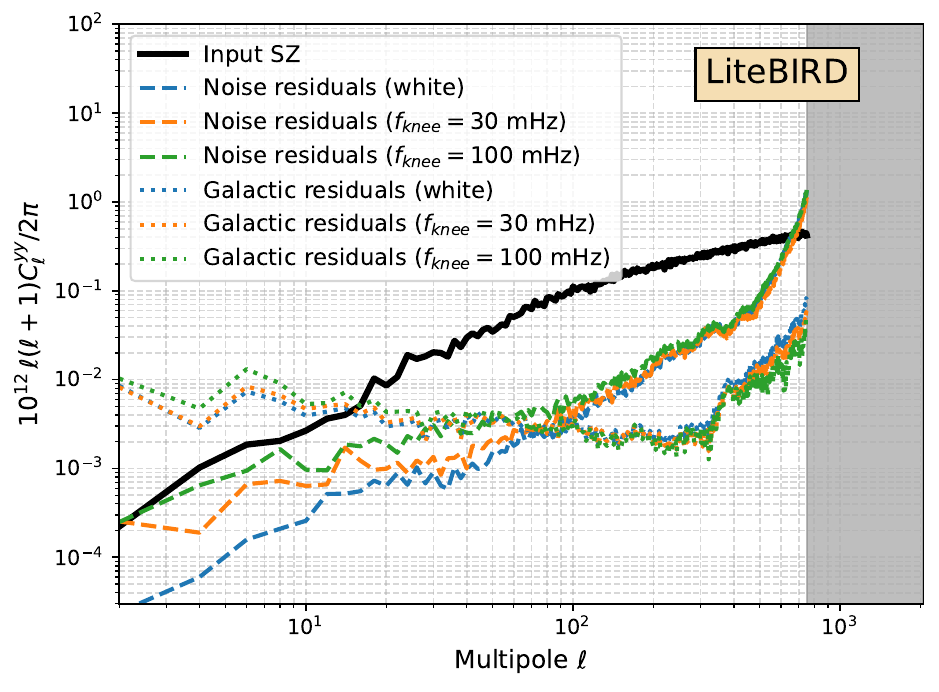}
\caption{\label{fig:1f} Impact of $1/f$ noise on the \lb\ SZ $y$-map power spectrum without destriping.  The power spectra of residual noise contamination (\emph{dashed}) and residual Galactic foreground contamination (\emph{dotted}) in the $y$-map, following component separation, are compared to the input $y$-map power spectrum (\emph{solid black}). Results are presented for white noise (\emph{blue}), white noise + $1/f$ noise with $f_{\rm knee}=30$\,mHz (\emph{orange}), and white noise + $1/f$ noise with $f_{\rm knee}=100$\,mHz (\emph{green}). 
}
\end{figure}

Our results are summarised in \cref{fig:1f}, showing the impact of \lb\ $1/f$ noise on the reconstructed $y$-map power spectrum at low multipoles, for knee-frequencies ${f_{\rm knee}=30}$\,mHz (orange dashed) and ${f_{\rm knee}=100}$\,mHz (green dashed). The ideal scenario with white noise is also shown for comparison (blue dashed). The presence of $1/f$ noise in \lb\ intensity maps leads to a noticeable increase of noise contamination (dashed lines) in the reconstructed $y$-map power spectrum at low multipoles $\ell < 50$.
Within the multipole range $\ell = 2\rm{-}20$, the average noise amplification factor is $3.4$ for ${f_{\rm knee}=30}$ mHz (orange dashed) and $5.3$ for ${f_{\rm knee}=100}$ mHz (green dashed), compared to the white-noise scenario (blue dashed). 
The additional variance introduced in the intensity maps by $1/f$ noise presents a minor challenge for mitigating Galactic foreground contamination at lower multipoles using NILC. This is evident in \cref{fig:1f}, where the inclusion of $1/f$ noise induces a slight increase of residual Galactic foreground contamination (dotted orange/green lines) compared to the white-noise situation (dotted blue line) at multipoles $\ell < 20$. 
As a result, this leads to a slight increase of power in the estimated thermal SZ signal at $\ell < 20$.

Nevertheless, even without destriping, the impact of $1/f$ noise on the reconstructed \lb\ $y$-map at low multipoles remains relatively insignificant. Indeed, the noise contamination (dashed lines) is still effectively mitigated by NILC, remaining well below the thermal SZ signal (solid black) at all multipoles. Furthermore, the marginal increase in power at $\ell < 20$ in the reconstructed SZ signal due to residual Galactic contamination (dotted lines) remains within the cosmic variance limits, which are notably large at these multipoles due to the non-Gaussian contribution from the SZ trispectrum \citep{Komatsu2002,Horowitz2017,Bolliet2018}.

It is worth noting that the presence of $1/f$ noise, although minor in the context of thermal SZ signal reconstruction with \lb, could still have implications for diffuse SZ science at large angular scales because the higher noise variance at low multipoles inevitably results in increased residual Galactic foreground contamination in the $y$-map. For instance, in the search for the two-halo term contribution to the SZ power spectrum at low multipoles, a strategy proposed by ref.~\citep{Rotti2021} involves masking the resolved clusters in the $y$-map, leaving only the diffuse SZ signal. This approach aims to eliminate non-Gaussian contributions to the SZ signal and reduce cosmic variance at low multipoles, a strategy that is also relevant to increase the detection significance of the expected cross-correlation between the thermal SZ and integrated Sachs-Wolfe (ISW) effects \citep{Taburet2011}. Within this approach, it would be valuable to assess the impact of $1/f$ noise and residual foregrounds on \lb's ability to measure the two-halo term contribution in the thermal SZ power spectrum at low multipoles or detect the ISW-SZ cross-correlation.  However, this is a topic beyond the scope of our present study and is left for future investigations.

\subsection{Cosmological parameter constraints}\label{subsec:cosmo} 

 We now evaluate how the residual foreground and noise contamination of the reconstructed $y$-maps from either \pl, \lb\ or the joint \lb-\pl\ data sets impacts the recovered uncertainties on cosmological parameters.

As the input $y$-map of the simulation incorporates real clusters alongside simulated ones (see \cref{subsubsec:tsz}), the actual scaling of the input $y$-map power spectrum with cosmological parameters deviates from the anticipated scalings of a pure  (i.e. without additional real clusters) SZ simulation. Consequently, in this section, we substituted the power spectrum of the input $y$-map with that of a theoretical thermal SZ model based on \pl\ 2018 best-fit cosmology, but we supplemented it with the actual residual foreground and noise power spectra obtained from NILC for each data set. Given that our focus is on uncertainties rather than central parameter values, such a substitution of the thermal SZ signal does not alter the overall conclusion regarding the relative performance of each data set in terms of cosmological parameter uncertainties, as long as residual foregrounds and noise emanate from the actual component-separation output.

\begin{figure}[tbp]
\centering 
\includegraphics[width=\textwidth,clip]{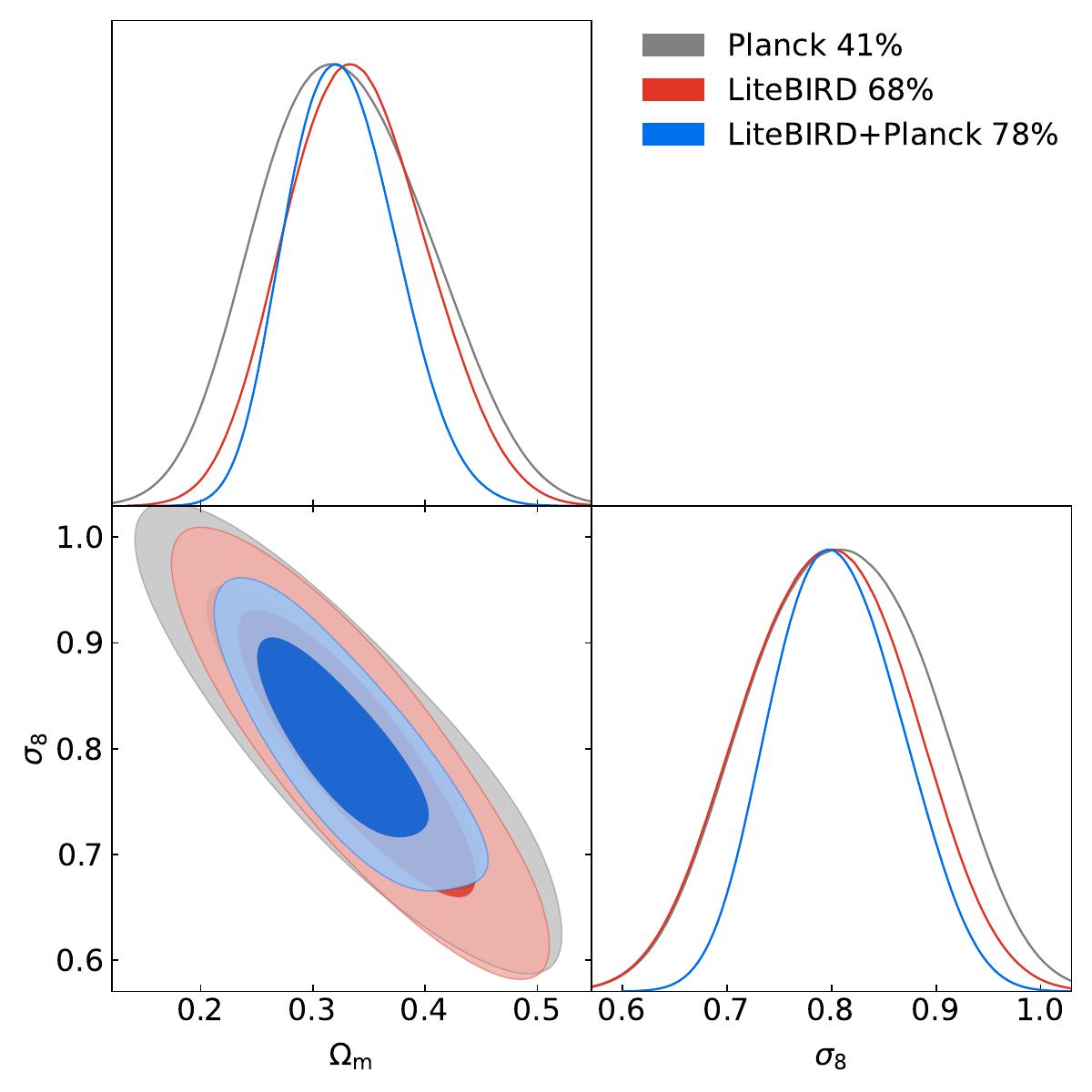}
\caption{\label{fig:triangle1} Comparison of cosmological constraints from the \pl, \lb\ and joint \lb-\pl\ $y$-maps. The percentages indicate the respective sky fractions of uncontaminated sky available from each $y$-map.
 }
\end{figure}

The resulting power spectra from the individual \pl, \lb, and combined \lb-\pl\ data sets are passed through an MCMC pipeline to infer cosmological parameters, specifically the matter density ($\Omega_{\rm m}$) and the amplitude of dark matter fluctuations ($\sigma_8$).
Our approach closely follows the procedures outlined in ref.~\cite{planck2014-a28}, with the exception that we utilise the thermal SZ model emulator developed in ref.~\cite{Douspis2022} instead of the full halo model. The emulator computes the thermal SZ angular power spectrum given a set of cosmological parameters ($\Omega_b$, $\Omega_{m}$,$H_0$, $\sigma_8$, $n_s$) and cluster parameters ($B_{\rm SZ}$, the hydrostatic mass bias), to which residual Galactic, CIB and  point-source spectra are added with respective amplitudes ($A_{\rm gal}$, $A_{\rm CIB}$, $A_{\rm PS}$):
\begin{align}
   C_{\ell}^{\rm Model} = C_{\ell}^{\rm SZ}\left(\{\Omega_b, \Omega_{m}, H_0, \sigma_8, n_s, B_{\rm SZ}\}\right) + A_{\rm gal} C_{\ell}^{\rm gal} + A_{\rm CIB} C_{\ell}^{\rm CIB} + A_{\rm PS} C_{\ell}^{\rm PS} + C_{\ell}^{\rm noise}\,.
\end{align}
The resulting spectrum is compared to the "observed" power spectra as estimated from the NILC $y$-maps  (\pl\ alone, \lb\ alone and \lb+\pl) in an MCMC sampler assuming a Gaussian likelihood:
\begin{align}
-2\ln \mathcal{L} \propto \sum_{\ell \leq \ell'} \left(C_{\ell}^{\rm NILC} - C_{\ell}^{\rm Model} \right) M_{\ell\ell'}^{-1} \left(C_{\ell'}^{\rm NILC} - C_{\ell'}^{\rm Model} \right)\,,
\end{align}
where $C_{\ell}^{\rm NILC}$ is the reconstructed $y$-map power spectrum from either \pl, \lb, or \lb-\pl\ data set, and $M_{\ell\ell'}$ is the covariance matrix collecting the errors on the power spectrum.
In addition to the sample variance estimated from the power spectrum of the reconstructed $y$-maps, the covariance matrix includes the non-Gaussian contribution to the error via the thermal SZ trispectrum \citep{Cooray2001}, as described in refs.~\cite{Komatsu2002,Horowitz2017,Bolliet2018}, which is also computed by the emulator.
 Note that this additional noise dominates at large scales ($\ell<1000$) because it is  inversely proportional to the fraction of the sky available. Since the thermal SZ alone cannot constrain all cosmological parameters we consider additional Gaussian priors on $n_{\rm s}$ coming from ref.~\cite{planck2016-l01} ($n_{\rm s}=0.9649\pm0.0042$) and a prior on the hydrostatic mass bias $B_{\rm SZ}=0.65\pm0.1$. We also estimated the error propagation on the uncertainty on the foregrounds model amplitudes (Galactic dust, CIB and point sources) to be of the order of 20\,\% on the residuals and thus consider respective Gaussian priors : $\{A_{\rm gal}, A_{\rm CIB}, A_{\rm PS}\} = 1\pm 0.2$.

In \cref{fig:triangle1}, we present a comparison of cosmological constraints obtained from the \pl, \lb, and combined \lb-\pl\ $y$-maps. It should be noted that the trispectrum contribution to uncertainties was omitted in the \pl\ data analysis \citep{planck2014-a28}, which resulted in a smaller uncertainty on $\sigma_8$ compared to posterior studies, such as refs.~\citep{Bolliet2018,Salvati2018}, and the current analysis on \pl\ simulations.
Thanks to the increased fraction of uncontaminated sky (from $f_{\rm sky}=41\,\%$ to $78\,\%$), the constraint on $S_8 = \sigma_8 \left(\Omega_{\rm m}/0.3\right)^{0.5}$ derived from the \lb-\pl\ combined $y$-map (blue) exhibits a 15\,\% improvement in precision compared to the constraint derived from the \pl\ $y$-map (grey). This translates in an increase on the figure of merit on $\left(\Omega_{\rm m}, \sigma_8\right)$ by 1.46.

\subsection{Thermal SZ effect from patchy reionisation}\label{subsec:reionisation} 

In this section, we consider a possible contribution to the thermal SZ effect that would be induced by inhomogeneous (\emph{patchy}) reionisation. Detecting such a thermal SZ signal would provide valuable insights into the reionisation epoch in the early Universe \citep{Namikawa2021}. To investigate this, Gaussian map realisations representing the thermal SZ emission resulting from patchy reionisation were generated, correlated with realisations of the inhomogeneous optical depth. These maps were then added to our existing \lb\ simulation, in addition to the low-redshift thermal SZ signal from galaxy clusters, and we conducted the component analysis once more to reconstruct the $y$-map.

The map of inhomogeneous optical depth ($\tau$-map) due to patchy reionisation can, in principle, be reconstructed using quadratic estimators on CMB temperature and polarization anisotropies \citep{Namikawa2021}, relying on the high angular resolution and sensitivity of upcoming ground-based telescopes. Here, we simply use the input $\tau$-map of the simulation and assume a reconstruction noise associated with the quadratic estimator, in accordance with expectations for CMB-S4 \citep{Namikawa2021}. Moreover, we take a simplified approach by neglecting low-redshift correlations between the $y$-map and the $\tau$-map in our simulation, allowing us to focus on the signal from patchy reionization and forecast the signal-to-noise ratio for the high-redshift contribution, assuming the low-redshift contribution can be mitigated, for example, through cluster masking. As a by-product, we note that this low-redshift contribution could in principle be detected with a higher signal-to-noise ratio, but this is beyond the scope of this study.

\begin{figure}[tbp]
\centering 
\includegraphics[width=\textwidth,clip]{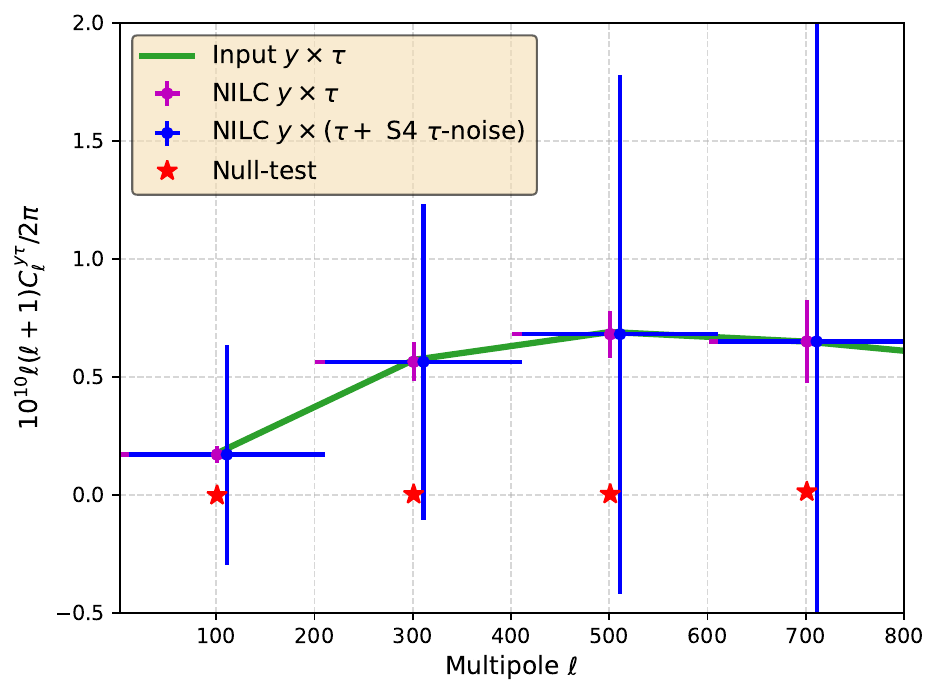}~
\caption{\label{fig:szreio} Cross-power spectrum between the thermal SZ $y$-map and the inhomogeneous optical depth $\tau$-map from patchy reionisation (average over twenty realisations of $\tau$ and $y$). \emph{Green line}: input $y$-map cross input $\tau$-map. \emph{Purple dots}: \lb\ $y$-map cross input $\tau$-map. \emph{Blue dots}: \lb\ $y$-map cross input $\tau$-map with CMB-S4 reconstruction noise. \emph{Red stars}: null-test, i.e. \lb\ $y$-map cross input $\tau$-map when thermal SZ effect from reionisation is turned off in the simulation.}
\end{figure}

\Cref{fig:szreio} presents the cross-power spectrum $C_\ell^{\,y\tau}$ (in blue) between the reconstructed \lb\ $y$-map and the inhomogeneous optical depth map ($\tau$-map) arising from patchy reionisation, averaged over twenty realisations of the reionisation $\tau$ and $y$ fields. Purple error bars, computed from twenty realisations, account for the reconstruction noise in the $y$-map but not in the $\tau$-map. Blue error bars include the expected noise associated with the quadratic-estimator reconstruction of the $\tau$ field from CMB-S4, and in this case were derived analytically using \cref{eq:ytauerror} below.
Our analysis successfully recovers the expected $y\tau$ cross-correlation signal (in green), albeit with relatively large error bars. This result stands out distinctly from the null-test outcome (red stars), where we intentionally excluded the thermal SZ contribution from patchy reionisation in the simulation.

By integrating over all the multipole bins, we measure the $y\tau$ correlation due to patchy reionisation with a signal-to-noise ratio of
\begin{align}
    \frac{S}{N} = \sqrt{\sum_\ell \left(\frac{C_\ell^{\,y\tau}}{\sigma_\ell^{\,y\tau}}\right)^2}\simeq 1.6\,,
\end{align}
where the error on the cross-power spectrum $C_\ell^{\,y\tau}$ is computed analytically in each bin as
\begin{align}
\label{eq:ytauerror}
    \sigma_\ell^{\,y\tau} = \sqrt{\frac{\left(C_\ell^{\,y\tau}\right)^2+C_\ell^{\,yy}\,\left(C_\ell^{\tau\tau}+N_\ell^{\tau\tau}\right)}{\left(2\ell+1\right)\,\Delta\ell\,f_{\rm sky}}}\,,
\end{align}
with $C_\ell^{\,yy}$ and $C_\ell^{\tau\tau}$ being the auto-power spectra of the \lb\ $y$-map (which includes foreground residuals and noise) and the $\tau$-map, respectively, $N_\ell^{\tau\tau}$ the noise power spectrum associated with CMB-S4 reconstruction of the $\tau$ field, $\Delta\ell$ the size of the multipole bins, and $f_{\rm sky} = 0.4$ the sky fraction associated with CMB-S4.

Our results show that \lb\ can provide suggestive evidence of the thermal SZ effect resulting from patchy reionisation, after foreground cleaning with NILC. Moreover, the results show that the signal-to-noise ratio for this detection will only be limited by the resulting noise in the $\tau$-map that can be achieved with a high-resolution ground-based experiment such as CMB-S4 (see blue versus purple error bars in \cref{fig:szreio}).

\section{Conclusions}\label{sec:conclusion}

Our study has thoroughly evaluated the capabilities of the \lb\ mission in mapping the distribution of hot gas in the Universe through the thermal SZ effect. Utilising comprehensive sky simulations that accounted for various sources of Galactic and extragalactic foreground emission, alongside specific instrumental characteristics of the \lb\ mission, such as inhomogeneous sky scanning and $1/f$ noise, and a dedicated component-separation pipeline, we successfully mapped the thermal SZ Compton $y$-parameter across 98\,\% of the sky.

Despite \lb's lower angular resolution for galaxy cluster science, our analysis highlighted the mission's key strengths, including full-sky coverage, enhanced sensitivity, and additional frequency bands as compared to \pl. This enabled reconstructing an all-sky thermal SZ Compton $y$-map from \lb\ channels, demonstrating reduced foreground contamination at large and intermediate angular scales compared to the \pl\ $y$-map. The $1/f$ noise from \lb\ temperature channels was shown to be mitigated below the level of the thermal SZ signal at all multipoles after component separation. Combining \lb\ and \pl\ channels in the component-separation pipeline further yielded an optimal Compton $y$-map, capitalising on the advantages of both experiments, with the higher angular resolution of \pl\ channels enabling the recovery of compact clusters beyond \lb's beam limitations, while the numerous sensitive \lb\ channels effectively mitigate foregrounds.

The added value of \lb\ was evident in the examination of maps, power spectra, and one-point statistics of various sky components, as well as in the improved uncertainties on cosmological parameters. Notably, the cosmological constraint on $S_8 = \sigma_8 \left(\Omega_{\rm m}/0.3\right)^{0.5}$ obtained from the combined \lb-\pl\ $y$-map power spectrum demonstrates a noteworthy 15\,\% reduction in uncertainty compared to constraints derived from \pl\ alone. 

Furthermore, in the context of patchy reionisation in the early Universe, \lb\ is identified as a valuable mission for providing preliminary evidence of the faint thermal SZ effect induced during the reionisation epoch. This could be achieved with a modest signal-to-noise ratio of 1.6 through the cross-power spectrum of the \lb\ $y$-map with the inhomogeneous optical depth map obtained from CMB-S4.

Several additional effects could possibly be detected with the improved all-sky map of the thermal SZ Compton $y$-parameter from \lb, thanks to reduced foreground and noise contamination at large and intermediate scales:
The quadrupole-like thermal SZ effect caused by structures in the local Universe, such as the Milky Way and the local supercluster \citep[e.g.,][]{Abramo2003,Dolag2005};
The dipole anisotropy of the thermal SZ effect induced by the motion of the solar system with respect to the CMB frame \citep[e.g.,][]{Chluba2005};
The dipole-modulated CMB anisotropies that have a thermal SZ-like SED, which could be detected through cross-correlation between the $y$-map and the CMB map with higher significance than previously reported \citep{planck2020-LVI};
The expected cross-correlation between thermal SZ and ISW effects \citep[e.g.,][]{Taburet2011};
The excess power at low multipoles from the two-halo contribution to the diffuse SZ emission \citep[e.g.,][]{Rotti2021}.
We envision exploring these effects in future work.

In summary, our study showcases the anticipated significant impact of \lb\ on SZ science, highlighting its valuable potential for advancing our understanding of the large-scale structure of the Universe.

\acknowledgments
The authors thank Jean-Baptiste Melin for useful discussions on SZ simulations.
%
This work is supported in Japan by ISAS/JAXA for Pre-Phase A2 studies, by the acceleration program of JAXA research and development directorate, by the World Premier International Research Center Initiative (WPI) of MEXT, by the JSPS Core-to-Core Program of A. Advanced Research Networks, and by JSPS KAKENHI Grant Numbers JP15H05891, JP17H01115, and JP17H01125.
The Canadian contribution is supported by the Canadian Space Agency.
The French \textit{LiteBIRD} phase A contribution is supported by the Centre National d’Etudes Spatiale (CNES), by the Centre National de la Recherche Scientifique (CNRS), and by the Commissariat à l’Energie Atomique (CEA).
The German participation in \textit{LiteBIRD} is supported in part by the Excellence Cluster ORIGINS, which is funded by the Deutsche Forschungsgemeinschaft (DFG, German Research Foundation) under Germany’s Excellence Strategy (Grant No.~EXC-2094 - 390783311).
The Italian \textit{LiteBIRD} phase A contribution is supported by the Italian Space Agency (ASI Grants No.~2020-9-HH.0 and 2016-24-H.1-2018), the National Institute for Nuclear Physics (INFN) and the National Institute for Astrophysics (INAF).
Norwegian participation in \textit{LiteBIRD} is supported by the Research Council of Norway (Grant No.~263011) and has received funding from the European Research Council (ERC) under the Horizon 2020 Research and Innovation Programme (Grant agreement No.~772253 and 819478).
The Spanish \textit{LiteBIRD} phase A contribution is supported by MCIN/AEI/10.13039/501100011033, project refs. PID2019-110610RB-C21, PID2020-120514GB-I00, PID2022-139223OB-C21 (funded also by European Union NextGenerationEU/PRTR), and by MCIN/CDTI ICTP20210008 (funded also by EU FEDER funds).
Funds that support contributions from Sweden come from the Swedish National Space Agency (SNSA/Rymdstyrelsen) and the Swedish Research Council (Reg.~no.~2019-03959).
The UK  \textit{LiteBIRD} contribution is supported by the UK Space Agency under grant reference ST/Y006003/1 -- "LiteBIRD UK: A major UK contribution to the LiteBIRD mission -- Phase1 (March 25)."
The US contribution is supported by NASA grant no.~80NSSC18K0132.
%
%
We also acknowledge the support of the Spanish Ministry of Science and Innovation through grants PID2022-140670NA-I00 and PID2021-126616NB-I00. 

\bibliographystyle{JHEP}
\bibliography{main,Planck_bib}

\end{document}